\documentclass[epsfig]{article}
\usepackage{graphicx}

\newcommand{\refcompcomp}{2.1}

\begin{document}

\title{\bf {\sc tlusty} and {\sc synspec} Users's Guide IV:
Upgraded Versions 208 and 54}
\author{I. Hubeny\footnote{University of Arizona, 
Tucson; USA; ihubeny.astr{\tt @}gmail.com}, 
C. Allende Prieto\footnote{IAC, La Laguna, Tenerife, Spain; callende{\tt @}iac.es}, 
Y. Osorio\footnote{IAC, La Laguna, Tenerife, Spain; yeisson{\tt @}iac.es},
and T. Lanz\footnote{Observatoire de Cote d'Azur, Nice, France; thierry.lanz@oca.eu}}
\date{\today}
\maketitle

\begin{abstract}
We present a brief description of the newly upgraded versions of the programs
{\sc tlusty}, version 208, and {\sc synspec}, version 54. {\sc tlusty} is
used to generate model stellar atmospheres or accretion disks, and  
{\sc synspec} produces detailed synthetic spectra and/or opacity tables. This paper complements 
published guides that provide a detailed description of earlier
versions of the codes, {\sc tlusty205}, and {\sc synspec51}. The main 
upgrades include the flexible construction of opacity tables in {\sc synspec}, and their
use in producing hybrid models with {\sc tlusty}, with important species treated in NLTE, 
while the bulk of opacity of atomic and molecular lines and continua are considered
in LTE using a pre-calculated opacity table.

There is also a number of additional changes and upgrades that increase the versatility
and flexibility of these codes.
\end{abstract}

\tableofcontents

\newpage

\bigskip
\bigskip

\section{Introduction}
The programs {\sc tlusty}, version 205, and {\sc synspec}, version 51, 
were described comprehensively in
a series of three papers by Hubeny \& Lanz (2017a, b, c; hereafter referred to as
Paper I, II, and III, respectively). In this paper, we document recent
changes in these codes. 
The user is strongly encouraged to consult the previous parts of the Guide, at least Paper~I,
since without them the present description cannot be fully understood.

\section{TLUSTY}

There are several basic differences between the newly developed
version 208 and the previously released version 205:
\begin{itemize}
\item A modified and upgraded treatment of opacities and chemical abundances.
\item Upgrades for working with opacity tables.
and related changes in the treatment of explicit atoms, levels, and transitions.
\item An update of the molecular equation of state, and related changes in
the format of the model output.
\item An improved treatment of convection.
\item A new and upgraded treatment of collisional rates.
\item An update of the Rybicki scheme for solving the linearized structural equations,
which now works for accretion disks, and even works well for NLTE models.
\item A ``fixed-density" option, useful for solving NLTE line formation and
ionization balance for snapshots of independent hydrodynamical calculations.
\end{itemize}
These changes are most relevant for computing model atmospheres of cool stars,
while the last change may be applicable to all types of stellar atmospheres or
accretion disks.

In addition, there are various other small changes of the layout of the standard output,
which we will briefly describe in \S\,\ref{outchan}.

%------------------------------------------------------------

\subsection{Outline of opacities and chemical abundances}
\label{outl}

There are three classes of opacity sources considered
by {\sc tlusty}, namely:
\begin{description}
\item[Explicit opacities] -- bound-free and bound-bound
transitions between explicit levels, and free-free transitions of explicit ions.
They are computed on the fly, and these are the only ones that can be considered
in NLTE.
\item[Additional opacities] -- any other opacity source that is computed on the fly.
This option existed from the very early versions of {\sc tlusty}, but has been upgraded
and made less error-prone in the present version. This is described below.
\item[Opacity from a pre-calculated table]  -- This option was introduced 
in version 205, and has been upgraded in the present one, as described in this
document.
\end{description}
Historically, the additional opacities, in the above meaning, were the following:
H${}^-$ bound-free and free-free opacity, H${}_2^+$ bound-free and free-free opacity, 
hydrogen Rayleigh scattering opacity, plus an outdated option for including bound-free
opacity of additional, higher, levels of He~I and He~II. The corresponding switches.
activated in the non-standard (keyword) parameter file,
are called IOPHMI, IOPH2P, IRSCT, IOPHE1, and IOPHE2. The default values of
these switches were set to 0, so these sources were not considered unless specifically
required by setting the corresponding switch to a non-zero value, typically 1. 

\begin{table}
\caption{Additional opacities - keywords and their values}
\begin{center}
\begin{tabular}[h]{|l|c|l|c|c|}
\hline
\hline
Opacity source &  Ref & Keyword & Default  & Set automatically to 0 if \\
\hline
\hline
H- b-f and f-f & a & IOPHMI & 1 & H- is already an explicit ion \\
\hline
 H${}_2^+$ b-f and f-f & a  & IOPH2P &  1  & IFMOL=0 or $T >$ TMOLIM \\
\hline
 H${}_2^-$ f-f & b & IOPH2M &   1  & IFMOL=0 or $T >$ TMOLIM\\
 \hline
He- b-f and f-f & a & IOPHEM &  1  &  \\  
\hline 
CIA  H${}_2$-H${}_2$  & c & IOH2H2 & 1 &  IFMOL=0 or $T >$ TMOLIM\\
\hline 
CIA  H${}_2$-He  & e &  IOH2HE &  1  &  IFMOL=0 or $T >$ TMOLIM\\
\hline 
CIA  H${}_2$-H   & f  &  IOH2H &  1  &  IFMOL=0 or $T >$ TMOLIM\\
\hline 
CIA  H-He   & d &  IOHHE &  1  &  IFMOL=0 or $T >$ TMOLIM\\
 \hline
CH opacity & a & IOPCH  &  1 & IFMOL=0 or $T >$ TMOLIM \\
\hline
OH opacity & a & IOPOH  &  1 & IFMOL=0 or $T >$ TMOLIM \\
\hline
\hline
H Rayleigh scat. & a & IRSCT & 1 & \\
\hline
He Rayleigh scat. & a & IRSCHE & 1 & \\
\hline
H${}_2$ Rayleigh scat. & a & IRSCH2 & 1 & IFMOL=0 or $T >$ TMOLIM \\
\hline
\end{tabular}
\\[2.5pt]
{\small
References: (a)~Kurucz (1970); (b)~Bell (1980); (c)~Borysow et al. (2001);
(d)~Gustafsson \& Fromhold (2001); (e)~Jorgensen et al. (2000);
(f)~Gustafsson \& Fromhold (2003).
}
\end{center}
\end{table}

We have now deleted the outdated switches IOPHE1 and IOPHE2, and added 
more additional opacity sources. The new opacity sources include the H${}_2^-$ free-free opacity,
and four different {\em collision-induced opacity} (CIA) sources, namely those for the
H${}_2$--H${}_2$, H${}_2$--He, H${}_2$--H, and H--He collisions -- see Table~1.
{\em The default values of all
of them are set  to 1, that is, all sources are considered unless specifically excluded by setting
the corresponding switch to 0}. In this way, one avoids accidentally omitting a potentially
important opacity source, e.g.  the H${}_2^+$ or Rayleigh scattering opacity for cool stars.\footnote{This
was even done in some test cases presented in Paper~III, where Rayleigh 
scattering was omitted despite it provides a non-negligible contribution to the total opacity.}

Another potential source of error was the treatment of the H${}^-$ opacity. If one sets 
IOPHMI=1, and at the same time H${}^-$ is specified as an explicit ion, its opacity is
included twice. It was entirely up to the user to check the internal consistency of
the input parameters on this regrad. The situation when using opacity tables is even more delicate,
in the sense that some of these additional sources may have already been considered there.

Therefore, we have decided to make the overall treatment of additional opacities
much less error-prone by letting {\sc tlusty} check and automatically correct possible inconsistencies.
Table 1 summarizes the relevant switches and corrective actions.\\[-4pt]

\noindent
{\bf Important note:} In all cases, a particular keyword is reset to 0 if the given opacity source 
was already considered in the opacity table, unless a new keyword parameter
KEEPOP is set to a nonzero value. This option is included for pedagogical purposes,
or for investigating what would be the error introduced in a model by considering 
a particular opacity source twice.

We also stress that when {\sc synspec} works in the table-generating mode, the
additional Rayleigh scattering opacity sources are disabled, because the resulting
opacity table should only consider the true absorptive opacity.

\smallskip

We have also increased a flexibility in specifying the chemical abundances of elements.
As explained in Paper~III, \S\.4.2, the chemical abundances are specified in the second 
block of the standard input to {\sc tlusty} or {\sc synspec}, namely through the parameter
ABN. The second block of the standard input has now the following structure:

The first line contains one or two numbers:
\begin{description}
\item[NATOMS] -- as before, the highest atomic number of an element that is
considered (explicitly or non-explicitly).
\item[IABSET] -- if set, it specifies the adopted set of the {\em standard solar abundances} to be used.
If it is not set, the program will assign it to 0. The individual sets are the following:\\
$\bullet\,=0$ -- the adopted set of the standard solar abundances is that of Asplund et al. (2005),
which is essentially the same as that of Grevesse et al. (2007) used in the recent large MARCS 
(Gustafsson et al. 2008) and Kurucz (Meszaros et al. 2012) model atmosphere grids.\\
$\bullet\,=1$ -- the adopted set of the standard solar abundances is a combination of 
meteoritic/photospheric abundances from Asplund et al. (2009) adopted by Husser et al. (2013) 
in their extensive grid of PHOENIX model atmospheres and synthetic spectra.\\
$\bullet\,=2$ -- the adopted set of the standard solar abundances is that of Grevesse \& Sauval (1998),
which was the default in the previous versions of {\sc tlusty} and {\sc synspec}.\\
$\bullet\,$ -- any other number: IABSET is reset to 0.
\end{description}
Next, there are NATOMS records, each containing three items: MODE, ABN, and MODPF. The parameter
MODE will be described in \S\,\ref{basic}, and MODPF is now outdated, so it can be set to 0. Here
we describe the parameter ABN. Its meaning is essentially the same as in the previous versions,
the only difference is that instead of dealing with a hardwired set of solar abundances, the parameter
ABN now {\em relates the intended abundances to the adopted \em standard solar abundances}, 
specified through the parameter IABSET.
\begin{description}
\item[ABN]  a specification of the abundance of the given species:\\
$\bullet\,=0$ -- the {\em standard solar abundances} is assumed;\\
$\bullet\,<$ 0 -- a non-standard abundance is assumed, abs(ABN) has now the meaning of
         the abundance expressed as a multiple of the standard
         solar abundance (i.e. $-0.1$ means 1/10 of the standard solar abundance, $-5$ means 5 times
         the standard solar abundance, etc.);\\
$\bullet\,>$ 0 -- a non-standard abundance is assumed, expressed as 
         $N({\rm elem})/N({\rm ref})$, i.e. relative by number
	 to the reference species.  The reference atom is H by default,
	 but the reference element can be changed by means of the
	 optional parameter IATREF (see Paper~III, \S\,7.4.4).\\
$\bullet\,>10^6$ -- non-homogeneous (depth-dependent) abundance is assumed. In this
  case, the immediately following $N\!D$ lines should be added that
  contain the individual values of the
  abundance (relative to hydrogen by number), for all depth points 
  $d=1,\ldots,N\!D$.
\end{description}
For convenience, we list the adopted values of the chemical abundances for all the three
standard solar abundances sets in Appendix A. Although using more recent sets 0 and 1 of
Asplund et al. (2005, 2009), and Grevesse et al. (2007) is preferable, we include the old
set of Grevesse \& Sauval (1998) for comparison with earlier results.

%------------------------------------------------------------

\subsection{Working with opacity tables}

\subsubsection{Basic concepts}
\label{basic}
In version 205, one could either use an opacity table describing the total opacity,
or not use an opacity table at all. 
The present version introduces a third option,
already mentioned in Papers~I and III, of working with partial
opacity tables. This option consists in using an opacity table that contains opacities
for a subset of chemical species, while the opacity of other species is considered
explicitly, typically in NLTE (they can be considered in LTE, but this possibility does not
offer any advantages, since in such a case all species could have been included in
the opacity table). An example of using a partial opacity table is, for instance, 
the construction of a solar-type model atmosphere with 
H, Mg, and Ca in NLTE, while including the opacity from all other species, atomic as well
as molecular, in the opacity table. Such a table will be referred to as a {\em partial opacity table}. 
We call this approach to compute model atmospheres a {\em hybrid opacity-table/NLTE} (OT/NLTE) approach.

The basic control keyword parameter is IOPTAB, already described in Paper~III \S\, 7.4.1.
For convenience, we repeat its description here:
\begin{description}
\item[IOPTAB] -- a switch that specifies the use of pre-calculated 
\index{IOPTAB keyword parameter}
opacity tables:\\
$\bullet\,= 0$ -- classical option -- no pre-calculated opacity table is used;\\
$\bullet\,< 0$ -- all opacities are included by means of a pre-calculated opacity
table. In this case, one does not specify any explicit atoms, ions, and levels.
Only LTE models can be computed with this option. There are two possibilities here:
\begin{itemize}
\item =-1 -- a full opacity table is used, but the equation of state and thermodynamic parameters are
computed on the fly.\\
\item =-2 -- in addition to the opacity, the equation of state and thermodynamic parameters
are given by pre-calculated tables.
\end{itemize}
$\bullet\,>$ 0 -- hybrid option, using a partial opacity table -- one still selects 
explicit atoms, ions, and levels, for
which the opacity is computed on the fly, while for the remaining species one
uses an appropriate  pre-calculated opacity table.
This option was not operational in version 205, but it is in the present version 208.\\
DEFAULT: IOPTAB=0 
\item[IFRAYL] -- this parameter takes effect only when using opacity tables
in the {\em full opacity table approach}, that is if IOPTAB $<0$. Since all the (thermal) opacity 
is included in the table, and opacity is not calculated on the fly, one needs to make sure that the
scattering opacity is included as well. 
Possible values of IFRAYL are:\\
$\bullet\,>0$ -- the Rayleigh scattering opacity is computed on the fly, based on the values
of the Rayleigh scattering parameters IRSCT, IRSCHE, IRCSH2 for the scattering on H, He, and H${}_2$,
respectively.
Since these parameters have default values of 1 (the source is included), the Rayleigh scattering
opacity is included by default unless it is explicitly rejected.\\
$\bullet\,<0$ -- the Rayleigh scattering opacity is read from a special table {\tt rayleigh.tab}.
[This option is implemented, but the table {\tt rayleigh.tab} is not set up, so this option cannot
be used at the present time.]\\
$\bullet\,=0$ --  Rayleigh scattering is neglected completely (which is wrong, but the option may be 
used to assess the importance of this scattering mechanism).\\
DEFAULT: IFRAYL=1
\end{description}

There are some subtle points here. In the case of a full opacity table, the frequency
points selected for the run of {\sc tlusty} are either the same frequency points as those used
in the table, or a different set but chosen to be within the range covered by the table.
Let $\nu_{\rm min}$ and $\nu_{\rm max}$ be the lowest and highest frequencies in
the table. In the hybrid option, one may consider frequencies that are higher
than $\nu_{\rm max}$. For example, an opacity table used for solar-type stars has typically
the maximum frequency around the  Lyman limit,
$\nu_{\rm max} \approx 3.28\times 10^{15}$s${}^{-1}$, while the ionization 
frequency of some
ions treated explicitly may be higher. For instance, when magnesium is treated explicitly,,
one would include Mg~I, Mg~II, and the ground state of Mg III. The ionization frequency
for the ground state of Mg~II is $\approx 3.6\times 10^{15}\,$s${}^{-1}$, so in order to
evaluate the radiative rate for the Mg~II ionization, needed for NLTE calculations of the Mg level
populations,
one needs to specify the background opacity going at least up to $\approx 7\times 10^{15}$s${}^{-1}$.

To this end, we have introduced two new options\footnote{Plus two more options, MODE=4 and 5, specific to 
{\sc synspec } -- see \S\,\ref{semiex} and \ref{quasex}} for treating the chemical species, controlled
by the parameter MODE, which is contained in the block of data for chemical 
elements of the standard input, described in Paper~III, \S\,4.2. Here is its 
description:
\begin{description}
\item[MODE] -- a specification of the mode of treatment of a given species:\\
$\bullet\,= 0$  -- the element is not considered at all (is completely ignored).\\
$\bullet\, \not= 0$  -- regardless of the actual value,  the element is allowed to
participate in the equation of state, namely to contribute to the
total number of particles, to the total charge, and to the formation of molecules.
\begin{itemize}
\item $= 1$  -- the element is treated {\em implicitly}. In this case, 
its opacity is not computed on the fly, i.e. it is either neglected (in the classical approach
without an opacity table; i.e., with setting IOPTAB=0), or is contained in the opacity table 
(IOPTAB$\not=0$). 
\item $= 2$  -- the element is treated {\em explicitly}, i.e., selected energy levels 
of the included ionization states are considered explicitly; i.e., their
populations are determined by solving the corresponding kinetic equilibrium equations.
The opacity of this element is computed on the fly, using the current level populations.
\item $= 3$ -- the element is viewed as explicit, but its opacity is allowed to contribute only
for frequencies $\nu > \nu_{\rm max}$, i.e., beyond the range of the opacity table.
If the opacity table is not used (i.e., IOPTAB=0), then  $\nu_{\rm max}$ is set to 
automatically to $0$, so that in this case there is no difference between modes 2 and 3.
\item $= 4$ --  the element is viewed as {\em semi-explicit} -- see \S\,\ref{semiex}. 
This option takes effect only for
{\sc synspec}; for {\sc tlusty} it is equivalent to MODE=2, i.e., the normal explicit mode.
\item $= 5$ --  the element is viewed as {\em quasi-explicit} -- see \S\,\ref{quasex}. 
This option takes effect only for
{\sc synspec}; for {\sc tlusty} it is equivalent to MODE=2, i.e., the normal explicit mode.
\item $= -2$ --  the element is viewed as explicit, in the sense that its opacity is computed 
on the fly, but the populations of all explicit levels of all its explicit ions are held fixed to the
values from the input model atmosphere.
\item $= -3$ -- analogous to option with MODE=3, plus the populations of 
all its levels are held fixed.
\end{itemize}
\end{description}

The last two options allow a correct evaluation of the total opacity beyond
the frequency range of the opacity table. For instance, if a current run intends to
consider Mg in NLTE, and the hydrogen opacity is part of the opacity table,
then one sets MODE=2 for Mg, and MODE=3 for H. In this case, the H opacity will
be considered only for  $\nu > \nu_{\rm max}$, which will be satisfactory because
it is the dominant source of the background opacity in the Lyman continuum. Since
H is being set as an explicit element, its level populations will also be treated in NLTE.
One can also set MODE=$-3$ for H, in which case the opacity is evaluated as before,
but the H level populations are being held fixed to the input values (LTE or NLTE,
depending on the input model).

Finally, we remind the reader that the only way {\sc tlusty} can consider molecular line opacity
is through the opacity table. There is no mechanism in {\sc tlusty} to compute 
molecular line opacity on the fly, or treating molecular level populations in NLTE.

\subsubsection{Reading and initialization of the opacity table}
\label{fort15}

This is controlled by the specific input file {\tt fort.15}. The file contains one or two lines:\\[6pt]
First line (mandatory): OPTABLE, IBINOP, where:
\begin{description}
\item[OPTABLE] -- a character string that contains the filename of the opacity table.
\item[IBINOP] -- a switch indicating the global format of the table:\\
$\bullet = 0\,$ -- the table is stored in text format;\\
$\bullet  \not= 0\,$ -- the table is stored in binary format.
\end{description}
Second line (optional): ISTEPT, ISTEPR, ISTEPF:
\begin{description}
\item[ISTEPT] -- the step for taking into account the tabulated values of 
temperature. If ISTEPT=1, all the temperature
values are considered; if ISTEPT=2, every second temperature value is taken into
account, etc.
\item[ISTEPR] -- an analogous step for density;
\item[ISTEPF] -- an analogous step for the wavelength (frequency) grid.
\end{description}
If these step parameters are not specified, they are set to their default values of 
ISTEPT=ISTEPR=ISTEPF=1.

%-----------------------------------------------------------

\subsubsection{Fine points of computing NLTE models with opacity tables}
\label{ot/nlte}

It should be stressed that there are two basic types of OT/NLTE hybrid models.
A simple one is to take the structure from a previously computed model atmosphere,
and to solve for NLTE line formation of selected species, while keeping the
atmospheric structure ($T$ and $N$) fixed -- the so-called {\em restricted NLTE problem}.
Another possibility is to allow the atmospheric structure to change as well, thus
studying the influence of NLTE effects in one or several species on the global
atmospheric structure. In the latter case, it is recommended to use the Rybicki scheme,
set by the keyword parameter IFRYB=1 or 2. In the present version of the code, the default scheme,
the hybrid CL/ALI method, does not work properly for this purpose.

The numerical performance of each method is demonstrated in \S\,\ref{tesop} on two simple examples; both
use an LTE model generated with a full opacity table as a starting one,
namely:\\
-- {\tt g55nres}  -- computes a restricted NLTE model; that is a model in which the structure
(temperature, density, electron density) is held fixed, and one computes NLTE level populations 
of H, Mg, and Ca levels, using a reduced opacity table that excludes these elements. The model is
computed using the standard Complete Linearization/Accelerated Lambda Iteration (CL/ALI)
method (see Paper~II, Paper~III, and Hubeny \& Lanz 1995);\\
-- {\tt g55nryb}  -- a similar model that computes a full NLTE model in which departures form LTE are allowed for
H, Mg, and Ca, but where also the model structure is allowed to change as a response to NLTE effects
in these species. The Rybicki scheme (see Paper~II) is used here.

The former model represents a traditional procedure  to examine NLTE effects in individual
chemical elements, while the latter model is more rigorous. From a numerical point of view,
both schemes have their advantages and drawbacks. The {\tt g55nres} model linearizes the
kinetic equilibrium equations for individual level populations, alongside with the radiative
transfer equation, so the changes of the NLTE level populations are determined self-consistently
with the changes of the radiation field in the iteration process. However, 
since in this case one deals with over 40,000 discretized frequency points, most of them are
treated via ALI in the CL/ALI scheme 
[for details, refer to Hubeny \& Lanz (1995), or Hubeny \& Mihalas (2014, \S\,18.4)], 
while only some 50 points are being linearized (frequencies close to the edges of 
the ground-state H~I, Mg~I, Mg~II, and Ca~II continua, and in the centers of the strongest lines
(L$\alpha$, L$\beta$, L$\gamma$ and H$\alpha$ of hydrogen, and first resonance lines of
Mg~I, Mg~II and Ca~II). The convergence of this model is not fast, but acceptable. requiring 
about 30 iterations of the CL/ALI scheme to reach a maximum relative change in all quantities below $10^{-3}$.
This relatively slow convergence is caused by the fact that the large majority of frequency points
are not linearized. Due to a similar reason, the Cl/ALI scheme has presently serious convergence 
problems when uses to compute a fully self-consistent model where all structural equations
are being solved (that is, in addition to the radiative transfer and kinetic equilibrium, also the hydrostatic.
charge, and radiative/convective equilibrium equations are solved). In this particular case, the interplay between 
convection and the atmospheric structure is the main source of convergence problems.

In contrast, the Rybicki scheme, which essentially linearizes the radiation field in all frequencies,
and solves the radiative transfer equation together with the radiative/convective equilibrium equation,
avoids such convergence problems. However, the kinetic + charge equilibrium equation and the hydrostatic
equilibrium equations are not a part of the linearization process, so that they are essentially treated by
means of a Lambda iteration (that is, these equations are being solved only after a completed iteration
of the global linearization scheme). This is not a big problem for the hydrostatic equilibrium equation
because it involves the radiation intensities only if radiation pressure is important, which is not
the case for cool star model atmospheres. However, this may be a problem for the kinetic equilibrium, 
that is for determining NLTE level populations, because it is well known that the Lambda iteration
converges very slowly, or fails to converge at all, for NLTE level populations. For these reasons,
the Rybicki scheme was never applied to compute NLTE level populations or
NLTE model atmospheres.

However, {\sc tlusty} does not use a traditional Lambda iteration in the formal solution step (that is,
just alternating between solving for NLTE populations with the radiation field fixed, and solving
the radiative transfer equation with the level population fixed), but instead uses the ALI-based
preconditioning scheme of Rybicki \& Hummer (1991, 1992); see also Paper~II and Paper~III for
additional numerical details. To make the action of this scheme efficient, one needs to set the
number of the ``Lambda" iterations, the keyword parameter NLAMBD, to a relatively large
value, such as 8 as in the present test case.  This leads only to a modest increase in computing
time, while the accuracy of the NLTE level populations improves dramatically.
If this is not done, and the parameter NLAMBD is left at its default value, namely 2, the model
formally converges very fast (in 5-6 iterations), but the resulting NLTE level populations may still
be quite different from the exact solution (0.5, up to even 0.7 dex in some cases). 
In contrast, when using NLAMBD$=8$, and also to be on the safe side by decreasing the 
convergence criterion CHMAX to $10^{-4}$ or $10^{-5}$,
the NLTE level populations are indistinguishable from
those obtained by the CL/ALI scheme (obviously, for the temperature/density structure being held
at the structure determined by the Rybicki scheme).

The Rybicki scheme was thus shown to work well even for NLTE model atmospheres.  However,
slight modifications were necessary, which are mostly transparent to the user. The only visible
modification is a somewhat extended meaning of the keyword parameter IFRYB, which we give below:
\begin{description}
\item[IFRYB] -- a switch controlling the basic iteration scheme:\\
$\bullet\, = 0$ -- a standard CL/ALI scheme is used;\\
$\bullet\, = 1$ --  a simple Rybicki scheme is used; to be used for LTE models;\\
$\bullet\, = 2$ --  an extended Rybicki scheme, to be used for NLTE models.\\
DEFAULT: IFRYB=0 (standard CL/ALI scheme)
\end{description}
%

%------------------------------------------------------------

\subsection{New format of the opacity table}
\label{optab}

To make the computation of model atmospheres with opacity tables less error-prone,
we have introduced a new format for the opacity table. The old format can still be
used, but only in the full opacity table mode (i.e., with IOPTAB$<0$), and if a new
keyword parameter IOPOLD is set to  a non-zero value. In any case, the bulk of the table
(that is, the actual opacity as a function of frequency for all temperature-density pairs)
is analogous in both cases; the only difference is in the header of the table. Here we describe the
new header in detail. Any opacity table constructed with {\sc synspec54} will have
the new format; the possibility of using the old format is kept only for downward
compatibility with previous versions and existing opacity tables.

The purpose of adding new information to the opacity table is to avoid
inconsistencies between the parameters used in constructing the opacity 
table, and those used for running {\sc tlusty}. 
For instance, we can now avoid including accidentally an 
opacity source twice, or forgetting
an opacity source that may be important, as outlined in the previous section.

The new format of the opacity table still comes in two flavors. The traditional way is using
NTEMP temperature values (for details, refer to \S\,\ref{infort2}), and for each temperature NDENS
values of density (or some other density-like parameter, see \S\,\ref{optabl}), with their numbers and 
values being the same for all temperatures. Alternatively, one may choose different numbers and different values 
of density for each temperature. We  will call such tables {\em irregular tables}.
This will be explained in detail in \S\,\ref{infort2}. We stress that the format
of the table, while being generated by {\sc synspec]} by setting appropriate input parameters, is {\bf completely 
transparent to the {\sc tlusty} users}. The program recognizes which format of the table is being dealt with.
However, if the user intends to generate a graphical interface program to visualize the data from the
opacity table, these details are important, so that we will point them out below. 
For better visibility, they are given in {\it italics}.

Here is the outline of the structure of the new opacity table:
\begin{enumerate}
\item A table of abundances of the first 92 elements, used when generating the table
by {\sc synspec54}.  There are two sets of abundances 
listed there, namely those used in the equation of state, and those
used when computing detailed opacities. In the case of using a full opacity table, that is with
{IOPTAB~$\!<\!0$}, these two abundances are identical, while for a partial opacity table
(IOPTAB~$>0$) the abundances of some species in the second column are set to 0,
i.e. their opacities are removed from the opacity table.
\item A list of values for the keyword parameters IFMOL and TMOLIM used when
generating the opacity table.
\item A list of additional opacity switches, referring to the opacities of H-, H${}_2^+$, H${}_2^-$, He-,
the four CIA opacities, CH, and OH. A value of 1 means that the source was included when constructing the
opacity table.
\item As in the old format, the numbers of frequencies, temperatures, and densities for the table.\\
{\it For irregular tables, the number of densities (now formal) is negative. Then there is an additional input 
record that for all temperatures specifies the number of density values.}
\item As before, a list of temperatures and densities in logarithmic scale.\\
{\it For irregular tables, there is a list of temperatures as usual, and then {\rm NTEMP} records with the individual
values of the densities.}
\item A new list of electron densities (again, in logarithmic scale) corresponding to the individual
temperature-density pairs listed above.\\
{\it For irregular tables, the format is analogous to that of listing the density values.}
\item Exactly as before, for each frequency point, the value of the frequency, and the
logarithms of the opacity (per gram) for all temperature-density pairs.
The inner loop is over temperatures, and the outer loop over densities.\\
{\it For irregular tables, the order of loops is reversed.}
\end{enumerate}

\medskip

For completeness, we summarize below the new keyword parameters
used to describe the treatment of additional opacities and opacity tables:
\begin{description}
\item[IOPHMI, IOPH2P, IOPH2M, IOPHEM, IOPCH, IOPOH] -- switches for additional opacity
sources -- see Table 1
\item[IOH2H2, IOH2HE, IOH2H, IOHHE] -- switches for additional, CIA, opacity
sources -- see Table 1
\item[IRSCT, IRSCHE, IRSCH2] -- switches for additional scattering sources -- see
Table 1
\item[IOPOLD] -- a switch for considering the old format for the opacity table
(only for the full-table mode, IOPTAB $<0$), by setting its value to 1.\\
DEFAULT: IOPOLD=0 (new format)
\item[KEEPOP] -- a switch for disabling the automatic correction of the additional opacity
switches to avoid considering opacity sources that may have already been included in the
opacity table. If this parameter is set to a non-zero value, the input additional opacity
switches IOPHMI, IOPH2P, IOPH2M, IOPHEM, IOH2H2, IOH2HE, IOH2H, IOHHE,
IOPCH, and IOPOH are not being 
checked for compatibility with the values used when constructing the opacity table.\\
DEFAULT: KEEPOP=0 (an automatic correction is performed)
\end{description}

%------------------------------------------------------------

\subsection{Setting frequency points}
\label{freq}

In previous versions of {\sc tlusty}, choosing the frequency points was done in two possible ways:
\begin{itemize}
\item For classical model atmosphere construction, not using an opacity table 
(that is, setting IOPTAB=0), the frequency points are
set by the standard procedure, described in Paper~III, \S\,4.1 and 7.4.2 (for continuum frequencies),
and \S\,11.3 (for line frequencies). The actual frequencies are set by the program; the corresponding
input data are used to determine their number and values. The number of continuum frequencies is determined 
based on the input parameter NFREAD, and their actual setting is given through the keyword parameters 
FRCMIN, FRCMAX, {CFRMAX}, {NFTAIL}, and DFTAIL
(Paper~III, \S\,7.4.2). Basically, the program sets two frequency points close to the edge of all
continua (bound-free transitions) arising from explicit levels, and places some points 
in between the edges and in the region of frequencies higher
than the highest-frequency edge (determined by FRCMAX, NFTAIL, and DFTAIL). 

The line frequencies are selected for each line between explicit levels, based on the
specific input in the corresponding atomic data file, as described in Paper~III,
\S\,11.3. Once this is done for all lines, the complete set of frequency points is checked for 
points corresponding to different lines that lie very close and are thus redundant,  and such
frequency points are eliminated.
\item For a pure opacity-table run (IOPTAB$<0$), there are no explicit transitions, and the frequency
points are set in three possible ways:\\
(i) to be logarithmically equidistant between FRCMIN and FRCMAX (set if the input parameter NFREAD
is negative); or\\
(ii) to be exactly equal to the frequencies of the opacity table (set by the keyword parameter IFRSET=0,
independently of the value of NFREAD).\\
(iii) to be logarithmically spaced between the minimum and maximum frequencies in the table, set by
a value of IFRSET$> 0$\footnote{Obviously, IFRSET should not be smaller than 
the number of frequencies of the opacity table by large factors, in order not to deteriorate the accuracy
of resulting models considerably.}.
\end{itemize}
However, this procedure is not satisfactory for the hybrid opacity-table mode (IOPTAB=1). We have therefore
upgraded the procedure in the following way:
\begin{itemize}
\item In this case (IOPTAB=1), one sets the input parameter NFREAD to a positive value, so that the frequency points corresponding to both bound-free and bound-bound transitions between explicit levels
are determined as in the classical mode (IOPTAB=0).
\item One then adds the frequency points corresponding to the opacity table (and therefore describing the
opacity of all non-explicit species); these frequencies are either exactly the same as in the opacity table,
or are logarithmically equidistant between two limiting frequencies. In the latter case, the number of
table-related frequencies and their range are optional, and are given by the input parameters.
\end{itemize}
Here is a detailed description of the corresponding parameters:
\begin{description}
\item[NFREAD] (4th line of the standard input) -- has essentially the same meaning as before, namely:\\
$\bullet > 0$ -- the standard mode, i.e., the program sets the frequency points for all
explicit continua and lines;\\
$\bullet \leq 0$ -- the subset of frequency points that are set independently of the actual explicit 
continua and lines is constructed. This subset represents either a full set of frequencies --
in the case of the full opacity-table mode  (IOPTAB $< 0$); or it represents a subset of frequencies
(typically, the set of table-related frequencies) to be merged with the previously determined
frequency points.The actual number of such frequency points depends on the keyword parameter IFRSET.
\item[IFRSET] (one of the keyword parameters) -- a table-related frequency setting parameter:\\
$\bullet = 0$ -- the table-related frequency points subset is exactly the same as the frequency points
used in the table;\\
$\bullet > 0$ -- the table-related frequency set is composed of IFRSET points, logarithmically equidistant
between the minimum and maximum frequency of the opacity table, and the values of opacities in
the new set of frequencies are interpolated from the original table.
In other words, this option allows
one to reduce the number of frequencies treated through the opacity table, without severely limiting
the accuracy of the overall procedure. This option is similar, but more flexible, than a reduction of the
number of frequencies of the table described earlier in \S\,\ref{fort15}, because the present option
enables consider an arbitrary number of frequency points for the table-related subset, while the
former option only allows to consider one half, one third, etc. frequencies from the original table.
On the other hand, the latter option may be advantageous from the point of view of reducing the
total memory requirement of the code, given that one can use a very large opacity table and immediately
remove a portion of frequencies before their corresponding opacities are stored in memory.\\
$\bullet < 0$ -- This option is not very useful, but opens the possibility of testing the
sensitivity of a resulting model atmosphere to the frequency range considered in the calculations.
It is invoked only if NFREAD $<0$, and sets the table-related frequency points to contain abs(NFREAD)
points, logarithmically equidistant between FRCMIN and FRCMAX.  It is similar to the option IFRSET$>0$,
only that the minimum and maximum frequencies do not have to be equal to the minimum and maximum frequency 
of the opacity table.\\
DEFAULT: IFRSET=0
\end{description}
This might seem complicated, but in fact it is not. To help the user, we give some simple examples below.
In all examples, let us assume that the opacity table contains 100,000 frequencies.

\smallskip
(1) To compute a model atmosphere in a full opacity-table mode, and using the exactly same frequencies
as in the opacity table. In this case, one needs to set:\\
IOPTAB $=-1$   in the keyword (non-standard) parameters file;\\
NFREAD $= 0$  (or, in fact, any number NFREAD $<0$) -- in the standard input;\\
IFRSET$=0$ -- in the keyword parameter file (but it need not be specified because  0  is the default value).

\smallskip
(2) Analogous as before, but using only 30,000 frequencies from the table, logarithmically spaced between
the minimum and maximum tabular values:\\
IOPTAB $=-1$\\
NFREAD $=0$\\
IFRSET $=30000$

\smallskip
(3) To compute a model atmosphere in the hybrid opacity-table mode, including all frequencies from
the table, in addition to those associated to explicit transitions, one sets:\\
IOPTAB $=1$\\
NFREAD $=50$ (for instance, or some other non-zero value -- see below)\\
IFRSET $=0$ (again, this does not have been specified because it is a default).

\smallskip
(4) Analogous as above, but again using only a subset of 30,000 frequencies from the opacity table:\\
IOPTAB $=1$\\
NFREAD $=50$ \\
IFRSET $=30000$ 

\medskip
These examples represent some of the most useful cases of actual model construction. A note about
the value of NFREAD in the case of hybrid opacity-table mode (IOPTAB=1): For classical models,
this value is usually chosen to be larger because one needs to reproduce the continuum opacity accurately.
However, when using an opacity table, one has plenty of frequency points included anyway,
so a moderate value of NFREAD is sufficient, mostly to set up frequencies in the tail regions beyond
the span of the opacity table.

%------------------------------------------------------------

\subsection{Line broadening}
\label{lbroad}
\subsubsection{Hydrogen lines}
The adopted treatment of hydrogen lines is described
in detail in Paper~III, \S\,12.1. Here we only provide a short description, highlighting differences from
the previous versions.  

In both {\sc tlusty} and {\sc synspec}, there are several options available.
A control of these options is somewhat different in {\sc tlusty} and {\sc synspec}.\footnote{The reasons for that 
are both historical and functional. With {\sc tlusty}, one can compute very simple models, with only a relatively
small number of frequency points considered. Moreover, for simple model atmospheres one may even use
a Doppler profile for hydrogen lines. However, {\sc synspec} is used primarily for computing detailed spectra,
where many frequency points are set up, and where considering for instance a Doppler profile for hydrogen lines does 
not make much sense.}
In {\sc synspec}, the choice
is fully controlled by the keyword (non-standard) parameter IHYDPR -- see below. In {\sc tlusty}, this choice is 
controlled, as for any other element, by the corresponding atomic data file, namely by the parameter 
IFANCY, 
that sets the mode of treating the line profile (see Paper~III, \S\,11.3.), namely:
\begin{description}
\item[IFANCY]  -- a mode of treatment of the absorption profile :\\
$\bullet\,= 0$  --  Doppler profile;\\
$\bullet\,= 1$ or $-1$ --  Voigt profile;\\
$\bullet\,= 2$ or $-2$  -- approximate Stark (+ Doppler) profile for hydrogenic lines
after Hubeny et al. (1994);\\
$\bullet\,= 3$ or $-3$  -- hydrogen line profiles given by Lemke's (1997) tables. \\
$\bullet\,= 4$ or $-4$  -- hydrogen line profiles given by Tremblay \& Bergeron (2009) tables. 
\end{description}
However, if the
keyword parameter IHYDPR is set to 1 or 2, 
it overwrites the setup specified by IFANCY. This is useful to avoid the need of modifying manually the 
atomic data file for hydrogen.
\begin{description}
\item[IHYDPR] -- a mode of treatment of the hydrogen line broadening. \\
$\bullet\, = 0$ -- in {\sc tlusty}, the specification of the mode of line profile evaluation remains to
be defined by the values of IFANCY from the hydrogen atomic data file.\\
-- In {\sc synspec},
hydrogen line broadening is treated using an approximate Stark
broadening after Hubeny et al. (1994).\\
$\bullet\, =1$ -- hydrogen line broadening is computed using the Stark broadening tables 
constructed by Lemke (1997),
for all Lyman and Balmer lines with a principal quantum number of the
upper level $\leq 10$.
The table contains data for higher lines as well, but these are not recommended 
because they turned out to be incorrect.\\
$\bullet\, =2$ -- hydrogen line broadening is described using the
Tremblay \& Bergeron (2009) data. The available data file, kindly suppled by P.-E.Tremblay, 
is extended with respect  to the one used in the previous versions. 
The current file contains data for the first 20
lines of the Lyman and Balmer series, the first 19 lines of the Paschen series, and the first 
10 lines of the Bracket series. (The previous file contains data for only the first 10 members 
of the Lyman and the Balmer series.)\\
DEFAULT: IHYDPR=0
\end{description}

\subsubsection{Metal lines}
Traditionally, line broadening was treated differently in {\sc tlusty} and {\sc synspec}.
The rationale for this approach is that in {\sc tlusty} one generally does not require
detailed knowledge of the 
spectrum, and line broadening only enters through its
influence on evaluating radiative rates in the bound-bound transitions, and for evaluating
integrals needed to handle the radiative equilibrium. In both cases, the quantity of interest is an
integral of the absorption profile coefficient times the mean intensity of radiation, and
such integrals are dominated by the line core, where the profile coefficient is well
approximated by the Doppler profile. This is the reason why the large
majority of explicit lines are treated with a Doppler profile.

{\sc tlusty} has an option  to treat selected lines with a Voigt profile, 
but in this case the corresponding damping
parameters should be included  in the atomic data file - see Paper~I, \S\,11.3.
The atomic data files provided in the standard  {\sc tlusty} distribution contain explicit 
damping parameters for only very few lines, typically strong resonance lines of the
most important atoms/ions.

Such a  treatment, however, leads to a somewhat peculiar situation in the case of hybrid Opacity
Table/NLTE (OT/NLTE) approach. The background opacity, treated through the opacity table
generated by {\sc synspec}, is computed assuming Voigt profiles for all lines, while the lines
treated explicitly by {\sc tlusty}, which are supposedly more important, would be treated with a Doppler
profile. Moreover, the Doppler profiles are usually assumed to be depth-independent -- see Paper~III, \S\,11.3.

It would be impractical to modify all the available atomic data files. Instead, {\sc tlusty} reads
the original atomic data files, and in the OT/NLTE approach (IOPTAB=1),
modifies the setup of the basic broadening parameters for explicit lines in the following way:
\begin{itemize}
\item If a given line was already set for a treatment with a Voigt profile, that is the input parameter
IFANCY (later renamed to IPROF) is set to 1, then 
\begin{itemize}
\item the parameter LCOMP is set to {\tt .TRUE.} - i.e., assuming a depth-dependent profile (which was very likely
already done the original data), and the frequency-point setting parameter INTMOD is set to 3, which
invokes a frequency grid appropriate for the modified Simpson integration rule\footnote{Defined here 
as a set of three-point integrations, starting at the line center; 
each consecutive one having a doubled interval between points, that is with frequency displacements from
the line center expressed in units of Doppler width as $0, x_0, 2x_0, 4x_0, 6x_0, 10x_0, 14x_0, 22x_0, 30x_0, \ldots$. Generally, for $2n+1$ points, the total extent of integration is $(2^{n+1}-2)x_0$.
Conversely, $x_0= {\rm X\!M\!A\!X}/[2^{({\rm N\!F-1})/2+1}-2]$.}.
\item the number of frequency points per half-profile (NF) is set to: \\
-- $\max({\rm NF}, 25)$ 
\item the extent of coverage of frequency points from the line center expressed in units of fiducial
Doppler width, XMAX, is set to \\
-- $\max({\rm X\!M\!A\!X}, 4000.)$ for resonance lines%, and to\\
\end{itemize}
The damping parameters are left at the values supplied in the original atomic data file.
\item If a given line was set to a Doppler (Gaussian) profile in the original file (IFANCY=0), 
then it is automatically reset to IFANCY=1, i.e.
for a treatment with a Voigt profile,  again LCOMP is set to {\tt .TRUE.} and INTMOD=3, and 
with NF and XMAX reset as follows: If $i$ is the index of the lower level of the transition
 in the numbering of explicit levels of the
given ion, that is, $i=1$ is a ground state, then
\begin{itemize}
\item for $i=1$, ${\rm NF} \rightarrow \max({\rm NF}, 23)$;
and ${\rm X\!M\!A\!X}\rightarrow\max({\rm X\!M\!A\!X}, 4000.)$,
\item for $i\le 3$, ${\rm NF} \rightarrow \max({\rm NF}, 21)$;
and ${\rm X\!M\!A\!X}\rightarrow\max({\rm X\!M\!A\!X}, 2000.)$,
\item for $i\le 5$, ${\rm NF} \rightarrow \max({\rm NF}, 17)$;
and ${\rm X\!M\!A\!X}\rightarrow\max({\rm X\!M\!A\!X}, 600.)$,
\item for $i\le 10$, ${\rm NF} \rightarrow \max({\rm NF}, 13)$;
and ${\rm X\!M\!A\!X}\rightarrow\max({\rm X\!M\!A\!X}, 200.)$,
\item otherwise, ${\rm NF} \rightarrow \max({\rm NF}, 9)$;
and ${\rm X\!M\!A\!X}\rightarrow\max({\rm X\!M\!A\!X}, 60.)$.\footnote{Such a choice of 
parameters means that in all 
cases the setting of frequency points is very similar. Resonance lines are allowed to contribute 
to a larger frequency range than other lines, but in all cases the closest points to the line center
are separated by about $1/2$ of the fiducial Doppler width, so that the radiative rates are evaluated
with sufficient accuracy.}
\end{itemize}

\item 
Damping parameters are set to (for an explanation, refer to Paper~III, \S\,11.3):
GAMAR=0, STARK1=0, VDWH=1, i.e. the damping parameters are assumed to be given
by classical expressions -- see Paper~I, Appendix A.
For convenience, we give the classical expressions for the natural. Stark,  and Van der Waals broadening here
as well.
\begin{itemize}
\item Natural broadening:
\begin{equation}
\Gamma_{\rm rad} = 
2.67\times 10^{-22}\nu_0^2,
\end{equation}
\item{Stark broadening}
\begin{equation}
\Gamma_{\rm Stark} = 
10^{-8} n_{\rm eff}^{5/2}\, n_{\rm e},
\end{equation}
where $\nu_0$ is the frequency of the line center,
$n_{\rm eff} \equiv Z_I^2 [E_H/(E_{I}-E_{j})]^{1/2}$ is the effective
quantum number of the upper level, $j$, of the transition in ion $I$, with the 
excitation energy $E_{j}$ and the ionization energy $E_I$,
$Z_I$ is the effective  charge ($Z_I=1$ for neutrals) of the ion $I$,  $E_H$ 
is the ionization energy of hydrogen, and $n_{\rm e}$ is the electron density.
\item{Van der Waals broadening}
\begin{equation}
\Gamma_{\rm vdW} =
4.5\times 10^{-9} (2.5\, n_{\rm eff}^4/Z^2)^{0.4}\,c_w,
\end{equation}
with
\begin{equation}
\label{vdw}
c_w=(N_{\rm H}+0.42 N_{\rm He}+ 0.85 N_{{\rm H}_2}) (T/10^4)^{0.45},
\end{equation}
where $N_{\rm H}$, $N_{\rm He}$, and $N_{{\rm H}_2}$ are the total number densities of neutral
hydrogen, neutral helium, and molecular hydrogen, respectively.
The contribution of the hydrogen molecule may be quite important for cool  stars.
\end{itemize}

We recall that we use here a classical formalism, where for each broadening mechanism 
$\Gamma$ represents a half-intensity width corresponding to the Lorentz profile, 
expressed in circular frequencies,  in the atom's rest 
frame [see, e.g. Hubeny \& Mihalas (2014, \S\,8.1)]. In the ordinary frequency units, this width
becomes $\Gamma/2\pi$. Specifically, the normalized profiles are given by
\begin{equation}
\phi(\omega) = \frac{\Gamma/2\pi}{[(\omega-\omega_0)^2 + (\Gamma/2)^2]},\quad {\rm and} \quad
\phi(\nu) = \frac{\Gamma/4\pi^2}{[(\nu-\nu_0)^2 + (\Gamma/4\pi)^2]},
\end{equation}
where $\omega$ is the circular frequency, $\omega_0$ the circular frequency of the line center,
and analogously $\nu$ and $\nu_0$ are ordinary frequency and the line-center frequency; their relation
being $\nu=\omega/2\pi$. We also note that the usual Voigt parameter, that represents the damping parameter
expressed in the units of Doppler widths, is given by $a=\Gamma/(4\pi\Delta\nu_D)$, where
$\Delta\nu_D$ is the Doppler width.

\end{itemize}

%----------------------------------------------------------------------

\subsection{Equation of state}
\label{eostlu}

The treatment of the equation of state was described in Paper~II, \S\,2.7. However, the present 
versions, {\sc tlusty208}, as well as {\sc synspec54}, contain a number of improvements in the treatment 
of molecules, both regarding the solution of the chemical equilibrium, 
and the evaluation of the partition functions.

Chemical equilibrium is treated by the Tsuji (1973) formalism. Consider, for instance, a molecule 
$X_lY_mZ_n$,  formed by $l$ nuclei of atom $X$, $m$ nuclei of atom $Y$, and  $n$ nuclei of
atom $Z$. The partial pressures of the individual constituents satisfy
\begin{equation}
\label{kp}
K_P = \frac{P_X^l P_Y^m P_Z^n}{P_{X_lY_mZ_n}} ,
\end{equation}
where $P_X$, $P_Y$, $P_Z$, and $P_{X_lY_mZ_n}$ are the partial pressures of $X$, $Y$, $Z$,
and $X_lY_mZ_n$, respectively, and $K_P$ is the equilibrium constant. 
A generalization for  a different number of constituents is straightforward. For negative ions,
electrons are considered as an additional component. For positive ions, electrons are again taken as
additional component, but Eq. (\ref{kp}) is modified to read
\begin{equation}
\label{kpp}
K_P = \frac{P_X^l P_Y^m P_Z^n}{P_{(X_lY_mZ_n\!)^{\!+}} P_e} ,
\end{equation}
where $P_e$ is the partial pressure of electrons.

For the purpose of numerical
evaluation, the equiiibrium constant  is approximated by the expression
\begin{equation}
\log K_P = \sum_{i=0}^4 a_i \theta^i,
\end{equation}
where $\theta=5040/T$, and
the fitting coefficients $a_i$ are given by Tsuji (1973). This approach was used in the previous version
of {\sc tlusty}. For the present version one of us (YO) used calculations from Barklem \& Collet (2016) to compute
new fitting coefficients for most of molecular species considered by Tsuji, and added new molecular species,
including a number of positive and negative molecular ions. In total, the chemical equilibrium is solved for 503
molecular species, in contrast to the original  356 species considered by Tsuji, and 38 atomic species.

Similarly, the partition functions for the molecular species were evaluated using the Irwin (1981) tables,
wherever available, or using the Tsuji (1973) data for molecules not considered in the  Irwin tables. Again,
a new Irwin-like table was created (by YO) by fitting Barklem \& Collet's results. Many more molecular
species are considered in the new table (324) in contrast to the original Irwin's table (66). 
However,
we stress that {\sc tlusty} uses them only for an evaluation of the entropy  (see \S\,\ref{convec} and Appendix C),
but not for a computation of the molecular line opacity which is is taken from a pre-calculated opacity table,
generated by {\sc synspec} (or by another code with a table-generating capability). The molecular partition
functions are therefore much more important in {\sc synspec} than in {\sc tlusty}. In {\sc synspec}, we
have also implemented partition functions from the EXOMOL project\footnote{http://www.exomol.com/data} --
see \S\,\ref{eossyn}., but not in {\sc tlusty} because the available Barklem \& Collet data are
sufficiently accurate for this purpose [as is seen from Eq. (\ref{entr}), partition functions enter only as
logarithms].

To make a model atmosphere calculation more accurate and flexible, as well as to
allow for a downward compatibility,
we have introduced several new keyword parameters:
\begin{description}
\item[IFMOL] -- its meaning is exactly the same as in version 205, i.e., it
represents a  switch for including molecular formation in the equation of state. 
If it is set to a non-zero value, then the molecules are taken into account in the
equation of state, but not necessarily at all depths. This is determined by the following, 
newly introduced keyword parameter TMOLIM.\\
DEFAULT: IFMOL=0 (but in {\sc synspec} IFMOL=1 is a default!)
\item[TMOLIM] -- sets the division temperature (K) for including molecules in the equation of state.
For $T < $  TMOLIM, the equation of state includes molecules, and 38 neutral
and singly ionized atomic species, as described in 
Paper II, \S\,2.7.3 and above; while for $T \geq $  TMOLIM molecular formation is not considered,
and the equation of state is treated in the standard way, as described in Paper II, 
\S\,2.7.1.\\
DEFAULT: TMOLIM = 9000.
\item[MOLTAB] -- a switch to select the molecular equilibrium (Tsuji-like) table of fitting coefficients to
evaluate the equilibrium constants\\
$\bullet\,=0$ -- the original Tsuji (1973) is used (kept for comparison purposes);\\
$\bullet\,\not= 0$ -- an improved table based on Barklem \& Collet results is used\\
DEFAULT: MOLTAB=1 (i.e., the new, improved table)
\item[IIRWIN] -- a switch to consider Irwin (1981) partition function {\em for atoms and ions} 
as a default for $T<16,000$~K.\\
$\bullet\,=0$ -- the default for atoms/ions are the Hamburg (Traving et al. 1966) partition functions.\\
$\bullet\,=1$ --  the partition functions for atoms/ions for $T<16,000$ K for neutrals, once, and  
twice ionized atoms are evaluated using Irwin (1981) data.\\
DEFAULT: IIRWIN=1
\item[IRWTAB] - analogous to MOLTAB; a switch to select the table of fitting coefficients to evaluate 
the molecular partition functions. \\
$\bullet\,=0$ -- the original Irwin's table is used.
Since the improved table is obviously preferable, we keep this option for possible comparisons with old approaches.\\
$\bullet\,\not= 0$ -- an improved table, based on Barklem \& Collet (2016) data for diatomic molecules, is used.\\
DEFAULT: IRWTAB=1
\item[IPFEXO] - a switch to set up an evaluation of the molecular partition functions using the EXOMOL data.
It is relevant only to {\sc synspec} -- see \S\,\ref{eossyn}.
\end{description}
We stress that in version 205 when setting IFMOL=1 the molecular equation of state
was considered for all depths in the atmosphere, including the deepest, and usually hottest,
layers, where some higher ions may be formed, which were however disregarded in
the molecular state equation routine. This problem is now relieved by introducing the keyword parameter
TMOLIM. We have still upgraded our equation of state solver, which now takes into account molecules, 
neutral, once and twice ionized atoms, negative ions of some atoms and molecules, and electrons.
In the previous versions, twice ionized atoms were not considered, which could cause some inaccuracies at 
temperatures close to that specified by TMOLIM.

When NLTE is considered in a model where molecular formation is allowed
for, the total number of atoms of a given element treated in NLTE (a quantity that enters
the right-hand-side
of the closing relation of the set of kinetic equilibrium equations) is updated consistently,
i.e. the nuclei that have already been sequestered
in molecules have been subtracted. This was not true in version 205.

Another consequence of introducing an option to make the equation of state with molecules
more consistent is that the previous basic output parameters of the model, namely the temperature,
$T$, the electron density, $n_{\rm e}$, and the mass density, $\rho$,
are no longer sufficient to describe the atmospheric state. 
One needs to add an additional state parameter, the total particle number density, $N$.
This is explained in more detail in Appendix~B.

In order to make this change painless for users, and particularly for
various graphical tools that use the output form previous versions of {\sc tlusty},
the output on unit 8 (described in Paper~III, Chap. 8) has now the 
following structure:

\smallskip
\noindent $\bullet$ {\bf 1st line}: two numbers, ND  and  NUMPAR, where
\begin{description}
\item[ND] -- number of depth points
\item[NUMPAR] - number of input model parameters for each depth\\
$\bullet$ $ > 0$  -- has the same meaning as in the previous versions, that is
the total particle number density $N$ is {\em not} part of the set of
output parameters. The first three parameters are $T$, $n_{\rm e}$, $\rho$, 
and there may be additional parameters for disks (vertical coordinate $z$), and/or
for NLTE models (the set of populations of all explicit levels).\\
$\bullet$ $ < 0$ -- $N$, the total particle number density, is part of the set of output 
parameters, with their total number being $-$NUMPAR. The first four parameters are now
$T$, $n_{\rm e}$, $\rho$, and $N$; the rest are the same as above.\\
This option is switched on automatically when the treatment of molecules is
set, i.e., when IFMOL=1.
\end{description}
\smallskip
\noindent $\bullet$ {\bf subsequent lines}: analogous to the previous versions of {\sc tlusty}, that is
with a line or more for each atmospheric layer, namely:\\
-- for LTE models -- just $T$, $n_{\rm e}$, $\rho$, and for models with molecules also $N$;\\
-- for NLTE models -- these 3 or 4 parameters, and the populations of all explicit levels
(in cm${}^{-3}$).

%------------------------------------------------------------

\subsection{Convection}
\label{convec}

The new version offers a somewhat modified treatment of convection, which is however
essentially transparent for the user. The main difference from version 205 is a change in the
default evaluation of the adiabatic gradient, controlled by the keyword parameter IFENTR --
see below. The adiabatic gradient is now evaluated by default using entropy instead of
internal energy, which was the default in the previous versions. We have made sure
that both approaches yield very similar results. The reason of preferring an evaluation through 
the entropy is that one needs to compute numerically fewer partial derivatives, which
leads to better accuracy. For more details, and expressions for the entropy
of the material, see Appendix~C. 

We have also introduced two new keyword parameters for controlling the evaluation of partial
derivatives of the corresponding thermodynamic parameters. We describe them below, and 
we repeat for convenience the description of parameter IFENTR.
\begin{description}
\item[IFENTR] -- a switch for evaluating the adiabatic gradient:\\
$\bullet$ $ = 0$  -- the adiabatic gradient $\nabla_{\rm ad}$ is evaluated through the internal
energy -- see Paper~I, Eqs. (346) - (348).\\
$\bullet$ $ > 0$  -- the adiabatic gradient $\nabla_{\rm ad}$ is evaluated through the entropy --
see Appendix C.\\
DEFAULT: IFENTR=1
\item[DIFT] -- numerical value of $\Delta T/T$ used for evaluating a partial derivative of any
thermodynamic quantity ($f$) with respect to $T$, namely
\begin{equation}
\nonumber
\frac{\partial f}{\partial T} = \frac{f[T(1+\Delta T/T)] - f[T(1-\Delta T/T)]}{2 T\, \Delta T}
\nonumber
\end{equation}
DEFAULT: DIFT=0.01
\item[DIFP] -- analogous quantity for the derivatives with respect to the gas pressure\\
DEFAULT: DIFP=0.01
\end{description}

%------------------------------------------------------------

\subsection{Collisional rates}

This modification  offers two
significant improvements in the treatment of the collisional rates of specific ions
and transitions; details will be published elsewhere (Osorio et al., in prep.). These are:

(i) Including collisions not only with electrons, but also with protons and neutral hydrogen atoms. 

(ii) Collisional rates may be supplied as tabular values instead of analytic expressions. 
In all cases, the rates tabulated are {\em upward} rates, and downward rates are evaluated 
internally via the detailed balance relations.

\medskip

There is a subtle point here. In order to properly formulate the condition of detailed
balance, and thus to evaluate the downward rates, 
we need first to consider several issues.

Atomic physics supplies a {\em cross-section}, $\sigma_{ij}(v)$, for producing the transition 
$i\! \rightarrow\! j$ by collisions
with a certain colliding particle, or projectile (electron, proton, or neutral hydrogen), 
moving with velocity $v$ relative to the atom. Then the total number of transitions is
equal to the level population times the collisional rate $C_{ij}$,
\begin{equation}
n_i C_{ij} = n_i n_P \int_{v_0}^\infty\! \sigma_{ij}(v) f(v) dv \equiv n_i n_P\, q_{ij},
\end{equation}
where $v_0$ is the velocity corresponding to the energy of the transition, $E_{ij}$, 
$n_P$ is the number density of projectiles, 
and $f(v)$ their velocity distribution, assumed here to be Maxwellian. The quantity
$q_{ij}$, the velocity-averaged cross-sections, is also called the {\em collisional rate coefficient}. This
quantity is the actual input to {\sc tlusty}.
In order to evaluate the rate of the inverse transition $j \rightarrow i$, one invokes the detailed
balance relation, which stipulates that in LTE the number of transitions $i\rightarrow j$ is equal to
number of transitions $j \rightarrow i$.

\subsubsection*{Collisional excitation}
Regardless of the nature of the projectile, the detailed balance relation reads:
\begin{equation}
\label{dbexc}
n_i^\ast n_P^\ast C_{ij} = n_j^\ast n_P^\ast C_{ji}, \quad \Rightarrow\quad 
C_{ji}/C_{ij} = (n_i/n_j)^\ast = (g_i/g_j) \exp(E_{ij}/kT),
\end{equation}
where the asterisk indicates an LTE population, $n_i$ and $n_j$ are the populations
of the states $i$ and $j$ of the atom under study, the $g$'s are the statistical weights, and
$E_{ij}$ is the energy of the transition $i \rightarrow j$.
Since the nature of projectiles remains unchanged during the collision, their LTE
number density cancels out. This is the standard expression widely used in astrophysical work..

\subsubsection*{Collisional ionization}
We consider the three possible projectiles in turn:
\smallskip

\noindent $\bullet$ {\em Ionization by electrons}\\
Again, this is a standard case, analogous to Eq. (\ref{dbexc}),
\begin{equation}
\label{invel}
C_{ji}/C_{ij} = (n_i/n_j)^\ast = n_{\rm e} \Phi_0 T^{-3/2} (g_i/g_j) \exp(E_{ij}/kT),
\end{equation}
where the level $j$ now corresponds to  the next higher ion, usually, but
not necessarily, its ground state. 
The ratio of LTE populations is determined through the Saha-Boltzmann relation, 
where $\Phi_0 = (1/2)(2\pi m_{\rm e}k/h^2)^{-3/2} = 2.0706\times 10^{-16}$, and $T$
is the temperature.  For details, refer e.g. to Hubeny \& Mihalas (2014, \S\,4.2 and 9.3).

\medskip

\noindent $\bullet$ {\em Ionization by neutral hydrogen}\\
From the physical point of view, the ionization processes 
can be either a true collisional ionization, that is a reaction
\begin{equation}
{\rm A}_i + {\rm H}(1s) \leftrightarrow {\rm A}^{\!+}_j + {\rm H}(1s) + e, 
\end{equation}
with an inverse reaction being a three-body recombination;
or a charge exchange reaction
\begin{equation}
{\rm A}_i + {\rm H}(1s) \leftrightarrow {\rm A}^{\!+}_j + {\rm H}^-, 
\end{equation}
where the inverse process is called {\em mutual neutralization}.
Here, A${}_i$ denotes an atom A at state $i$, and A${}^{\!+}_j$  the corresponding ion at state $j$
(again,  usually, but not necessarily, its ground state). One can in principle consider collisions
with hydrogen in excited states, but the cross-sections for such collisions are poorly known, and
in any case such a process is much less important, so it is neglected here. 

As shown e.g. by Barklem (2016), the later reaction, namely a charge exchange with neutral hydrogen,
is much more probable than the true ionization process, and therefore in {\sc tlusty} we only consider
this process.
The corresponding detailed balance relation is therefore
\begin{equation}
n_i^\ast\, n_{{\rm H}}^\ast\, q_{ij}^H = n_j^\ast\, n_{\rm{H}^{\!-}}^\ast\, q_{ji}^H,
\end{equation}
where 
$n_{\rm H}$ and $n_{\rm{H}^{\!-}}$ are the populations of the ground state of
neutral hydrogen and of H${}^-$, respectively. The superscript $H$ indicates that the rate corresponds
to interaction with neutral hydrogen.
Consequently 
\begin{eqnarray}
\label{dbh}
q_{ji}^H/q_{ij}^H &=&  (n_i/n_j)^\ast (n_{\rm H}/n_{\rm{H}^{\!-}})^\ast   \nonumber \\
 &=&[n_{\rm e} \Phi_0 T^{-3/2} (g_i/g_j) \exp(E_{ij}/kT)] \, [n_{\rm e}\Phi_0 g_{1\!s}^{-1} T^{-3/2} 
 \exp(E_{{\rm H}^{\!-}}/kT)]^{-1} \nonumber \\
&=&(g_i/g_j) g_{1\!s} \exp[(E_{ij} - E_{{\rm H}^{\!-}})/kT].
\end{eqnarray}
where 
$E_{{\rm H}^-}$ is the
dissociation energy of H${}^-$, 
$E_{{\rm H}^-}=0.7552$ eV.

The collisional rates for ionization by neutral hydrogen are therefore
\begin{equation}
\label{colrats}
C_{ij}^H = n_{\rm H}\, q_{ij}^H,\quad\quad
C_{ji}^H = n_{\rm{H}^{\!-}}\, q_{ji}^H,
\end{equation}
where $n_{\rm H}$ and $n_{\rm{H}^{\!-}}$ are now {\em actual}, i.e. generally NLTE, populations
of the ground state of hydrogen and of H${}^-$.

Equation (\ref{colrats}) is used in {\sc tlusty} for the numerical evaluation of the rates of 
collisional ionization by neutral hydrogen. We also mention that a relation for the downward
collisional rate, analogous to a relation for the downward velocity-averaged cross-section,, Eq. (\ref{dbh}), 
can be written as
\begin{eqnarray}
\label{colinv}
C_{ji}^H &=& n_{{\rm H}^{\!-}}\, q_{ji}^H = n_{\rm H}\, (n_{{\rm H}^{\!-}}\!/n_{\rm H})\, q_{ji}^H \nonumber \\
&=& n_{\rm H} (n_{{\rm H}^{\!-}}\!/n_{\rm H}) (n_{\rm H}/n_{{\rm H}^{\!-}})^\ast (n_i/n_j)^\ast q_{ij}^H\nonumber  \\
&=& C_{ij}^H (n_i/n_j)^\ast (n_{{\rm H}^{\!-}}\!/n_{{\rm H}^{\!-}}^\ast) (n_{\rm H}^\ast/n_{\rm H})\nonumber  \\
&\equiv& C_{ij}^H (n_i/n_j)^\ast (b_{{\rm H}^{\!-}}\!/b_{\rm H}),
\end{eqnarray}
where $b_{{\rm H}^-}$ and $b_{\rm H}$ are the NLTE departure coefficients ($b$-factors)
for H${}^-$, and the ground state of hydrogen, respectively. In the case when H${}^-$ and the ground state
of hydrogen are in LTE, $b=1$, the inverse rate is given by the same simple expression
as for collisional ionization with electrons,  
Eq. (\ref{invel}),
but this no longer applies in a general situation.

\medskip

\noindent $\bullet$ {\em Ionization by protons}\\
Analogously, the collisions with protons are dominated by the charge exchange reaction
\begin{equation}
{\rm A}_i + {\rm H}^{\!+} \leftrightarrow {\rm A}^{\!+}_j + {\rm H}(1\!s).
\end{equation}
The corresponding detailed balance relation is
\begin{eqnarray}
q_{ji}^p/q_{ij}^p &=&  [n_i/n_j]^\ast [n_{{\rm H}^{\!+}}/n_{\rm{H}}]^\ast  \nonumber \\
&=& [n_{\rm e} \Phi_0 T^{-3/2} (g_i/g_j) \exp(E_{ij}/kT)] \, [n_{\rm e} \Phi_0 g_{1\!s} T^{-3/2}
 \exp(E_{\rm H}/kT) ]^{-1} 
\nonumber \\
&=& g_i/(g_j g_{1\!s}) \exp[(E_{ij} - E_{{\rm H}})/kT].
\end{eqnarray}
where $E_{{\rm H}}$ is the ionization energy of hydrogen,  $E_{{\rm H}}=13.595$ eV. The superscript $p$
indicates collisional rates with protons.

The collisional rates are given, analogously to Eq. (\ref{colrats}), by
\begin{equation}
\label{colratsp}
C_{ij}^p = n_{{\rm H}^{\!+}}\, q_{ij}^p,\quad\quad
C_{ji}^p = n_{\rm{H}}\, q_{ji}^p,
\end{equation}
Finally, a relation for the inverse collisional rate is, analogously to Eq. (\ref{colinv}),
\begin{equation}
C_{ji}^p = C_{ij}^p (n_i/n_j)^\ast (n_{\rm H}/n_{\rm H}^\ast) (n_{{\rm H}^{\!+}}^\ast/n_{{\rm H}^{\!+}}),
\end{equation}
which again differs from the standard relation, Eq. (\ref{invel}). 
 
\bigskip
 
From the user's point of view, this upgrade involves a simple modification
of the input atomic data files, described in detail in Paper~III, Chap.~11.
 
The standard input record for each atomic transition, both bound-free and bound-bound,
may contain, in addition to the previous 9 parameters (II, JJ, MODE, IFANCY, ICOL,
IFRQ0, IFRQ1, OSC0, CPARAM), another one, NCOL. {\sc tlusty} detects\footnote{This is detected
and properly handled 
by {\sc synspec} as well, although {\sc synspec} does not need this information, but to keep the input
of data for bound-free transitions consistent and free from confusion.}
whether this parameter is present, so the code works smoothly for both the traditional atomic 
data files, as well as the upgraded ones. If the parameter NCOL is present, and has a non-zero
value, then a number of subsequent parameters are read. The meaning of the new parameters
is explained below.
\begin{description}
\item[NCOL] -- number of tables of collisional rates (corresponding to different 
collisional processes); NCOL $\leq 3$. Subsequently, there are NCOL blocks of input values, each containing:
\end{description}
\noindent $\bullet$ {\bf 1st record}:
\begin{description}
\item[ITYPE] -- the type of the collision process, namely for bound-bound transitions:\\
$\bullet =1$ -- collisions with electrons\\
$\bullet =2$ -- collisions with protons\\
$\bullet =3$ -- collisions with neutral hydrogen atoms\\
while for bound-free transitions,\\
$\bullet =1$ -- collisions with electrons -- electron collisional ionization\\
$\bullet =2$ -- collisions with protons -- charge exchange with protons\\
$\bullet =3$ -- collisions with hydrogen atoms -- charge exchange with hydrogen
\item[NCTEMP] -- number of temperature values for the subsequent table
\end{description}
\noindent $\bullet$ {\bf following records}
\begin{description}
\item[CTEMP(I),I=1,NCTEMP] -- values of the temperature [K]
\item[CRATE(I),I=1,NCTEMP] -- values of the collision rate coefficient [cm${}^3$ s${}^{-1}$]
\end{description}
%

%------------------------------------------------------------

\subsection{Fixed-density models}
In version 205 and all earlier versions of {\sc tlusty}, the state parameters that could be held fixed, and the corresponding
structural equation not being solved, were:

$\bullet$ total particle number density ($N$) -- not solving the hydrostatic equilibrium

$\bullet$ temperature ($T$) -- not solving the radiative/convective equilibrium

$\bullet$  electron density ($n_{\rm e})$  -- not solving the charge conservation.

\noindent Another option was also to keep fixed all three parameters, in which case only the radiative
transfer, together with kinetic equilibrium equation, is solved (the so-called restricted NLTE problem). 

In version 208, we have added another option, namely keeping the mass density, $\rho$, fixed, while
$N$ and $n_{\rm e}$ can change.
This option is useful for producing NLTE level populations and electron density for snapshots of
hydrodynamical simulations that supply temperature and mass density. Therefore, in this
option one usually considers the temperature to be fixed as well.

The option is controlled by the following two keyword parameters:
\begin{description}
\item[IFIXDE] -- If set to a non-zero value, the option to keep fixed density is
switched on. In this case, the fixed  temperature switch, INRE, is set to zero.\\
DEFAULT: IFIXDE=0
\item[TFLOOR] -- represents the floor value of the input temperature. It is included because in
some hydro simulations, for instance for accretion shocks in young cool stars, the ambient atmosphere
can have very low $T$, say around 3,000 K or so, while the accretion shock can have temperatures
of the order of $10^5$ or $10^6$ K. Since one is typically interested in these hot layers, the ambient
temperature can be artificially increased to avoid numerical problems associated with
dealing simultaneously with very high and very low temperatures.\\
DEFAULT: TFLOOR = 6000.
\end{description}

%------------------------------------------------------------

\subsection{Changes of the output}
\label{outchan}

There are three output files that are changed or added with respect to the previous versions,
namely (i) changes of the standard output file;  (ii) changes of the output file {\tt fort.7} -- a condensed
model atmospheres, which were described in \S\,\ref{eostlu}, and (iii) an additional  output file {\tt fort.22}
that contains the absolute $b$-factors -- see below in \S\,\ref{outchan22}. 

Another change is an introduction of several keyword parameters that control the amount optional tables produced in
 the standard output.
In the previous versions, there was a single parameter IPRINT that controlled a number of output options,
so its meaning and an actual use were quite confusing. We list the new keyword parameters below. In all cases,
if a parameter is set to zero, the corresponding output is not produced. All parameters have a six-character
name, IPxxxx, where xxxx is an abbreviation of what kind of output it controls. The default values
of all these parameters are set to 0 (i.e., not producing the corresponding output). These pieces of output 
are interesting or useful essentially only if there are convergence or some other problems.

The first parameter controls an output of an additional table in the initialization step:
\begin{description}
\item[IPTRAN] - input parameters for explicit {\it TRAN}sitions. The table is self-explanatory. 
It should be invoked judiciously because it may be large.
\end{description}
The next two parameters control an output of tables at each iteration step of the global iteration scheme:
\begin{description}
\item[IPCONF] - output of {\it CON}vective {\it F}lux, for all depths, at each iteration.
The individual columns are:
depth index; Rosseland optical depth $\tau_{Ross}$; temperature $T$ (K); logarithmic temperature gradient
$\nabla\equiv d\ln(T)/d\ln(P)$; adiabatic gradient $\nabla_{\rm ad}$; relative convective flux,
$F_{\rm conv}/F_{\rm tot}$; relative radiative flux, $F_{\rm rad}/F_{\rm tot}$; and finally $F_{\rm rad}+F_{\rm conv}$,
which should become, at later iterations, to be equal or at least close to 1.
\item[IPRYBH] - an information about a recalculation of the hydrostatic equilibrium equation and a determination 
of an updated gas pressure and density in the case of Rybicki scheme in the formal solution
step, done by subroutine {\it RYBH}eq, described in Appendix E.
The individual columns are:
depth index, column mass $m$ (g cm${}^{-2}$, current mass density $\rho$ (g cm${}^{-3}$,
radiation pressure $P_{\rm rad}$ (cgs units), the ration of the current radiation pressure to the equilibrium one,
$P_{\rm rad}/(3\sigma T^4/4c)= P_{\rm rad}/(2.5213\times 10^{-15} T^4)$, 
where $\sigma$ is the Stefan-Boltzmann constant;
gradient of the radiation pressure $dP_{\rm r}/dm$; gravity acceleration, In the case of disks, there is an
additional column, the  vertical distance form the central plane $z$ (cm).
\end{description}
The last two parameters control a production of two auxiliary tables at the final iteration:
\begin{description}
\item[IPELCH] - a table of the {\it EL}ectron density {\it CH}eck in the case of opacity tables -- see below
\item[IPELDO] - a table of the contribution of the individual {\it EL}ectron {\it DO}nors -- see below.
\end{description}

\subsubsection{Standard output}
\label{outchan6}

Here we describe the structure of the standard output, highlighting the changes 
made in the present version. We label the individual items with ``{\it old}" or ``{\it new}",
referring to whether they existed in the previous versions, or are newly introduced here. The
individual items are listed consecutively; the {\it old} items were described in some
detail in Paper~III, Chap.9.
\begin{itemize}
\item {\it old} -- the basic header; namely the values of $T_{\rm eff}$ and $\log g$; 
\item {\it new} -- a verbatim copy of the content of the keyword parameter file
(whose name being specified as {\tt FINSTD}  in the standard input file); 
\item {\it old} -- a list of chemical elements and their adopted abundances;
\item {\it old} -- a table of explicit ions and their basic parameters;
\item {\it old} -- a table of explicit energy levels and their basic parameters;
\item {\it new}, produced only in the case when an opacity table is used (IOPTAB $\not= 0$) 
-- a block containing a detailed information about the adopted opacity table:
\begin{itemize}
\item the name of the file containing the opacity table;
\item the numbers of frequencies, temperatures, and densities in the table;
\item a list of chemical abundances. There are three columns. The first one lists
the abundances used by {\sc tlusty} in the current run (the same ones as listed above).
The next two columns list the abundances used by {\sc synspec} to generate the opacity
table; one for the equation of state, and the other one for evaluating the opacity, as described
in \S\,\ref{optab}. These two abundances differ only in the case of a partial opacity table
where some chemical species are not included in evaluating the opacity;
\item the values of the keyword parameters IFMOL and TMOLIM used by {\sc synspec} for
constructing the opacity table, and by {\sc tlusty} in the current run;
\item information about the individual additional opacity sources, namely whether
they are included, or possibly rejected because the opacity table already contains
them;
\end{itemize}
\item {\it old} -- several auxiliary quantities (number of overlapping transitions; 
an assessment of the accuracy of the frequency integrations; number of frequencies;
a list of explicit frequencies);
\item {\it old} -- a list of the most important keyword parameters used in the current run;
\item {\it new} and {\it old} -- several auxiliary tables of intermediate results, not particularly important unless
something goes wrong. The appearance of these tables is controlled by the keyword parameters
IPCONF and  IPRYBH -- see above.
\item {\it new\/} and {\it old} -- final section, entitled {\tt FINAL RESULTS},
which contains three tables:
\begin{itemize}
\item {\it new} -- if the keyword parameter IPELCH is $\geq 1$, then a table entitled {\tt CHECK OF ELECTRON DENSITY}
is produced, which contains, for each depth point, the following quantities:
\begin{itemize}
\item depth index,
\item temperature,
\item actual electron density produced by {\sc tlusty} in the current run,
\item electron density evaluated by the LTE equation of state for the actual values of temperature 
and density at the given depth, $n_{\rm e}(T,\rho)$. Its value may be generally different
from the actual $n_{\rm e}$ for NLTE models. In the case of partial opacity table
(IOPTAB$>0$), the difference between these two values indicates a degree of consistency
of the overall calculation because {\sc synspec}
uses the same procedure for determining electron density for the tabular values
of $T$ and $\rho$. In the case these two values are significantly different, the opacities
following from the opacity table are not consistent because they were evaluated using
an inconsistent electron density;
\item electron density evaluated by an interpolation
from the values listed in the opacity table. A comparison of this value to the previous
one indicates an accuracy of interpolating in the opacity table, because the same
interpolation procedure is being used to evaluate opacity for actual values of $T$ and
$\rho$ from the tabulated opacity at the set of discrete values of $T$ and $\rho$
used in the table;
\end{itemize}
\item {\it new} -- if the keyword parameter IPELDO $\geq 1$, then a table entitled \\
{\tt   RELATIVE CONTRIBUTIONS OF THE INDIVIDUAL ELECTRON DONORS}\\ 
is produced,
which lists, for each depth, relative contribution to the total number of electrons 
from the most important species (H, He, C, N, O, Na, Mg, Al, Si, S, Ca, and Fe). The first
three columns of the table are the depth index, temperature, and the value of
$n({\rm H-})/n_{\rm e}$, representing a sink of electrons due to H- (a negative contribution);
\item {\it old} -- a table entitled \\ 
{\tt    FINAL MODEL ATMOSPHERE} \\
that presents
a list of the basic state parameters of the resulting model. The table is self-explanatory.
\end{itemize}
\end{itemize}

\subsubsection{New output file {\tt fort.22}}
\label{outchan22}

As explained in more detail is \S\,\ref{inpnlt}, in some applications of {\sc synspec} it is advantageous
to use as input when dealing with NLTE level populations  not the values of the populations themselves, but rather the 
``absolute" $b$-factors, defined as  a ratio of NLTE and (absolute) LTE populations, 
\begin{equation}
\label{bast}
b_i^\ast \equiv n_i/n_i^{\ast\ast},
\end{equation}
where $n_i$ is a NLTE population of level $i$, and $n_i^{\ast\ast}$ the {\em absolute} LTE population,
obtained by solving the Saha-Boltzmann equations for all levels of all considered ionization
stages of the element.

In contrast, the traditional $b$-factors (stored as before in output file {\tt fort.12}) are defined by
\begin{equation}
\label{btrad}
b_i \equiv n_i/n_i^{\ast} = n_i/[n_{\rm e} n_1^+ \Phi_i(T)],
\end{equation}
where $n_{\rm e}$ is the electron density, $n_1^+$ the {\em actual} population of the ground state of the
next ion, and $\Phi_i(T)$ is the Saha-Boltzmann factor -- see Paper~II, \S\,2.3, or Hubeny \& Mihalas
(2014, Chap.\,9). This definition is useful from the point of view of understanding NLTE line
formation from the theoretical point of view, while the definition we adopt here, Eq. (\ref{bast}) is
useful from the point of view of abundance analysis, which in fact is one of the principal uses
of {\sc synspec}.

The traditional $b$-factors are stored in the output file {\tt fort.12}, as in the previous versions, while the
absolute $b^\ast$-factors are stored in the new output file {\tt fort.22}, which has exactly the same format as
{\tt fort.12}, and in fact also as {\tt fort.7} -- see \S\,\ref{eostlu}. 
To repeat, all the three files have the same header: the 1st line contains the number of depths
and the number of physical parameters per depth; then several lines that contain the mass-depth coordinate
(the column mass in g cm${}^{-2}$), and then for each depth the temperature, electron density, mass density,
total particle density (for models with molecules), and then for all explicit levels either the level populations ({\tt fort.7}).
or traditional $b$-factors ({\tt fort.12}), or absolute $b^\ast$-factors ({\tt fort.22}).
Finally, we stress that the two latter files, {\tt fort.12} and {\tt fort.22}, are generated only for NLTE models,
because for LTE models all the $b$-factors are identically equal to unity.

\subsubsection{Specific intensities}

There is a new option in {\sc tlusty} to generate and store not only the emergent flux, but also the 
direction-dependent specific intensities\footnote{Before, the specific intensities were only produced by {\sc synspec}, if specifically required.}.
In the past, the number of 
frequency points considered by {\sc tlusty} was often relatively low, and in any case {\sc synspec} was usually
used as a post-processor to produce a detailed spectrum. With the introduction of opacity tables in {\sc tlusty},
and the corresponding ease to produce models with large number of frequencies considered, it became
useful to obtain angle-dependent specific intensities  at the end of a run of {\sc tlusty}. 
This is controlled by a new keyword parameter:
\begin{description}
\item[INTENS] - a switch for producing emergent specific intensities:\\
$\bullet\, = 0$ -- no specific intensities are generated; \\
$\bullet\, > 0$ -- specific intensities are generated for INTENS number of directional cosines $\mu$, linearly
spaced between 0.1 and 1.\\
DEFAULT: INTENS=0 
\end{description}
The specific intensities 
are stored
in the output file {\tt fort.18}, which contains, for each frequency point considered by {\sc tlusty}, 
the wavelength (\AA), and
INTENS values  of the specific intensity $I_\nu(\mu)$ (erg cm${}^{-2}$s${}^{-1}$Hz${}^{-1}$),
starting with $\mu=0.1$.

%%%%%%%%%%%%%%%%%%%%%%%%%%%%%%%%%%%%%%%%%

\section{SYNSPEC}
\label{synspec}

There are several main changes between versions 51 and 54:
\begin{itemize}
\item Changes in the line list.
\item An option to construct opacity tables.
\item A modified and upgraded treatment of additional opacities.
\item A modified algorithm of selecting lines from the line list.
\item Changes in the molecular equation of state analogous to those made
in {\sc tlusty}.
\item An improved methodology for the association of NLTE energy levels between
{\sc tlusty} and {\sc synspec}.
\item Subtle changes in the computation of the line opacity.
\item Allowance for spectral synthesis for LTE model atmospheres with the additional
input of NLTE level populations (more accurately, $b$-factors) taken from a different model.
\end{itemize}

\subsection{Changes in the line list}
\label{linlist}
We stress that the structure of a line list was described in some detail in Paper~I, \S\,5.3.3, but only for 
a list providing data for lines of atoms and ions, while the molecular line lists, which have 
a somewhat different format, were not described. Since several more 
options were introduced recently, we will describe the structure of the line lists in detail.
We also mention that, unlike previous versions, there is now the option 
of having any of the lists in binary format, which may save a considerable amount of computer 
time, in particular for evaluating opacity tables -- see below.

\subsubsection{Atomic line lists}
\label{linlist_at}
While molecular lines can be input by means of several separate files
(for instance for some individual molecular species), {\sc synspec} is designed to work with a single
atomic line list. Originally, the atomic line list had to be transmitted to {\sc synspec} with a fixed 
filename, namely {\tt fort.19}, but now it can have any name, specified in the input file {\tt fort.3} 
(see below).

Currently, there are two variants of an atomic line list, the {\em standard} one, used in all previous versions
of {\sc synspec}, and an {\em extended} one, which is preferable for NLTE model atmospheres 
because it provides a better association between atomic energy levels used in {\sc tlusty} and 
{\sc synspec} -- see \S\,\ref{nltlev}.

\medskip
\noindent
({\it i}) The {\em standard line list} contains one input record for each spectral line. In fact, it could
contain more records, but this option is cumbersome and is rarely used.
Each input record contains the following variables:

\begin{verbatim}
     alam anum gf excl ql excu qu agam gs gw inext
\end{verbatim}
and if the continuation indicator {\tt inext} is non-zero, then the next record is
\begin{verbatim}
     wgr1 wgr2 wgr3 wgr4 ilwn iun iprf
\end{verbatim}
These quantities have the following meaning:\\ [2pt]
$\bullet$ {\tt alam} -- wavelength [nm]. One follows the convention that for
$\lambda < 200$ nm the wavelengths are taken in vacuum, while for $\lambda \geq 200$ nm the
wavelengths are for standard air (they can however be modified to vacuum wavelengths by setting 
the parameter {\tt alast} in {\tt fort.55}---see below---to a negative value); \\ [1pt]
$\bullet$ {\tt anum} -- numerical code of the element and ion (using Kurucz's 
convention, namely atomic number, decimal point, and degree of ionization:  
for instance 2.00 = He I, 6.03 = C IV, 26.01  = Fe II, etc.)\\ [1pt]
$\bullet$ {\tt gf} -- $\log g\!f$ ($g$ is the statistical weight of the lower level, and $f$ the oscillator strength)\\ [1pt]
$\bullet$ {\tt exclu} -- excitation potential of the lower level [cm${}^{-1}$]\\ [1pt]
$\bullet$ {\tt ql} -- the $J$ quantum number of the lower level\\ [1pt]
$\bullet$ {\tt excu, qu} -- analogous quantities for the upper level\\ [1pt]
$\bullet$ {\tt agam} -- if non-zero, $\log\Gamma_{\rm rad}$ for radiation damping  [s${}^{-1}$]\\ [1pt]
$\bullet$ {\tt gs} -- if non-zero, $\log\Gamma_{\rm stark}$ for Stark broadening [s${}^{-1}$]\\ [1pt]
$\bullet$ {\tt gw} -- if non-zero, $\log\Gamma_{\rm vdW}$ for Van der Waals broadening  [s${}^{-1}$]\\ [1pt]
$\bullet$ {\tt inext} -- if =1, next record is needed, where:\\ [4pt]
$\bullet$ {\tt wgr1 wgr2 wgr3 wgr4} -- Stark broadening values from Griem (1974)
tables, namely the values for $T=(5, 10, 20, 40) \times 10^3$ K, respectively\\ [1pt]
$\bullet$ {\tt ilwn} -- manual setting of the index of the lower level as 
used by {\sc tlusty}\\ [1pt]
$\bullet$ {\tt iun} -- manual setting of the index of the upper level as
used by {\sc tlusty}\\ [1pt]
$\bullet$ {\tt iprf} -- if non-zero, special procedure for evaluating the Stark broadening
(at present ony for He I -- see Paper~I,  Appendix A).\\ 

In any of the parameters {\tt agam, gs, gw} is set to zero, approximate
formulae are adopted for evaluating the corresponding broadening parameter. For details refer 
to Paper I, Appendix A; see also here, \S\,\ref{lbroad}.

\medskip
\noindent
({\it ii}) The {\em extended line list}.
The entry for each spectral line contains, in addition to the standard 11 parameters
(ALAM, ANUM, GF, EXCL, QL, EXCU, QU, AGAM, GS, GW, INEXT -- see above, 
the set of 6 following new  parameters:

\begin{verbatim}
     isql ilql ipql isqu ilqu ipqu
\end{verbatim}
where\\ [2pt]
$\bullet$ {\tt isql} -- the quantum number $2S+1$ of the lower level\\ [1pt]
$\bullet$ {\tt ilql} -- the quantum number $L$ of the lower level\\ [1pt]
$\bullet$ {\tt ipql} -- the parity of the lower level\\ [1pt]
$\bullet$ {\tt isqu, ilqu, ipqu} -- analogous values for the upper level.\\ [2pt]
These parameters are explained in detail in \S\,\ref{nltlev}.

We stress that  {\sc synspec} automatically detects whether the line list is in the standard,
old, format or whether it has the extended format, with these 6 new parameters. If they are present, one can still 
force {\sc synspec} to ignore the new information for all ions by setting the parameter INLIST 
(in the input file {\tt fort.55}) to a value $\geq 10$ -- see also \S\,\ref{linlist_spec}.. 
One can also force {\sc synspec} to
ignore this information for a particular ion by setting the parameter NONSTD in the
standard input file to 10, or, more precisely, if a non-zero value of NONSTD was intended, 
setting NONSTD to the desired NONSTD+10.

\subsubsection{Molecular line list}
\label{linlist_mol}

There are three variants of the molecular line list; the {\em standard} (classical-broadening) list;
its variant called a {\em minimalistic} list; and the
EXOMOL-{\em type} list, differing in the extent of the
set of the supplied pressure (Van der Waals) line broadening parameters,
and consequently, in the subsequent treatment of  the pressure broadening in {\sc synspec}. The latter addition
was motivated by recent progress in evaluating the broadening parameters, and an ever
increasing availability of data (e.g., Barton et al., 2017a,b). We describe them below.

\medskip
\noindent
({\it i}) {\em Standard list}. Its structure is analogous, albeit simplified, to a standard atomic list. Each input record
contains the following parameters for each line:
\begin{verbatim}
     alam anum gf excl agam gs gw 
\end{verbatim}
The meaning of the variables is exactly equivalent to those for the atomic line list. Since the molecular lines
one always assumed to be formed in LTE, the parameters for the upper level (EXCU and QU) are not needed. 
The largest
source of uncertainty in treating lines whose data are provided by such a list is in the evaluation of 
pressure (Van der Waals) broadening. We stress that Stark broadening is typically negligible under
the conditions where molecules are formed in significant amounts to make molecular lines contribute
appreciably to the total opacity, and natural broadening is usually well approximated by the classical
formula. The Van der Walls broadening parameter is taken by a generalization of Eq. (4) in Appendix~A of
Paper~I; see also Eq. (\ref{vdw}) in this paper,
\begin{equation}
\label{vdwcla}
\Gamma_{\rm vdW} = 10^{\tt gw}\, (N_{\rm H}+0.42 N_{\rm He}+0.85 N_{{\rm H}_2})\, (T/10^4)^{0.45},
\end{equation}
where $N_{\rm H}$, $N_{\rm He}$, and $N_{{\rm H}_2}$ are the number densities of
neutral hydrogen, helium, and the hydrogen molecule, respectively, and $T$ is the temperature. 

\medskip
\noindent
({\it ii})  A {\em minimalstic list}. Its introduction was motivated by the observation that in a vast majority of
molecular line lists the parameters {\tt gs} and {\tt gw} have the same value for all lines, and moreover
for most molecular species. Therefore, for such molecules it is a waste of computer memory to store them
in the line list; instead they are assigned  by {\sc synspec} when initializing line data from the line list.
Similarly, the natural broadening in such lists has its classical value, which again does not have be 
stored in the line list, but is computed by {\sc synspec} instead. Consequently, the minimalistic line lists contain
only four values for each line:
\begin{verbatim}
     alam anum gf excl  
\end{verbatim}
which have the same meaning as above. The parameter {\tt agam} is computed as ${\tt agam}= 2.2e13/\lambda^2$,
with the wavelength $\lambda$ expressed in nm.. Parameters {\tt gs} and {\tt gw} can be specified through
the keyword (non-standard) parameters GSSTD and GWSTD, which have default values 3.1e-5 and 1.0e-7,
respectively. Therefore, if they are not specified in the input, the parameters {\tt gs} and {\tt gw} will have the same values as
in the usual standard lists.

\medskip
\noindent
({\it iii}) The EXOMOL-type list contains for each line the following parameters:
\begin{verbatim}
     alam anum gf excl agam  gvdwh2 gexph2 gvdwhe gexphe
\end{verbatim}
where the last 4 parameters describe the Van der Waals broadening following Barton et al. (2017a,b) as
\begin{equation}
\label{vdwexo}
\Gamma_{\rm vdW} = 10^{-6} ck\, [g_{{\rm H}2} (T/296)^{n_{{\rm H}2}} N_{{\rm H}_2} +
 g_{\rm He} (T/296)^{n_{\rm He}} N_{\rm He} ]
 \end{equation}
 where the parameters $g_{{\rm H}2} \equiv {\tt gvdwh2}$, $n_{{\rm H}2} \equiv {\tt gexph2}$, and
 analogously for the helium broadening. 
 Broadening by H atoms is  ignored.
 More details are given in Appendix F.
 
 We stress that {\sc synspec} recognizes which type of the list is used, so that no additional input
 parameter has to be specified.
 
 \subsubsection{Specification of the format of the line lists}
 \label{linlist_spec}
 
 In previous versions of {\sc synspec}, the file name of the atomic line list was hardwired as
 {\tt fort.19}, and the number and the names of the molecular line lists was set in the input file {\tt fort.55} --
 see Paper~I, \S\,5.3.1. The relevant lines in the file {\tt fort.55} are the 4th and 7th line. The fourth line
 contains
 \begin{verbatim}
       ifreq  inlte  icontl  inlist  ifhe2 
 \end{verbatim}
 \vskip-10pt
Here, only the parameter {\tt inlist} is relevant for the line lists, in that it specifies the format of the atomic line list:\\[4pt]
$\bullet$ {\tt inlist} = 0 or 10, all the line lists are in text format, while if \\
$\bullet$ {\tt inlist} = 1 or 11, all the line lists are in binary format.\\[-6pt]

Notice that {\tt inlist} has a dual meaning, because at the same time\\[4pt]
$\bullet$ {\tt inlist} $=0$ or 1 -- the additional quantum numbers (see \S\,\ref{linlist_at}), if present, are being
taken into accout;\\[1pt] 
$\bullet$ {\tt inlist} $=10$ or $11$, the additional quantum numbers, if present, are disregarded. \\

The 7th line, if it exists, contains parameters for the molecular line lists, namely\\[-16pt]
 \begin{verbatim}
       nmlist, (iunitm(i),i=1,nmlist)
 \end{verbatim}
 \vskip-12pt
 where;\\[3pt]
$\bullet$ {\tt nmlist} -- the number of  molecular line lists, \\ [1pt]
$\bullet$ {\tt iunitm(i),i=1,nmlist} -- unit numbers of the additional molecular line lists.\\[2pt]
In this case, if the unit number is {\it nn}, then the explicit filename is {\tt fort.}{\it nn}.

In the present version of {\sc synspec}, this approach to specify the filenames of the molecular line lists,
and the mode of the atomic line list, can still be used.
However, the user may  choose to use a {\em newly 
introduced input file} {\tt fort.3} that, if exists, overwrites the information obtained from {\tt fort.55}.

The structure of the file {\tt fort.3} is as follows. For each line list, including the atomic one, there is one
input record containing 2 (in the case of an atomic list) or 3 (in the case of a molecular line list) parameters:\\[2pt]
{\tt AMLIST, IBIN, TMLIM}    (that is, only {\tt AMLIIST} and {\tt IBIN} for atomic lists), where
\begin{description}
\item[AMLIST] -- a {\tt character}$\ast$40 variable that specifies the name of the list
\item[IBIN] -- an indicator of the type of file:\\
$= 0$ -- a text file.\\
$\not= 0$ -- a binary file.
\item[TMLIM] -- an upper limit of temperature for  which is the list being read and used. This parameter
only takes effect when evaluating the opacity tables\footnote{The reason why this parameter only 
takes effect for evaluating the opacity tables is that in this case one reads all lists repeatedly, namely
for each temperature/density pair, while for evaluating a synthetic spectrum for a given model atmosphere
the lists are read only once, but one encounters many different temperatures at the individual depths.
If the atmosphere is everywhere hot enough that, for instance, TiO is not formed, then one simply does
not take the TiO line list into account at all.}
- see next section.
\end{description}
This approach is more flexible than the one using the file {\tt fort.55}.

\medskip
\noindent
{\bf Important notes:} \\[2pt]
{\bf (1)} We would like to stress that an option of using an input from the file {\tt fort.55} to specify the line lists
and their mode is kept for downward compatibility. It is useful for users who have been using {\sc synspec}
in the past, and doing spectrum calculations either for atomic lines only, or for a single molecular
line list. In that case, one can use the original file {\tt fort.55}. In all other cases, in particular when
dealing with multiple line lists, it is safer and more flexible to set up the specifications for the line lists
using information in {\tt fort.3}.

\medskip

\noindent
{\bf (2)} When the file {\tt fort.3} is used, the parameter INLIST in the file {\tt fort.55}, discussed above, does no longer have 
a function to set the atomic line list in text or binary format, but it still influences the 
association of level indices between {\sc tlusty} and {\sc synspec}; that is, if INLIST $\geq$ 10, the additional
quantum numbers, if present in the atomic line list, are disregarded.

\medskip

\noindent
{\bf (3)} When the file {\tt fort.3} is used,  its first line has to correspond to the atomic line list, followed by several lines
corresponding to the individual molecular line lists. Their order is arbitrary, and their
number is independent of the value of the parameters {\tt nmlist} specified in the file {\tt fort.55}.

\medskip
\noindent
{\bf (4)} In the previous versions of {\sc synspec}, the basic mode IMODE (the first number in {\tt fort.55}) had to be
set to  a value $\geq 10$ in order molecules were considered at all. This is now obsolete, because of an
introduction of a general keyword parameter IFMOL,  transmitted to both {\sc tlusty} and {\sc synspec}
through the keyword parameter file. Specifically
\begin{description}
\item[IFMOL] -- as in {\sc tlusty}, it switches on the treatment of molecules in the
equation of state, independently of whether molecular lines are being considered
in the opacity or not. As for {\sc tlusty}, a non-zero value allows for considering molecules
in the equation of state, for local temperatures $T<$ TMOLIM; the latter being another
keyword parameter, with exactly the same meaning as for {\sc tlusty}.
\end{description}

At present, {\sc synspec} typically uses up to three molecular line lists: \\
(i) for basic diatomic molecules except TiO;\\
(ii) for TiO;\\
(iii) for H${}_2$O.

At present, the list for basic diatomics includes line data for the following molecules:
H${}_2$, CH, NH, OH, NaH, MgH, SiH, CaH, CrH, FeH, C${}_2$, CN, CO,  
MgO, AlO, SiO, CaO, and VO. The line data are so
far taken from the Kurucz website{\footnote {\tt http://kurucz.harvard.edu/molecules}}, although
we are working on replacing them by the EXOMOL data whenever available.

The number of lines in a full molecular line list may be  very large, typically several tens to hundreds of millions
of lines, but in fact most of them do not contribute appreciably to the total opacity. We have therefore
developed a procedure to eliminate weak lines from the line list from the outset. We stress that if 
weak lines are present in the line list, they are eliminated during the run of {\sc synspec} based on the
input parameter RELOP. However, working with extended line lists is cumbersome, because
(i) they take a lot of memory, (ii) they are harder to download, and (iii) even reading an extended 
line lists by {\sc synspec} takes a lot of unnecessary computer time. It is therefore desirable to
devise procedures to limit the number of lines in the line list, while keeping the accuracy of the
resulting total opacity reasonably high. This is described in some detail in Appendix G.

\subsubsection{Utility programs}

The standard distribution also contains two additional, very simple, programs for handling the line lists,
called {\sc list2bin} and {\sc redulis}.

\medskip
$\bullet$
The program {\sc list2bin} transforms a text line list to binary format. It is quite trivial, but is included
for convenience. Since there are altogether 5 possible formats of a line list (2 for atomic, and 3 for molecular lists),
the program automatically recognizes the format of the line list. The standard input is the text line list, and the
output binary list is on file {\tt fort.12}. It is advantageous to rename this file to have the same core name, with extension 
{\tt .bin}. For instance, assuming that the code was compiled as \\[-8pt] 

{\tt      gfortran  -o list2bin.exe list2bin.f}\\[2pt]
then to rewriting the contents of the file {\tt gfATO.dat} in binary format is simply accomplished by executing\\[-8pt]

{\tt      list2bin.exe <gfATO.dat}\\[-12pt]

{\tt      mv fort.12 gfATO.bin}

\medskip
$\bullet$
The program {\sc redulis} performs a reduction of the line list, as mentioned above, and is described in detail in 
Appendix G. The code works for all three types of molecular line lists. 
The standard input contains either just one number, $s_{\rm reject}$ (see Appendix G) (which is mandatory), 
or three more quantities, 
$T_{\rm char}$, $\lambda_{\rm char}$, and  $\log (N/U)_{\rm char}$ -- see again Appendix G. If these values are
not specified, the default values for these quantities, namely 3000, 10,000, and 10, respectively, are adopted.

The original list should be copied/linked to {\tt fort.20}, and the output is on {\tt fort.12}. The standard output 
contains information about how many lines of the individual molecular species are in the original list, and 
in the reduced one.

\subsection{Constructing opacity tables}
\label{optabl}

The option for calculating a grid of monochromatic opacities
is switched on by the parameter IMODE (1st number in the first line of the file
{\tt fort.55} -- see Paper~I, \S\,5.3.1), namely IMODE $=-3$ or $=-4$. Since there are some new possibilities, we
shall summarize the options for the parameter IMODE below:
\begin{description}
\item [IMODE] - the basic mode of operation of {\sc synspec}:\\
$=0$ -- normal synthetic spectrum;\\
$=1$ -- spectrum with only a few lines (obsolete);\\
$=2$ -- theoretical continuum, plus H (and possibly He~II) lines only;\\ 
$=-1$ -- the so-called ``iron-curtain option". This is now obsolete since the opacity table option 
(see below) is much more flexible;\\
$=-2$ -- a continuum-only variant of the iron curtain option; also obsolete;\\
$=-3$ -- a standard opacity-table option (total opacity);\\
$=-4$ -- evaluation of an opacity table with continua + H and He~II lines (analogous to IMODE=2)
 \end{description}

When the opacity-table mode is set up (IMODE=$-3$ or $-4$), the input values of temperature and electron or
mass density (thereafter called simply ``density") are specified
in unit 2 ({\tt fort.2}). 
The mass density $\rho$ is typically used as a density parameter.
The reason is that,
for cool models, the electron density is less suitable as a density parameter because
it is influenced by details in the  ionization equilibrium of various metals (Mg, Si, Fe, etc.) and,
moreover, for very cool models it decreases essentially to zero.

Numerically, each temperature/density pair is considered as an ``atmosphere'' represented 
by just 1 depth point, ND=1, with corresponding grid values of $T$ and $\rho$. 
{\sc synspec} then computes the opacity for all these ``atmospheres", i.e., $T/\rho$  pairs, 
consequently and independently of each other.

For each temperature/density pair, {\sc synspec} thus computes a detailed
table of opacity as a function of wavelength. The  internal wavelength grids
for the calculations are set by the program, and are generally different for each
temperature/density pair. 
In fact, the individual wavelength points are set by the same procedure as for
evaluating a synthetic spectrum -- see Paper~I.
In order to construct a universal opacity
table for many temperature/density pairs, the program transforms
the computed opacities to a universal grid of wavelength points that
is also specified in {\tt fort.2}. The wavelengths are logarithmically
equidistant between the starting and ending wavelengths specified as
input parameters (and therefore represent a scale with constant velocity steps). 
After finishing the calculation of opacities
for all temperature/density pairs, the program outputs the resulting
opacity table, in the format required by {\sc tlusty}. The name of this file
is also specified in {\tt fort.2}.

The  transformation of opacities from the internal to the universal wavelength grid 
can be done in two different ways:
\begin{enumerate}
\item By a simple interpolation in wavelength. This procedure resembles the idea
of opacity sampling used in constructing metal line-blanketed model atmospheres 
(see, e.g.,  Hubeny \& Mihalas 2014; \S\,18.5). A drawback of this procedure is 
that if the number of wavelengths in the universal grid is too low, some cores of the 
strong lines, as well as some continuum windows, may be missed.
\item By preserving (at least approximately) the integral of the opacity in the original and 
universal grids. Specifically, the opacity at the universal grid point $i$ with  wavelength
$\lambda_i$, denoted by $\kappa^{\rm tab}(\lambda_i)$, is given by 
\begin{equation}
\label{opacint}
\kappa^{\rm tab}(\lambda_i) = \frac{2}{\lambda_{i+1}-\lambda_{i-1}}
\int_{(\lambda_i+\lambda_{i-1})/2}^{(\lambda_i+\lambda_{i+1})/2}
\kappa(\lambda)\, d\lambda,
\end{equation}
where $\kappa(\lambda)$ is the (detailed) opacity computed for the internal wavelength
points.
Since the spacing of wavelengths in the internal grid is roughly equidistant, the above
equation is replaced by a mean value of opacities in the internal wavelength points that lie in
the wavelength interval\\ $[(\lambda_i+\lambda_{i-1})/2, (\lambda_i+\lambda_{i+1})/2 ]$.
\end{enumerate}

\medskip

There are two basic types of an opacity table, namely:
\begin{itemize}
\item a classical table, which also can come in three flavors, depending on how the values
of the density are selected for each temperature -- see below.
The opacity is constructed for NTEMP values of
temperature, and for each temperature for a several (generally different) values of density. 
\item a model-tailored table. 
Here,
one takes a given model atmosphere, with ${\rm ND}$ depth points, and constructs an
opacity table just for ${\rm ND}$ temperature/density pairs, corresponding to exact values 
of temperature and density of the input model, $(T_d, \rho_d),\ d=1,\ldots,{\rm ND}$.
Besides the temperature and density, {\sc synspec} also takes the electron density
from the input model, instead of evaluating it using the LTE equation of state.
This table may be useful for assessing the accuracy of the numerical
procedure of using pre-tabulated opacities, and for increasing the accuracy of computed
models.
\end{itemize}

To explain a potential utility of a model-tailored table,
let us take an example of computing a model atmosphere for a late-type star
with H and Mg treated explicitly, so that the opacity of these elements is computed on the fly, and
generally in NLTE\footnote{Their opacity may be treated in LTE as well, but such a treatment 
would not offer any advantage over using directly a full opacity table with all species included in it.},
while the opacity of remaining elements being
taken into account through an opacity table. The table was constructed for a discrete set of $T,\rho$
pairs. For each pair, the electron density was determined by an LTE equation of state,
$n_{\rm e}= n^\ast_{\rm e}(T,\rho)$ (where the asterisk indicates an LTE value), and the opacity 
was evaluated as
$\chi(\nu, T, \rho) = \kappa[\nu, T, n^\ast_{\rm e}(T,\rho)]/\rho$, where $\kappa$ is the absorption
coefficient per length (in cm${}^{-1}$), while $\chi$ is the opacity per mass (g cm${}^2$); the latter
is the quantity stored in the opacity table.
The electron density for the resulting model
may differ from its LTE value because of NLTE effects in Mg (and to a smaller extent in H). 
Consequently, the values of the opacity contained in the opacity
table may no longer be sufficiently accurate.
To assess the effects of such errors, one can compute a model-tailored opacity table including H and Mg in NLTE,
and rerun the model atmosphere with this table. If necessary, the overall procedure may be iterated, but we believe that such effects should be relatively small, so that just one recalculation of the atmospheric structure should 
be entirely adequate.

Another practical use is simply an evaluation of the interpolation errors for choosing the steps for the
opacity table.

% ------------------------------------------------------------------------

\subsubsection{Input file {\tt fort.2}}
\label{infort2}

The input file {\tt fort.2} has at least 5 lines, but in some cases (see below)	 more:

\medskip
$\bullet$ {\bf 1st line}: NTEMP, TEMP1, TEMP2:
\begin{description}
\item[NTEMP] --  number of temperatures, and the mode of setting the opacity table:\\
$\bullet$ $>0$ -- the opacity table is computed in the classical mode; NTEMP represents
the number of temperatures;\\
$\bullet$ $=0$ -- the opacity table is computed in the model-tailored mode.
In this case the temperatures, densities, and electron densities are read from input
model, stored in the file {\tt fort.8 } as in the standard run of {\sc synspec}. The remaining parameters
on this line and the two next lines do not take effect. 

\item[TEMP1] -- lowest temperature [K];
\item[TEMP2] -- highest temperature [K].
\end{description}
In the classical mode, the temperatures are set to be logarithmically equidistant.

\medskip
$\bullet$ {\bf 2nd line}: IDENS: 
\begin{description}
\item[IDENS] -- A switch that specifies
the treatment of density, and the nature of the density parameter. It has a dual meaning:\\
(i) the value of mod(IDENS, 10) determines the nature of the density parameter:\\
$\bullet$ mod(IDENS, 10)$=0$ -- (i.e. IDENS = 0, 10, 20) -- the density parameter has the meaning of electron density;\\
$\bullet$ mod(IDENS, 10)$=1$ -- (i.e. IDENS = 1, 11, 21) -- the density parameter has the meaning of mass density;\\
$\bullet$ mod(IDENS, 10)$=2$ -- (i.e. IDENS = 2, 12, 22) -- the density parameter has the meaning of gas pressure. 

At the same time, the value of IDENS determines \\
(ii) the actual values of the density parameter to be used in the opacity table. The next block of
input depends on this value:
\begin{itemize}
\item IDENS $<10$ -- (i.e. IDENS = 0, 1, 2) -- the values of the density parameters are the same for all
temperatures. In other words, the
pairs of temperature/density form a regular rectangle in the $\log T/\log\rho$ space (for simplicity, we take 
here the mass density for the density parameter, but the situation is the same for all possible choices of the
density parameter). In this case the block contains just one record:\\[4pt]
--- {\bf 3rd block -- one line}: NDENS, DENS1, DENS2, where:
\begin{description}
\item[NDENS] -- number of densities;
\item[DENS1] -- lowest density parameter [cgs units];
\item[DENS2] -- highest density parameter [cgs units].
\end{description}
\item IDENS $\geq 10$ and IDENS $<20$ -- similar to above, but the $T/\rho$ pairs occupy a slanted rectangle
in the $\log T/\log\rho$ space.
One specifies the limiting values separately for the lowest and for the highest
temperature. For other temperatures, the limiting vales are obtained by a logarithmic interpolation.
In this case, the next block also contains one record, namely:\\[4pt]
---{\bf 3rd block -- one line}: \\NDENS, DENSL1, DENSL2, DENSU1, DENSU2, where:
\begin{description}
\item[NDENS] -- number of densities;
\item[DENSL1] -- low density parameter limit for the lowest temperature [cgs units];
\item[DENSL2] -- high density parameter limit or the lowest temperature [cgs units].
\item[DENSU1] -- low density parameter limit for the highest temperature [cgs units];
\item[DENSU2] -- high density parameter limit or the highest temperature [cgs units].
\end{description}
\item IDENS $\geq 20$ and IDENS $<30$ -- the density limits and the number of densities are generally different
for each temperature. In other words, the $T/\rho$ pairs occupy a completely irregular area in the
$\log T/\log\rho$ space. In this case, the next block contains NTEMP input records, each containing
the limiting values as before:  \\[4pt]
\medskip
--- {\bf 3rd block -- NTEMP lines}: for each $T$, NDENS, DENS1, DENS2, where:
\begin{description}
\item[NDENS] -- number of densities for the given temperature;
\item[DENS1] -- lowest density parameter [cgs units] for this $T$;
\item[DENS2] -- highest density parameter [cgs units] for this $T$.
\end{description}
Again, the densities for each temperature are spaced logarithmically equidistantly.
\end{itemize}

\end{description}

We stress that
for cool stars, it is most advantageous to use mass density (e.g., IDENS=1, 11, or 21)
as the density parameter.

\medskip
$\bullet$ {\bf 4th block -- one line}: NLAMB, INTTAB, WLAM1, WLAM2:
\begin{description}
\item[NLAMB] -- number of wavelength points in the final (universal) opacity table;
\item[INTTAB] - a switch for determining the mode of transformation of opacities from the
internal to the universal wavelength grid points:\\
$= 1$ -- the opacities in the universal grid are determined by an interpolation in wavelength;\\
$\neq 1$ -- the opacities in the universal grid are determined by approximately preserving the integral
over wavelength, i.e., by Eq. (\ref{opacint});
\item[WLAM1] -- shortest wavelength in the opacity table [\AA];
\item[WLAM2] -- longest wavelength in the opacity table [\AA].
\end{description}

\medskip
$\bullet$ {\bf 5th block -- one line}: TABNAM, IBINGR:
\begin{description}
\item[TABNAM] -- a character variable specifying the name of the file with the opacity table;
\item[IBINGR] --  the format of the opacity table:\\
$\bullet$ $=0$ -- the table is stored in text format;\\
$\bullet$ $>0$ -- the table is stored in binary format.
\end{description}
\medskip
For example, a simple file {\tt fort.2} that can be used as a test, is
\begin{verbatim}
-----------------------------------------------------------------
11  3.0e3    1.5e4
1
7  1.0e-12  1.0e-6
100000  0  900. 110000.
'optab11_7f.dat'    0
-----------------------------------------------------------------
\end{verbatim}
which constructs a table for 11 temperatures between 3000 and 15000 K,
7 values of mass density between $10^{-12}$ and $10^{-6}$g\,cm${}^{-3}$, 
for 100,000 wavelengths between 900 and 110,000 \AA. Tabular values of the
opacity are determined by an approximate version of Eq. (\ref{opacint}).
The table is stored in the file \verb|optab11_7f.dat|, in text format.

If one intends to construct a table (as an example) with the same density limits as before for the
lowest $T$ (3000 K), but with limits $10^{-10}$ and $10^{-4}$ at the highest $T$, and with
interpolated values of the density limits in between, the first three lines of {\tt fort.2} will look like:
\begin{verbatim}
-----------------------------------------------------------------
11  3.0e3    1.5e4
11
7  1.0e-12  1.0e-6  1.e-10   1.e-4
 -----------------------------------------------------------------
\end{verbatim}
For completely irregular area covered by $T/\rho$ pairs, one can code for the first three
blocks, for instance: 
\begin{verbatim}
-----------------------------------------------------------------
11  3.0e3    1.5e4
21 
7  1.0e-12  1.0e-6
8  1.0e-12  1.0e-6
9  3.0e-13  3.0e-7
9  1.0e-13  1.0e-7
9  1.0e-13  1.0e-7
8  3.0e-13  3.0e-7
8  1.0e-12  1.0e-6
7  3.0e-13  1.0e-5
7  1.0e-11  1.0e-4
6  3.0e-10  1.0e-4
5  1.0e-10  1.0e-4
-------------------------------------------------------------------
\end{verbatim}

The next portion of the file {\tt fort.2} is optional. If it exists, it contains information
about changing chemical abundances or removing selected species.
It should be stressed that such changes concern only the evaluation of
opacities, and are meant as a tool for producing an opacity table for use in 
the hybrid NLTE/opacity-table mode of {\sc tlusty}, where some species are
treated explicitly, while the opacity of the remaining species enters
through a specific opacity table that does not contain opacity from
the species to be treated explicitly by {\sc tlusty}.

The equation of state, which essentially specifies a relation between
mass density and electron density and/or the total number density, takes
into account all the species.

We stress that the elemental abundances are specified in the standard
input file for {\sc synspec}. Recall that if nothing else is specified there,
the abundances are taken as the values from the chosen {\em standard solar abundances} set -- see
Sec.\,2.1.

If this portion of the file {\tt fort.2} is missing, all the species considered by {\sc synspec} 
(i.e., all species unless specifically removed by setting MODE=0 in the standard
input file -- unit 5; see \S\.2.2.1) are taken into account in
constructing the opacity table. If the additional part of {\tt fort.2} is present, it contains
a number of input records, one for each species that is going to be
removed (or its abundance changed). Each line contains two numbers:

 \smallskip
 IATOM      RELABN
 \smallskip
 
\noindent where
\begin{description} 
\item[IATOM ] -- atomic number of the species to be removed/modified;
\item[RELABN] -- a factor that multiplies the adopted standard solar abundance to that
used for computing the opacity. The most usual option is  RELABN=0. 
which removes the element completely.
\end{description}

For example, the file {\tt fort.2} may look like
\begin{verbatim}
-----------------------------------------------------------------
11  3.0e3    1.5e4
1
7  1.0e-10  1.0e-6
100000  0  900. 110000.
'optab11_7hmc.dat'    0
  1   0.
 12   0.
 20   0.
-----------------------------------------------------------------
 \end{verbatim}
for constructing an opacity table analogous to that above, but
removing hydrogen, magnesium, and calcium from the resulting
opacity table.

% ------------------------------------------------------------------------

\subsubsection{Input files {\tt fort.55} and {\tt fort.3}}

Let us look at an example of evaluating an opacity table, for the internal wavelength grid that extends 
from  900 to 110,000 \AA, with atomic as well as molecular lines. The molecular line data
data are transmitted through two line lists, for instance, one list for several important diatomic molecules, 
and the other one for TiO.
One can use either an original {\tt fort.55}, or the
new one, together with additional input file {\tt fort.3}. We shall give both examples below.

(i) Original approach:
The file {\tt fort.55} may look like this
\vspace{-4pt}
\begin{verbatim}
-----------------------------------------------------------------
   -3     1     0                        ! imode,idstd,iprint
    0     0     0     0                
    0     0     0     0     0         
    1     0     0     0     0         
    0     0     0                       
 900. -110000. 50.  2000   1.0e-3   0.15  
    2    20    21                        ! nmlist,iunit(ilist)
-----------------------------------------------------------------
\end{verbatim}
In this example, a negative value of the longest wavelength ($-110,000$)
signifies, as explained in Paper~I, that the wavelengths of all lines are taken in
vacuum (unlike the default approach adopted in {\sc synspec} that considers air wavelengths  for 
all lines with wavelengths larger than 2000 \AA). This setting is mandatory for constructing the
opacity tables.

The 4th number on the 4th line, INLIST=0, signifies that all the line lists are in ttext format.

Since this approach specifies only unit numbers for the molecular line lists, one should also
make symbolic links. For example, assuming that an atomic line list {\tt gfATO.dat} and two molecular
line lists {\tt gfMOL.dat} and {\tt gfTiO.dat}\footnote{As is described later on, the file {\tt gfTiO.dat}
is a line list for TiO lines taken from EXOMOL data, and reduced with the rejection parameter
$s_{\rm reject} = -8.5$, which contains about $8.3\times 10^6$ strongest lines out of the original number of
$99.9 \times 10^6$ lines. More details are given in  \S\,\ref{linlists}.}
are either copied or linked to the current directory, then
one has to make the following symbolic links
\begin{verbatim}
ln -s -f gfATO.dat fort.19
ln -s -f gfMOL.dat fort.20
ln -s -f gfTiO.dat fort.21
\end{verbatim}
In this case, all line lists are repeatedly read for all temperature/density pairs, even if, e.g., TiO lines
do not contribute at higher temperatures, and thus the TiO line list is read (and computer time is being spent)
in vain.

(ii) New approach.
In this case file {\tt fort.55} may remain unchanged, or its last line could be removed. As stressed above,
if the file {\tt fort.3} is present, it overwrites the line-lists-related information from {\tt fort.55}. File {\tt fort.3},
can then be, for instance
\begin{verbatim}
-----------------------------------------------------------------
'gfATO.dat'     0
'gfMOL.dat'     0    8000
'gfTiO.dat'     0    5000
-----------------------------------------------------------------
\end{verbatim}
In this case, the line list for TiO is read only for $T<5000$. Assuming now that one also has binary
versions of the molecular line lists, with a suffix {\tt bin}, then one may set up {\tt fort.3} as follows:
\begin{verbatim}
-----------------------------------------------------------------
'gfATO.dat'     0
'gfMOL.bin'     1    8000
'gfTiO.bin'     1    5000
-----------------------------------------------------------------
\end{verbatim}
Another advantage of using input from {\tt fort.3} is that, as mentioned above, the filenames of the
molecular line lists are printed at the standard output file, so one can have an explicit record of
what line list was used, in case there are several variants available.

. 

\subsubsection{Output files}
The basic output is the opacity table, in the format described above -- see \S\,\ref{optab}.
As described there, the file contains, besides the opacity values, a header that
specifies the chemical abundances and the values of the additional opacity switches 
used for constructing the table. This information is used by {\sc tlusty} to avoid possible
inconsistencies in treating opacities. 

The name of the file containing the opacity table is specified in the file {\tt fort.2}, as explained in \S\,\ref{infort2}.

Besides the {\sc tlusty}-compatible opacity table, the program produces
two additional output files that may be used for further inspection
of the results, or in the case there are problems, but these files do not have to be stored.
\begin{description}
\vspace{-2pt}
\item[fort.26] -- continuum opacity, namely a table of continuum wavelengths, versus
the opacity in the continuum, for each temperature/density pair.
\vspace{-4pt}
\item[fort.27] -- all values of opacity on the internal wavelength
grid for each temperature/density pair (i.e., before their interpolation
to the universal wavelength grid). This is a rather big file, and it is produced
only if the parameter IPRIN (specified in {\tt fort.55}) is larger than~2.
\vspace{-4pt}
\item[fort.29] -- an overview of the number of lines considered (atomic and molecular), and the 
computer time for each temperature density pair.
The table is more or less self-explanatory. Briefly, the individual columns are:
temperature index, density index, value of $T$ [K], values of $\rho$ [g cm${}^{-3}$], 
electron density, number of atomic lines selected, number of molecular lines from the individual
line lists (all in thousands), and the computing time spent to produce the opacities on the internal grid [s].
\vspace{-4pt}
\item[fort.63] -- a binary version of the opacity table. This can be used by {\sc tlusty} if
the input parameter IBINOP is set to a non-zero value -- see also \S\,\ref{fort15}.
\end{description} 

\subsubsection{The utility program {\sc ngrid}}

Since computing the monochromatic opacities on the internal wavelength
points is the most time-consuming part of the construction of an opacity
table, it is sometimes advantageous to generate and store the auxiliary output file
{\tt fort.27}, and subsequently use it for generating another opacity table with a different
universal wavelength grid and/or a different mode of transformation of opacities
from the internal to the universal grid. This can be done with the program {\sc ngrid}.

The program requires the following three input files:
\begin{enumerate}
\item Standard input file -- that contains three input records, the first two being analogous to the last two
records of the input file {\tt fort.2} used to generate the original table, namely

\medskip
$\bullet$ {\bf 1st line}: NLAMB, INTTAB, WLAM1, WLAM2, \\where
\begin{description}
\item[NLAMB] -- number of wavelength points in the new opacity table;
\item[INTTAB] - a switch for determining the mode of transformation of opacities from the
internal to the new universal wavelength grid:\\
$= 1$ -- the opacities in the new table are determined by an interpolation in the wavelengths;\\
$\neq 1$ -- the opacities in the new table are determined by an approximate version of Eq. (\ref{opacint});
\item[WLAM1] -- shortest wavelength in the opacity table [\AA];
\item[WLAM2] -- longest wavelength in the opacity table [\AA].
\end{description}

\medskip
$\bullet$ {\bf 2nd line}: TABNAM, IBINGR, \\where
\begin{description}
\item[TABNAM] -- a character variable specifying the name of the file with the new opacity table;
\item[IBINGR] --  the format of the new opacity table:\\
$\bullet$ $=0$ -- the table is stored in text format;\\
$\bullet$ $>0$ -- the table is stored in binary format.
\end{description}
\medskip

$\bullet$ {\bf 3rd line}:   OPTABLE, IBINOP, \\where
\begin{description}
\item[OPTABLE] -- a character variable specifying the name of the previously generated 
opacity table. 
\item[IBINOP] --  the format of the original (old) opacity table:\\
$\bullet$ $=0$ -- the table is stored in text format;\\
$\bullet$ $>0$ -- the table is stored in binary format.
\end{description}

\item File {\tt fort.27} -- as described above, this is the file that contains the set of opacities in the internal wavelength points, generated by a previous run of {\sc synspec}.

\item An old opacity table, whose name is specified by the input parameter OPTABLE. This file is needed
just for providing a header for the new opacity table, tracking the information from the original calculation.
\end{enumerate}

%-------------------------------------------------------------------------------------------
\subsection{Equation of state}
\label{eossyn}

All the improvements in the equation of state (EOS), and the corresponding keyword parameters controlling them,
used in {\sc tlusty208} and described in \S\,\ref{eostlu},
apply to {\sc synspec54} as well. In fact, the improvements in the equation of state are more important in {\sc synspec} than in {\sc tlusty}, because the most important effect of molecules, namely the
molecular line opacity, is evaluated by {\sc synspec} and enters {\sc tlusty} only as an input, pre-calculated,
opacity table. 

To evaluate the molecular line opacity, not only the molecular number densities are needed, 
but also accurate values of the partition functions. We will now describe the treatment of molecular partition functions in detail.

In  previous versions, the partition functions for the molecular species were evaluated using the Irwin (1981) tables,
whenever available, or using the Tsuji (1973) data for molecules not included in the  Irwin tables. In this version.
a new Irwin-like table was created (by YO) by fitting Barklem \& Collet's (2016) results. Many more molecular
species are considered in the new table (324) compared to the original Irwin's table (66).

We have also implemented partition functions provided by the EXOMOL project.\footnote{http://www.exomol.com/data}
In the current version, we include data for the partition functions of 29 diatomic molecules
-- AlO, C${}_2$, CH, CN, CO, CS, CaH, CaO, CrH, FeH, H${}_2$, HCl, HF, MgH, MgO, N${}_2$, NH, NO, NS, NaH,
OH, PH, SH, SiH, SiO, SiS, TiH, TiO, and VO; and 3 triatomic molecules -- H${}_2$O,  H${}_2$S, and CO${}_2$.  We stress
that our current molecular line lists do not yet contain data for some of these molecules, namely 
CS, HCl, HF, N${}_2$, NS, PH, SH, SiS, H${}_2$S, and CO${}_2$.

The EXOMOL tables contain values of the partition functions for temperatures between $T=1$ and $T=T_{\rm max}$
with a step of 1 K. The value of $T_{\rm max}$ varies for the individual species; ranging from 10,000 (for
H${}_2$O,  H${}_2$,, C${}_2$, and FeH) down to 3,000 (for CN, CS, CaH, CrH, MgH, and H${}_2$S), and
being typically 9,000, 8,000 or 5,000 for the rest. In order to provide values for $T>T{\rm max}$, we use the
following expression
\begin{equation}
U(T) = U^{\rm Irwin}(T) \frac{U^{\rm EXO}(T_{\rm max})}{U^{\rm Irwin}(T_{\rm max})}, \quad {\rm for} \quad T>T{\rm max},
\end{equation}
which scales the (available) Irwin values at $T>T_{\rm max}$ to match the EXOMOL data at $T_{\rm max}$.

The EXOMOL project provides the partition function for the individual isotopologues. Since {\sc synspec} works
in terms of the total number densities and partition functions for molecules without a distinction in isotopologues,
we average the individual partition function as follows, for example for a diatomic molecule AB,
\begin{equation}
U({\rm AB})_{\rm av} = \sum_i \sum_j U({}^{i\!}{\rm A}{}^{j\!}{\rm B})\, f^i_A f^j_B,
\end{equation}
where $U({\rm AB})_{\rm av}$ is the averaged partition function for species AB, $U({}^{i\!}{\rm A}{}^{j\!}{\rm B})$ are
the partition functions for the individual isotopologues, and $f^i_A$ is the fractional abundance of the isotope ${}^{i\!}{\rm A}$
of atom A. The summation extend over all physically possible values of $i$ and $j$.

\begin{description}
\item[IRWTAB] - a switch to select the table of fitting coefficients to evaluate 
the molecular partition functions;\\
$\bullet\,=0$ -- the original Irwin's table is used.
 Since the improved table is obviously preferable, we keep this option for possible comparisons with old versions.\\
$\bullet\,\not= 0$ -- an improved table, based on Barklem \& Collet (2016) data for diatomic molecules, is used.\\
DEFAULT: IRWTAB=1
\item[IPFEXO] - a switch to include the EXOMOL partition functions for the above specified molecules:\\
$\bullet\,=0$ -- the EXOMOL partition function are not considered;\\
$\bullet\,\not= 0$ -- the EXOMOL partition functions are considered, and  override other values when available. \\
DEFAULT: IPFEXO=1
 \end{description}

To get more insight into the results, and for comparing to results from other studies,
we have introduced four new output files, {\tt fort.51}, {\tt fort.52}, {\tt fort.53}, and {\tt fort.54}. 
We have also introduced a new global
option in {\sc synspec} that  only solves the chemical equilibrium and generates the files {\tt fort.51} -- {\tt fort.54}
without computing detailed opacities or the emergent radiation intensity. This option is switched on by the
keyword parameter IFEOS.
\begin{description}
\item[IFEOS] -- a switch to activate the ``EOS-only" option.\\
$\bullet \, =0$ -- the standard option to compute either a synthetic spectrum or an opacity table, depending on the input parameters;\\
$\bullet \, > 0$ -- an ``EOS-only" option, in which {\sc synspec} works as in the mode of evaluating the opacity
table for a set of temperature/density pairs, but instead of computing the opacities, it just solves the
global chemical equilibrium to obtain number densities of all atomic and molecular species.\\
$\bullet \, < 0$ -- the EOS quantities and output files are produced for an otherwise standard run (synthetic spectrum
or opacity tables), and the quantities are printed for a loop in depth indices ID=1,ND, and a step of abs(IFEOS). 
For the opacity tables, there is formally only one depth, so in order to print the EOS quantities 
one sets IFEOS$=-1$. For synthetic spectra, setting for instance IFEOS=-30 activates printing the EOS
quantities at depths 1, 31, 61, etc.
\end{description}
In the EOS-only option the input is similar to that for the opacity table option, but simplified because many
parameters used for evaluating the opacities are not needed here. Here is a description of the input files:
\begin{itemize}
\item Standard input (usually named {\tt fort.5}) -- remains the same as for all other applications of {\sc synspec}.
\item {\tt fort.2} -- only the first three lines are functional, the rest may be deleted (but no harm is
done if these are kept).
\item {\tt fort.55} -- not needed.
\item{Line lists} -- not needed.
\end{itemize}

The output files list some interesting 	quantities for all temperature/density pairs. Here is their description:

\subsubsection*{File {\tt fort.51}}
It includes a header that is almost self-explanatory. But to avoid confusion, here is a detailed explanation.
Each record contains:\\[-10pt]
\begin{tabbing}
$T\ \ \ \ \ \ \ \ \ \ $ \= --  \= temperature [K] \\
$\rho$  \> --  \> density [g cm${}^{-3}$]\\
$w_{\rm mol}$   \> --  \> mean molecular weight -- see Eq. (\ref{mu})\\
$N$   \> --  \> the total particle number density  [cm${}^{-3}$] \\
$N_{\rm e}$   \> --  \> electron density [cm${}^{-3}$] \\
$n(\rm{H}_{\rm tot})$   \> --  \> number density of hydrogen {\em nuclei}  [cm${}^{-3}$] \\
$N(\rm{H})\quad$  \> --  \> number density of neutral hydrogen  [cm${}^{-3}$]  \\
$N({\rm H}^-)$   \> --  \> number density of the negative ion of hydrogen  [cm${}^{-3}$] \\
$N({\rm H}_2)$   \> --  \> number density of the hydrogen molecule  [cm${}^{-3}$] \\
$N({\rm H}_2^-)$   \> --  \> number density of the negative ion of hydrogen molecule  [cm${}^{-3}$] \\
$N({\rm H}_2^+)$   \> --  \> number density of the positive ion of hydrogen molecule  [cm${}^{-3}$] \\
\end{tabbing}
 
\subsubsection*{File {\tt fort.52}}
The first three entries are the same as in {\tt fort.51}, then $N_{\rm e}/N$, $N(\rm{H}_{\rm tot})$ ,
$N(\rm{H})$, and $n({\rm H}_2)$, which is the number density of hydrogen {\em nuclei} sequestered in H${}_2$.
The next four entries, $a$(He), $a$(C), $a$(N), and $a$(O) represent a check of the total number densities 
of nuclei of He, C, N, O, in all important species they are constituents of, divided by the chemical abundance. These
numbers should be equal or very close to 1. 
Finally, there are three columns entitled {\tt molfr}(C), {\tt molfr}(N), {\tt molfr}(O), 
that represent a molecular fractions, i.e., thr fraction of C, N, and O nuclei, respectively,
sequestered in molecules.

\subsubsection*{File {\tt fort.53}}
This file summarizes the number densities of the 20 most important molecular species, namely those
for which we have data in the line lists. The printed quantities are $\log_{10}(N/U)$
[cm${}^{-3}$], because these are the quantities that enter the calculation of line opacities. The individual column are:
 temperature [K]. $\log_{10}\rho$ [g cm${}^{-3}$], and then 20 values of $\log_{10}(N/U)$ for the following molecules: 
 H${}_2$, H${}_2$O, OH, CH, CO, CN, C${}_2$, NH, SiH, SiO, TiO, VO, MgH,
CaH, NaH, MgO, AlO, CaO, CrH, and FeH. This file gives useful guidance for reducing the
number of lines in the molecular line lists.

\subsubsection*{File {\tt fort.54}}
The file is quite analogous to {\tt fort.53}; the only difference is that instead of $\log_{10}(N/U)$
one has $\log_{10}[N/n(H)]$, that is the ratio of the molecular number density to the number density
of hydrogen nuclei, for the same molecular species as in {\tt fort.53}.

%-------------------------------------------------------------------------------------------
\subsection{Additional opacities}
\label{syn_opadd}

Changes in the treatment of additional opacities are analogous to the corresponding
changes in {\sc tlusty}. Table 2 summarizes the additional opacity sources, the keyword
parameters that control them, their default values, and the conditions when {\sc synspec}
resets them. The table is analogous to Table 1 that describes the corresponding switches
and actions for {\sc tlusty}. As is easily seen, the only difference from their treatment in
{\sc tlusty} is that in {\sc synspec} the switches for additional Rayleigh scattering 
are reset to zero when an opacity table is constructed.

We underline that {\sc tlusty} will examine the header of any opacity table produced by {\sc synspec}
and  removes automatically any ``additional" opacity source that has already been included.

\subsection{Rejection of weak lines from the line list}

We stress at the outset that the rejection of lines considered here is quite different from {\em reducing} the
number of lines in the line list (mentioned earlier, and described in detail in Appendix G). The latter
reduces the line list itself, which then becomes smaller and easier and faster to deal with. But since
a given line list (reduced or full) may be used for quite different physical conditions, one still has to
reject lines that are weak in those particular conditions in order to make the overall procedure efficient;
that is, fast, yet sufficiently accurate.

As explained in Paper~I, \S\,5.3.1, the standard procedure of rejecting lines from
the line list that are deemed to be too weak 
to contribute to the total opacity is controlled by two input
parameters specified in the file {\tt fort.55}: IDSTD and RELOP. The parameter IDSTD
specifies the index of the depth point that is viewed as the {\em characteristic depth},
that is a depth where roughly $T \approx T_{\rm eff}$, so that the Rosseland optical
depth at this point is close to 1. In terms of depth indices, IDSTD is approximately equal
to (2/3) ND, unless the depth points of the input model atmosphere are set in a non-standard
way. Obviously, in the opacity-table mode, IDSTD=1 since ND is also =1.

\begin{table}
\caption{Additional opacities in {\sc synspec} -- keywords and their values}
\begin{center}
\begin{tabular}[h]{|l|l|c|c|}
\hline
\hline
Opacity source & Keyword & Default  & Set automatically to 0 if \\
\hline
\hline
H- b-f and f-f & IOPHMI & 1 & H- is already an explicit ion \\
\hline
 H${}_2^+$ b-f and f-f & IOPH2P &  1  & IFMOL=0 or $T >$ TMOLIM\\
 \hline
 H${}_2^-$ f-f & IOPH2M &  1  & IFMOL=0 or $T >$ TMOLIM\\
\hline
He- b-f and f-f & IOPHEM & 1  &  \\  
\hline 
CIA  H${}_2$-H${}_2$  & IOH2H2 & 1 &  IFMOL=0 or $T >$ TMOLIM\\
\hline 
CIA  H${}_2$-He  &  IOH2HE & 1  &  IFMOL=0 or $T >$ TMOLIM\\
\hline 
CIA  H${}_2$-H   &  IOH2H & 1  &  IFMOL=0 or $T >$ TMOLIM\\
\hline
CH opacity & IOPCH  & 1 & IFMOL=0 or $T >$ TMOLIM \\
\hline
OH opacity & IOPOH  & 1 & IFMOL=0 or $T >$ TMOLIM \\
\hline
\hline
H Rayleigh scat. & IRSCT & 1 & IMODE$=-3$  (opacity table) \\
\hline
He Rayleigh scat. & IRSCHE & 1 & IMODE$=-3$  (opacity table) \\
\hline
H${}_2$ Rayleigh scat. & IRSCH2 & 1 & IFMOL=0 or $T >$ TMOLIM \\
  &  &  &  or IMODE$=-3$  (opacity table) \\
\hline
\end{tabular}
\end{center}
\end{table}

The rejection parameter RELOP is the critical one. Its meaning is that a line is rejected if
the ratio of the line-center opacity to the continuum opacity {\em at the characteristic depth}
is less than RELOP. Such an approach was used in all previous versions of {\sc synspec},
and it worked quite well for hot and moderately warm atmospheres. However, for solar-type and cooler
stars the local conditions  at the characteristic depth may well be conducive to forming only a limited
number of molecular species, while many others are formed in cooler regions of the
atmosphere. Therefore, many molecular lines would be erroneously rejected if the rejection criterion is
based only on comparing opacities at the characteristic depth.\footnote{One can always compensate by
setting the parameter RELOP to a very low value, say $10^{-15}$ or even lower, but such an approach is 
inefficient and time-consuming because one would also select a number of lines that are weak everywhere
and should be rightfully rejected.}

We have therefore upgraded the previous treatment, essentially allowing for comparing 
line and continuum opacities at
more depth points. The criterion for including a line (i.e., not rejecting it) is that the ratio of
the line-center opacity to the continuum opacity at {\em any of the selected depth points exceeds the value of
the parameter RELOP}. The set of depth points at which the opacities are compared is
specified by the parameter IDSTD, which now has the following meaning:

\medskip

$\bullet$ IDSTD$ > 0$ -- a classical treatment. In this case IDSTD specifies the index of the characteristic depth,
and the opacity is compared at only this one depth point.
\smallskip

$\bullet$ IDSTD$ = 0$ -- The set of ``characteristic depth points" is composed of three depths, with indices
1, (ND-10)/2+1, and ND-9, i.e. id=1,ND,NDSTEP with NDSTEP=(ND-10)/2. This is the recommended option.
\smallskip

$\bullet$ IDSTD$< 0$ -- The characteristic depths are set again as id=1,ND,NDSTEP, but now with
NDSTEP=$-$IDSTD. This option allows for selecting more depths for comparing the opacities;
selecting IDSTD$=\!\!-1$ would lead to comparing opacities at all depth points. While this option
would represent an overkill for classical atmospheres, it may be useful in special circumstances, for 
example for models with a chromospheric temperature rise, atmospheres with shocks, etc.

%____________________________________________________________________

\subsection{Updated treatment of NLTE levels}
\label{nltlev}

This modification provides a better identification of the lower and upper levels of lines
specified through the line list, and the explicit (NLTE) levels specified through the
atomic data input files for {\sc tlusty} and {\sc synspec}. To this end, one needs:
\begin{enumerate}
\item an extended line list that contains more quantum numbers for the individual
levels, and
\item upgraded atomic data files that specify, for each explicit level, appropriate lower 
and upper limits
of quantum numbers. The notion of ``limits'' is invoked because an individual ``level'',
as considered in an atomic data file may in fact be a superposition of several genuine
levels. For a genuine level, the upper and lower  limits coincide. 
\end{enumerate}%
In the standard case, {\sc synspec} associates the energy levels only by means of their
energies (see Paper~I, Appendix C). If additional information is present
(as it is the case when an extended atomic line list is used),
then the program uses not only the energy, but also the quantum numbers $S$ and $L$,
as well as the level parity -- see below. This can be very useful for complex model atoms with
many levels, or with significant uncertainties in their energies.

In the following, we will describe these modifications in detail.
\subsubsection{Modification of the atomic line list}
These modifications were described in \S\,\ref{linlist_at}.
We stress that {\sc synspec} automatically detects whether a line list is in the standard
format or whether it has the 6 new parameters. If they are present, one can still 
force {\sc synspec} to ignore the new information for all ions by setting the parameter INLIST 
(in the input file {\tt fort.55}) to 10 or larger. For more details, refer to \S\,\ref{linlist_spec}.
One can also force {\sc synspec} to
ignore this information for a particular ion by setting the parameter NONSTD in the
standard input file to 10, or, more precisely, if a non-zero value of NONSTD was intended, 
setting NONSTD to the desired NONSTD+10. For details, refer to Paper~III, \S\,4.3.

\subsubsection{Modification of the atomic data file}
The entry for each explicit level contains, in addition to the standard 7 parameters
(ENION, G, NQUANT, TYPLEV, IFWOP, FRODF, IMODL -- see Paper~III, \S\,11.1), 
the following 8 new parameters:
\begin{description}
\item[ENION1, ENION2] -- lower and upper limit for the level energy (analogously to ENION,
they can be in any units -- erg, eV, cm${}^{-1}$, or as frequency, s${}^{-1}$);
\item[SQUANT1, SQUANT2] -- lower and upper limit of the $2S+1$ values;
\item[LQUANT1, LQUANT2] -- lower and upper limit of the $L$ values;
\item[PQUANT1, PQUANT2] -- lower and upper limit of the parities.
\end{description}
In analogy to reading the line list, {\sc synspec} detects whether the additional 
information is present in an atomic data file, and if not, it proceeds in the standard way.

\subsubsection{Additional input parameters}
There are several additional keyword parameters meant for global control of the use
of the energy and quantum number limits:
\begin{description}
\item[ERANGE] -- resets the energy limits in the case when they are set
too close to the actual energy of the level. Namely, if \\ 
(ENION1-ENION)/ENION$ < 10^{-6}$
then ENION1 is reset to\\ 
ENION1=ENION (1+ERANGE); and analogously for ENION2.\\
DEFAULT: ERANGE=0.1 (i.e., 10\%)
\item[ISPICK] -- if set to 0, then the $S$ quantum number is not used for associating the
levels from the line list to the explicit levels\\
DEFAULT: ISPICK=1 (i.e., the S value is being used)
\item[ILPICK] -- analogous switch for the $L$ quantum number\\
DEFAULT: ILPICK=1 (i.e., the L value is being used)
\item[IPPICK] -- analogous switch for the parity\\
DEFAULT: IPPICK=1 (i.e., the parity is being used).
\end{description}

\subsubsection{Input NLTE level populations}
\label{inpnlt}

In the previous versions of {\sc synspec}, the NLTE level populations of all levels 
of all explicit species were transmitted to {\sc synspec} as an input file, {\tt fort.8}.
This file is typically generated by a previous run of {\sc tlusty}\footnote{Such an 
input file could have also been generated by appropriately modifying output of
NLTE level populations generated by another code. While such procedure may be useful in
some cases, it might be cumbersome. It may prove to be more straightforward to rerun
{\sc tlusty} with appropriate atomic data files.} as an output file {\tt fort.7}.

In some applications, in particular in the context of cool stars, it may be useful to have as
input not the absolute values of NLTE level populations (in cm${}^{-3}$), but rather the
absolute $b^\ast$-factors, defined as ratios of NLTE and (absolute) LTE populations, 
as described in \S\,\ref{outchan22}. 
\begin{equation}
\label{bast2}
b_i^\ast \equiv n_i/n_i^{\ast\ast}
\end{equation}
where $n_i$ is a NLTE population of level $i$, and $n_i^{\ast\ast}$ the {\em absolute} LTE population,
obtained by solving the Saha-Boltzmann equations for all levels of all considered ionization
stages of the element.

The computation of a spectrum from an input NLTE model using the standard method (absolute NLTE populations)
or input $b^\ast$ factors will obviously give the same results. However, using a file that contains the 
$b^{\ast}$-factors allows the user to estimate reasonably well the NLTE level populations for a slightly 
different LTE structure without actually computing a new NLTE model.

This option is triggered by a new keyword parameter IBFAC, which acts as follows:
\begin{description}
\item[IBFAC] -- a switch defining the meaning of input NLTE populations on the input file {\tt fort.8}\\
$\bullet\,=0$ -- the input file contains the absolute level populations (in addition to the temperature, electron density,
mass density, and, for models with molecules, the total particle number density) -- the standard option;\\
$\bullet\,=1$ -- the input file fort.8 contains the $b^\ast$-factors instead of absolute NLTE populations.
When using this option, the user has to make sure that the file {\tt fort.8} indeed contains the absolute
$b^\ast$-factors, not the usual $b$-factors. The file is generated by {\sc tlusty} as {\tt fort.22} -- see \S\,\ref{outchan22}.\\
$\bullet\,=2$ -- In this case, one deals with an LTE model atmosphere with the additional input of $b^\ast$-factors,
with the aim to
provide an approximate NLTE line formation for LTE models, as further explained in \S\,\ref{pseudonlt}.
In that case the file {\tt fort.8} is an input LTE model atmosphere, while the $b^\ast$-factors file is to be
copied or linked to a file named {\tt bfactors}.\\
DEFAULT: IBFAC=0 (i.e., the standard option) 
\end{description}

%____________________________________________________________________

\subsection{Subtle points of computing line and continuum opacity}
\label{semiex}

As explained in Paper~I, computing a synthetic spectrum for an LTE model atmosphere
involves taking an information about bound-free transitions (continua) from the input file(s) used
for the construction if the corresponding model atmosphere (if a {\sc tlusty} model is used).
If the input model atmosphere is a Kurucz model, the analogous input files should be used.
More precisely, {\sc synspec} takes into account bound-free transitions from all explicit
levels, as specified by the standard input file, with the corresponding cross-sections taken
from the relevant atomic data files whose names are given in the standard input. The LTE level
populations of all explicit levels are evaluated by the standard Saha-Boltzmann expressions
using the corresponding data (level energies and statistical weights) given by the input
atomic data files.

This procedure is unambiguous for evaluating the continuum opacity. However, there is
a subtlety in evaluating the line opacity. For lines of non-explicit species, the line opacity
uses level populations of the lower and the upper levels with level energies and statistical 
weights taken from the {\em line list}.
The way lines of explicit species are treated depends on the input model atmosphere:
\begin{itemize} 
\item If the input model atmosphere is an LTE model, the input parameter INLTE
(input provided in the file {\tt fort.55}) is reset to 0, so that the line opacity is evaluated using 
data (level energies, oscillator strengths, broadening parameters) from the line list.
\item If an input model is NLTE,
the relevant parameters for the lower and upper level are taken from the
{\sc tlusty} atomic data files. A problem may arise if a ``level" referred to in an atomic data file
is in fact a representation of several actual energy states lumped together. In that case,
{\sc synspec} first associates a level referred to in the line list to that referred to in an
atomic data file, as described in the previous section, properly scales the level populations
by the statistical weights,
but still uses the energy from the {\sc tlusty} atomic data file to evaluate the corresponding
level populations. Since the latter level energy may represent an {\em averaged} energy of
a group of actual levels, the LTE population computed in this way may be different from the
LTE level population computed using actual level parameters from the line list. We stress that
similar problems may occur even for single (explicit) levels, simply because there is no
a priori guarantee that the level energies in the line list are exactly equal to those given in
the atomic data file.
\end{itemize}

The most significant problem may arise if an input model is nominally NLTE, and uses model
atoms that are reasonably sufficient for some species (say H, Mg for a solar-type star), 
but uses very simple model atoms for metals, essentially with only very few low-lying levels, 
with the aim just to capture the most important contributions to the
UV continuum opacity. While this approach would indeed reasonably describe the continuum opacity,
it could lead to significant inaccuracies in the line opacity of these species, as explained above.

\subsubsection{Semi-explicit species}
\label{semiex}

To avoid these potential problems, we introduced a hybrid option of ``semi-explicit" species.
These are those which are considered as {\em explicit} for evaluating the continuum opacity
(that is, with their level population and bound-free cross-section evaluated using data from the
{\sc tlusty} atomic data files), while considered as {\em non-explicit} when evaluating the line
opacity (that is, using data solely from the line list), and disregarding (potentially inaccurate)
NLTE populations of their levels. 

This option is invoked by setting the parameter MODE, which is a part of the second block of the
standard input file, described in detail in Paper~III, \S\,4.2, namely MODE=4. For convenience, 
we will provide a more detailed description of this parameter below.

\subsubsection{Quasi-explicit species}
\label{quasex}

We have introduced another, somewhat related but distinct, option of
``quasi-explicit" species. Their utility comes from the following observation. Let us assume that the
input model atmosphere was built using very detailed model atoms for, say, H, Mg, and Ca -- these are
true explicit species, and C, N, O, Na, and Si with simple model atoms -- these may be treated by 
{\sc synspec}  as semi-explicit. No other species were explicit in the input 
model.\footnote{We stress  that this is 
just an example used to explain the concept of the quasi-implicit species; not an endorsement 
to construct a model atmosphere in this way.}  However, we have realized, again as an example, that
continua of Al, P, S, Fe, and Ni may provide an important contribution to the total continuum opacity. 
Since these species were not explicit in the input model atmosphere,
their continuum opacity would not be considered by {\sc synspec} at all. In order to remedy this problem,
we have introduced the concept of {\em quasi-explicit species}. The user will provide atomic data files
for them, which may be relatively simple, just to capture the bound-free opacity from the most important 
levels. The populations of such levels will be computed in LTE, and their bound-free opacity {\em will}
be included with the cross-sections determined by the input parameters from the corresponding atomic
data files.

A more realistic example is computing a synthetic spectrum for models that
were computed in the Opacity Table/NLTE (OT/NLTE) mode, for instance models {\tt g55res} or {\tt g55ryb}
considered as test cases in \S\,\ref{tesop}. In these models, only H, Mg, and Ca are treated explicitly,
while the rest of the opacity is provided by a pre-calculate opacity table. When computing a synthetic
spectrum using the input for model, say {\tt g55res}, the only bound-free opacities that would be taken into account
would be those of H, Mg, and Ca. The opacity table was constructed (see \S\,\ref{synoptab}) using as
explicit the following species: He, C, N, O, Ne, Al, Si, S, and Fe. (the opacity of H, Mg, and Ca was excluded
for the use in hybrid models). In order to include the bound-free opacities of all species when computing
a spectrum for a hybrid model, He, C, N, O, Ne, Al, Si, S, and Fe
have therefore to be considered by {\sc synspec} as quasi-explicit species. For more details, see \S\,\ref{hyb}.

To avoid confusion, we stress that while an opacity table is being {\em constructed} by {\sc synspec} 
(in its opacity-table mode -- see above), it is {\em not being used} by {\sc synspec} for computing synthetic 
spectra. The obvious reason is that the opacity table is meant for use in {\sc tlusty}, and thus
does not necessarily have the wavelength resolution appropriate for truly  detailed spectrum synthesis.
In contrast, {\sc synspec} is designed precisely to provide a detailed spectrum (or opacities, 
in the opacity-table mode),  with all opacities computed from scratch.

\subsubsection{Summary of treating chemical species}
\label{sumsp}

Here we provide a description of the second block of the
standard input file, described in detail in Paper~III, \S\,4.2,
which specifies the selection of chemical elements
to be included in the calculations of model atmospheres or spectra, and provides the basic parameters
for these species. The block first contains one or two numbers, NATOMS and IABSET. NATOMS represents 
the highest atomic number of the elements considered, and IABSET, if set, specifies which set of 
the {\em standard solar abundances} is adopted.
If it is not set, the program will default to 0. For details, refer to \S\,\ref{basic}, and Appendix~A.
Subsequently, there are NATOMS input records, each containing three numbers:
\begin{description}
\item[MODE] -- a specification of the mode of treatment for the element:\\
$\bullet\,=0$ -- the element is not considered at all;\\
$\bullet\,=1$ -- the element is treated implicitly, that is it does not contribute to the continuum opacity
(but in the case of {\sc synspec} it does contribute to the line opacity);\\
$\bullet\,=2$ -- the element is considered fully explicitly; i.e. selected energy levels of the selected 
ionization stages are considered explicitly. In the case of {\sc synspec}, these populations
are either read from the input model atmosphere (in the case of NLTE models), or are
computed by {\sc synspec} (in the case of LTE models), using data provided by the atomic
data files.\\
$\bullet\,=4$ -- this is a new option, specific to {\sc synspec}, setting the element to the
{\em semi-explicit} mode, as described above.\\
$\bullet\,=5$ -- this is another new option, specific to {\sc synspec}, setting the element to the
{\em quasi-explicit} mode, described above.
\item[ABN] -- a specification of the chemical abundances of the individual species, in relation to the adopted 
standard solar abundance set,
with the following  convention:\\
$\bullet\,= 0$ -- the standard solar abundance  is assumed \\
$\bullet\,< 0$ -- a non-solar abundance is assumed, abs(ABN) has now the meaning of
         the abundance expressed as a multiple of the standard
         solar abundance (i.e. $-0.1$ means 1/10 of the standard solar abundance, $-5$ means 5 times standard
         solar abundance, etc.);\\
$\bullet\,> 0$ -- a non-solar abundance is assumed, expressed as 
         $N({\rm elem})/N({\rm ref})$, i.e. relative by number
	 to the reference species.  The reference atom is H by default,
	 but the reference species can be changed by means of the
	 optional parameter IATREF (see \S\,7.4.4. of Paper~III)\\
$\bullet\,>10^6$ -- non-homogeneous (depth-dependent) abundance is assumed. In this
  case, the immediately following $N\!D$ lines should be added that
  contain the individual values of the
  abundance (relative to hydrogen by number), for all depth points 
  $d=1,\ldots,N\!D$.
\item[MODPF] -- originally meant as a switch to set the mode of evaluation of the partition functions.  
However, the parameter MODPF is outdated because the evaluation of the partition 
functions (for atoms and ions) is now driven by the keyword parameter IIRWIN -see \S\,\ref{eostlu}.
\end{description}

%--------------------------------------------------------------------------------

\subsection{Summary of treating line and continuum opacities by {\sc synspec}}
\label{sumac}

The previously described treatment of lines and continua might seem complicated and confusing, so we
will summarize here various options and possibilities.
The basic difference is whether the input model is LTE or NLTE. We will describe these two
basic types of generating spectra in turn, with an additional possibility of using, in a given run, 
input data and model parameters for {\em both} an LTE as well as a NLTE model.
\subsubsection{Input LTE model atmosphere; with no additional input}
If the input model atmosphere is a {\sc tlusty}-generated model,
{\sc synspec} takes from the input model atmosphere only  3 or 4 basic parameters
(3 for models without molecules, 4 for models with molecules), namely $T$, $n_{\rm e}$, $\rho$, and,
for models with molecules, $N$.  

If the input model atmosphere is a Kurucz model, then {\sc synspec} takes from the input
$T$, $n_{\rm e}$, and
$P_{\rm g}$, the gas pressure. From these state parameters {\sc synspec} computes $N$ and
$\rho$. If molecules are absent (IFMOL=0), the procedure is trivial: $N=P_{\rm g}/kT$ and 
$\rho = \mu m_H (N-n_{\rm e})$, while if molecules are included, one has to solve the equation of state
as described in \S\,\ref{eossyn}. 

The basic building blocks of the spectrum synthesis are then treated
as follows:
\begin{description} 
\item[Atomic level populations:]  are generally determined by solving the Saha-Boltzmann relations.
\item[Continuum opacity:] is computed as a bound-free opacity for all explicit levels, 
a free-free opacity for all explicit ions, and various ``additional" opacities, as described earlier.
To this end, the standard input file may be identical to the standard input file for {\sc tlusty}
to generate the current model, or may be quite independent of it  -- if one uses for instance 
a Kurucz model atmosphere. In any case, it is important that the standard input specifies as explicit 
atoms/ions those that are expected to provide a non-negligible source of continuum opacity.
Examples of such files are given in Paper~III, and here in \S\,\ref{synoptab} and \ref{hyb}.
\item[Line opacity:] is quite independent of the standard input, and on the choice of explicit atoms and ions.
All elements are allowed to contribute to the line opacity, unless they are specifically excluded (MODE=0 in the
standard input). In other words, line opacity is computed for any non-excluded chemical elements,
regardless of the MODE, that is for MODE $\not= 0$
\end{description}
\subsubsection{Input NLTE model atmosphere}
The situation here is more tricky. But generally, since the point of computing NLTE model atmospheres is to
provide NLTE level populations of selected energy levels of particular atoms and ions, these level
populations are to be communicated to {\sc synspec}. They may, or many not be actually used
by {\sc synspec}. Usually they are, but we have explained
the reasons why they may not be used for the line opacity in \S\,\ref{semiex}.  
The important point is that the model input (unit 8) now contains, in addition to the 3 or 4 
parameters as in the case
of LTE, also the level populations of the energy levels treated explicitly by {\sc tlusty}. Therefore,
the standard input to {\sc synspec} {\em has to conform} to the standard input for {\sc tlusty} 
that generated the given model; in the sense that all species that were explicit in the {\sc tlusty}
input have to be explicit  (or semi-explicit) here, with MODE=2 or 4, and with the same number of
explicit levels. Otherwise, {\sc synspec} will get confused. The previous versions of {\sc synspec}
were much simpler in this respect -- one just used {\em exactly the same} standard input for both 
codes.  This can be done in the present version as well, but now {\sc synspec54} 
can offer a larger flexibility, and several additional
useful options, at the expense of a slight increase in complexity for setting the standard input.

{\bf Important note}: When we talk here about ``input NLTE level populations", we mean either
the traditional number densities (the standard approach), or ``absolute" $b^\ast$-factors -- 
see \S\,\ref{inpnlt}. Which one is included in the model input is determined by the keyword IBFAC, but
the approach to evaluate the line and continuum opacity is the same in both cases.

The basic building blocks of the spectrum synthesis
 are now treated as follows:
 \begin{description}
 \item[Atomic level populations:] Generally, populations of the explicit levels are taken from the model input,
 and others are computed in LTE.
 \item[Continuum opacity:] As mentioned several times before, a treatment of the continuum opacity was extended
 with respect to previous versions:
 \begin{itemize} 
 \item As before, the bound-free and free-free opacity are automatically included for all explicit levels and atoms.
 \item Since this may be insufficient,
 we introduced the  the ``quasi-explicit species", and these have to be
 specified in the standard input, with MODE=5. This is
 not for classical NLTE models, where one was supposed to take
 into account all important opacity sources by means of selecting appropriately the explicit atoms and levels,
 but rather for hybrid Opacity Table/NLTE (OT/NLTE) models where most of the opacity sources are
 included through the opacity table.
 The concept is discussed in \S\,\ref{quasex}, and an
 actual example is given in \S\,\ref{hyb}. The populations of quasi-explicit species 
 are computed in LTE, but their bound-free and
 free-free cross-section are calculeted exactly as for truly explicit species.
 \end{itemize}
 \item[Line opacity:] In this case, the evaluation of line opacity may depend on the parameters
 from the standard input, and obviously on the input NLTE level populations.
 The line opacity can be taken in two different ways:
 \begin{itemize}
 \item For explicit atoms (MODE=2), {\sc synspec} takes into account the NLTE level populations 
 in the input model by associating the level parameters given in the line list and those in the {\sc tlusty}. 
 standard input.
 The subtleties of this approach are discussed in Paper~I, and here in \S\,\ref{nltlev}.
 \item Some species, being taken as explicit in {\sc tlusty}, may be set to the ``semi-explicit" mode for {\sc synspec},
 simply by changing the MODE from 2 to 4. As a consequence, their NLTE populations are taken into
 account for the bound-free opacity, but the level populations for the line opacity are computed in LTE. 
 The rationale for this approach is explained above in \S\,\ref{semiex}.
 \end{itemize}
 \end{description}

 \subsubsection{LTE model atmosphere, with additional (NLTE) input}
 \label{pseudonlt}
 This is a new option that provides a reasonable approximation of NLTE line formation
 for otherwise LTE models. To this end, one uses two previously computed model atmospheres:
 one is a usual LTE model ({\sc tlusty} or Kurucz), and another one is a NLTE {\sc tlusty} model. The LTE
 model provides the atmospheric structure ($T$, $n_{\rm e}$, $\rho$, possibly $N$), while the
 NLTE model provides NLTE level populations. In this case, the level populations {\em have to be
 supplied} as $b^\ast$-factors. From them, {\sc synspec} computes actual NLTE populations using the
internal LTE populations based on the input LTE structure. Here are the basic
 ingredients:
 \begin{description}
 \item[Standard input:] Although the LTE model is a primary input, the standard input to {\sc tlusty}  for
 the current run must be set up in such a way that {\sc synspec} understands the indexing of the input
 $b^\ast$-factors in the second model, so that they can be correctly associated
 associates to the energy levels they were meant to. Since the standard input for LTE models
 is essentially arbitrary as far as the choice of explicit species is concerned, the standard input here
 is essentially the same as the current model were a NLTE one. In fact, the present setup
 represents a NLTE spectrum synthesis; the only 
 difference from a purely NLTE model described in the previous section is that the atmospheric 
 structure ($T$, $\rho$, etc.) is taken from a different, LTE, model.\footnote{Furthermore, this input model
 can in principle be a NLTE model, but such an option is not recommended since in this 
 case one could easily calculate a proper NLTE model and do a spectrum synthesis for it in a traditional way.
 This option is meant to provide a tool for spectrum synthesis for a large set of previously computed 
 LTE model atmospheres without a necessity of recomputing all in NLTE.}
 The standard input may also specify some species as quasi-explicit; that is, their level populations
 are computed in LTE (they are not a part of input NLTE level populations), 
 but their bound-free opacity is taken into account (see \S\,3.7.2). 
 
 The important point is that this option is switched on by setting the parameter IBFAC, and which is one
 of the keyword parameters (see \S\,3.6.4), to IBFAC$=2$.
 \item[Model input files:] Unlike the other options, here we have two input model files, and both
 have to be present:
 \begin{itemize} 
 \item Usual {\tt fort.8} file, which in this case represents the {\em input LTE model atmosphere}.
If this file contains level populations, they will be disregarded.
 \item File {\tt bfactors}, which represents the output file {\tt fort.22} ($b^\ast$-factors) from a 
 previous {\sc tlusty} run. An example is given later in \S\,\ref{hyb}.
 \end{itemize}
 \item[Atomic level populations:]  They are computed in two steps:
 \begin{itemize}
 \item First, LTE level populations are calculated for all non-excluded species (MODE $\not= 0$), 
 based on the standard input,  as for a typical LTE model.
 \item Second, for levels specified as explicit (MODE=2) or  semi-explicit (MODE=4), their
 $b^\ast$-factors were read from the secondary input (NLTE) model---file {\tt bfactors}, and their actual level populations
 are calculated based on these $b^\ast$-factors and the LTE level populations computed for the input LTE model
 structure -- see above.
 \end{itemize}
 \item[Continuum opacity:] 
 is computed as for a regular NLTE model, that is, the bound-free and
 free-free opacities are calculated for all explicit, semi-explicit, and quasi-explicit species.
 \item[Line opacity:] 
 Again, it is computed as for a traditional NLTE model, that is, using NLTE populations 
 for explicit species,
 while LTE populations are adopted for semi-explicit and quasi-explicit species, as described above.
 \end{description}

%%%%%%%%%%%%%%%%%%%%%%%%%%%%%%%%%%%%%%%%%

\section{Graphical user interface programs}
\label{gui}

As mentioned in Paper~I, there are several useful graphical interface programs for
analyzing results from {\sc tlusty}, as well as for running and making graphs with the results
from {\sc synspec}.

\subsection{Plotting output from  {\sc tlusty}}
\l;abel{guitlus}

We will first describe current graphical interface programs for {\sc tlusty}, available both
in the IDL and the Python languages.

\subsubsection{IDL}
As mentioned in Paper~I, the standard {sc tlusty/synspec} distribution contains two IDL programs,
{\tt pconv.pro} and {\tt pmodels.pro}.. Examples of their usage are presented in Paper~III,
\S\,6.1, and also here in Chap.\,5.

The program {\tt pconv} plots the contents of the convergence log, in file {\tt fort.9}, together with
the file {\tt fort.7} that supplies the column mass coordinate, and {\tt fort.69} that
provides the timing information. The program may be called simply as
\begin{verbatim}
IDL>  pconv
\end{verbatim}
in which case it assumes that the input files have their generic names, {\tt fort.9}, {\tt fort.7},
and {\tt fort.69}. If the file names were already modified, with a common core name, for
instance {\tt hhe35lt} (as in one of the test examples described later), namely
{\tt hhe35lt.9}, {\tt hhe36lt.7}, and {\tt hhe35lt.69},{\tt pconv} should be invoked as follows
\begin{verbatim}
IDL>  pconv,'hhe35lt'
\end{verbatim}
An example of a plot generated using this command is presented in Fig.\,1 of Paper~I.

The second basic program is {\tt pmodels.pro}, It plots one state parameter at a time
for one or more {\sc tlusty} model atmospheres, stored in unit 7. 
The first positional parameter is a list of core names of models to be plotted, the 2nd
parameter is an index of the state parameter to be plotted (that is, the order in which it 
appears on the unit 7 output, namely: $0=$ temperature, $1=$ electron density,
$2=$mass density, etc. The 3rd positional parameter is 0 to plot absolute values of the
given state parameter; or 1 to plot the difference of the state parameter relative to the first model.
Finally, the 4th positional parameter indicates that data should be overplot when it is set to 1.
The program also accepts and passes along any intended keyword for the IDL routine 
{\tt plot}, its (their) action is, however, limited
to the plot of the first model. The plots show a state parameter as a function of $\log m$, 
$m$ being the column mass. All state parameters, except temperature, are plotted as logarithms.

For example, let {\tt hhe35lt.7} and {\tt hhe35nl.7} are two computed model atmospheres (see \S\,5.2),
To plot temperature for these two models, one issues the command
\begin{verbatim}
IDL>  pmodels,['hhe35lt','hhe35nl']
\end{verbatim}
To include temperature for another model, say {\tt hhe35nc'} on the same plot, one issues
\begin{verbatim}
IDL>  pmodels,['hhe35nc'],0,0,1
\end{verbatim}
An example of a  plot generated by using these two commands is shown in Fig.\,2 of   Paper~III

To plot a difference (in logarithms) of the mass density between the second and
the first model, one types
\begin{verbatim}
IDL>  pmodels,['hhe35lt','hhe35nl'],2,1
\end{verbatim}
The program {\tt pmodels} also works for the {\sc tlusty} output files {\tt fort.12} and {\tt fort.22},which store the
$b$-factors and absolute $b^\ast$-factors, respectively. For instance, to plot the $b$-factors of the 
ground state of hydrogen for models {\tt hhe35nc} and {\tt hhe35nl}, one calls
\begin{verbatim}
|DL>  pmodels,['hhe35nc.12','hhe35nl.12'],3
\end{verbatim}
where now one needs to put the full name of the files. Using just the core names works only for the
$\ast${\tt .7} files.

\subsubsection{Python}

A new feature of the {\sc tlusty208} distribution is a development of the analogous utility programs
in PYTHON. The module {\tt tlusty.py} contains  the following routines:

(1) {\tt pconv} -- for plotting the convergence log contained in the output
from {\sc tlusty}, in this case from the files {\tt fort.9} and {\tt fort.69} only. The plot is quite
similar to that produced by the IDL program {\tt pconv.pro}; the only difference is that the relative
changes are plotted as a function of depth index instead of column mass.

(2) {\tt pmodel} -- extracting values of the column mass and a selected state parameter
from the {\sc tlusty} output file {\tt fort.7} (condensed model), {\tt fort.12}
($b$-factors for NLTE models), or {\tt fort.22} (absolute $b^\ast$-factors);

(3) {\tt pmodels} -- similar to {\tt pmodel}, but with an option to plot selected state
parameter for one or more {\sc tlusty} models, and a possibility of
plotting differences between individual models. This program is equivalent to the IDL
program {\tt pmodels.pro}.

(4) {\tt pmods} - a plot of a range of state parameters for one or more {\sc tlusty} models.

(5) {\tt pflux} - a plot of the emergent spectrum, either from {\sc tlusty} (unit 14 ), or from
{\sc synspec} (units 7 or 17).

\medskip
\noindent Here are some examples (in plain Python):

\noindent
-- To plot a convergence log of the last computed model:

\begin{verbatim}
>>> import matplotlib.pyplot as plt
>>> import numpy as np
>>> import tlusty as tl
>>> plt.ion()
>>> tl.pconv()
\end{verbatim}
\noindent$\bullet$
To plot a convergence log for a model  {\tt hhe35lt}
\begin{verbatim}
>>> pconv('hhe35lt')
\end{verbatim}
The resulting plot is shown in Fig.1.\\[4pt]
\noindent$\bullet$
To extract log(column mass) and temperature for the same model
\begin{verbatim}
>>> (m35,t35) = tl.pmodel('hhe35lt')
\end{verbatim}
\noindent
which can then be plotted or used for some other purpose.

\begin{figure}[h]
\begin{center}
\label{fig1}
\includegraphics[width=4in]{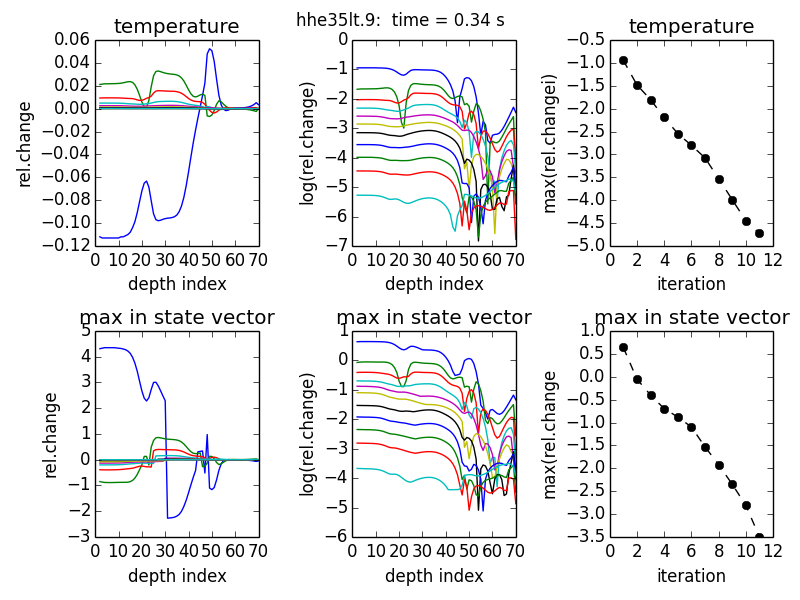}
\caption{Convergence log for the model {\tt hhe35lt}.
The upper three panels exhibit the relative changes in temperature, and the lower three
panels shows the maximum relative change of all state parameters. The left panels show the absolute
values of the changes as a function of depth index, the middle panels the same in the logarithmic scale,
and the right panels the maximum of relative changes at all depths as a function of iteration number.
The title contains the core name of the model, and the total computing time.}
\end{center}
\vspace{-1em}
\end{figure}

\noindent$\bullet$
To plot the temperature as a function of log(m) for the last computed
model, 
\begin{verbatim}
>>> tl.pmodels()
\end{verbatim}
\noindent$\bullet$
And to plot the temperature for three models, {\tt hhe35lt}, {\tt hhe35nc}, and
{\tt hhe35nl}, and then differences from the first model, do
\begin{verbatim}
>>> tl.pmodels(['hhe35lt','hhe35nc','hhe35nl'])
>>> tl.pmodels(['hhe35lt','hhe35nc','hhe35nl'],diff=True)
\end{verbatim}
The actual results of these two commands are presented in Figs. 2 and 3. For
a better resolution, the first plot was obtained by adding two commands,
\begin{verbatim}
>>> plt.xlim(-7,1)
>>> plt.ylim(20000,40000)
\end{verbatim}

%-------------------------------------
\begin{figure}[h]
\begin{center}
\label{fig2}
\includegraphics[width=4in]{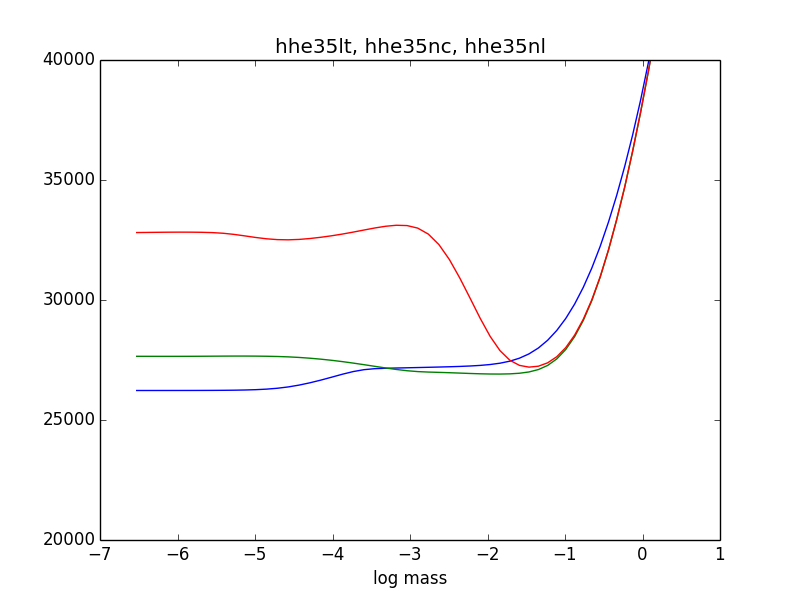}
\caption{Temperature [K] as a function of $\log m$ for three test H-He models
with $T_{\rm eff}= 35,000$ K, $\log g = 4$, namely
{\tt hhe35lt} -- blue line, {\tt hhe35nc} -- green line, and {\tt hhe35nl} -- red line.
For details, refer to \S\,\ref{teststlus}.}
\end{center}
\vspace{-1em}
\end{figure}
%-------------------------------------

%-------------------------------------
\begin{figure}[h]
\begin{center}
\label{fig3}
\includegraphics[width=4in]{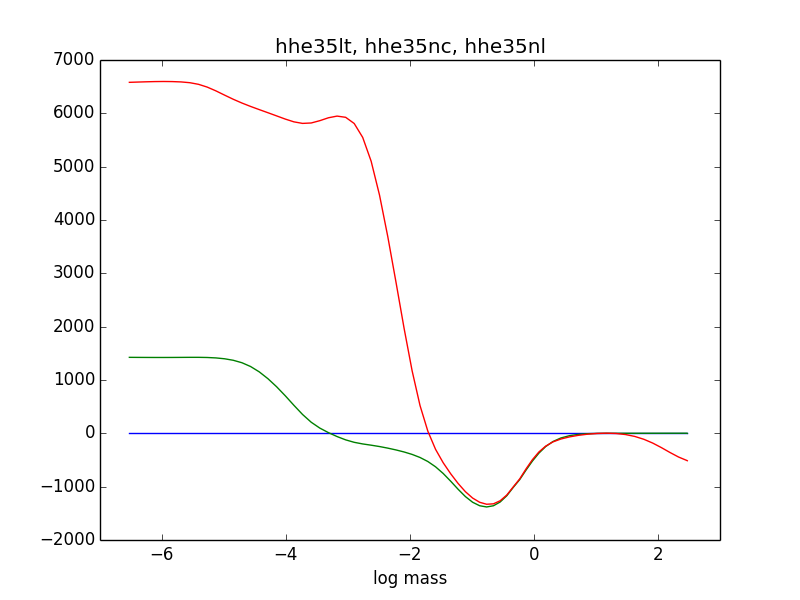}
\caption{Analogous to Fig.\,2, but showing differences of temperature from the
{\tt hhe35lt} model.}
\end{center}
\vspace{-1em}
\end{figure}
%-------------------------------------

%
\noindent$\bullet$
To plot level populations for the first five levels of hydrogen for the
model hhe35nl, 
\begin{verbatim}
>>> tl.pmods('hhe35nl',np.arange(3,8))
\end{verbatim}
The programs {\tt pmodel}, {\tt pmodels}, and {\tt pmods}  work also for {\sc tlusty} output files
 {\tt fort.12} and {\tt fort.22} that
store the $b$-factors, and absolute $b^\ast$-factors, respectively. One now needs to include 
the full file name, not just the core name, e.g.,
\begin{verbatim}
>>> tl.pmods(['hhe35nl.12'],np.arange(3,8))
\end{verbatim}
The resulting plot is shown in Fig.4.

%-------------------------------------
\begin{figure}[h]
\begin{center}
\label{fig3}
\includegraphics[width=4in]{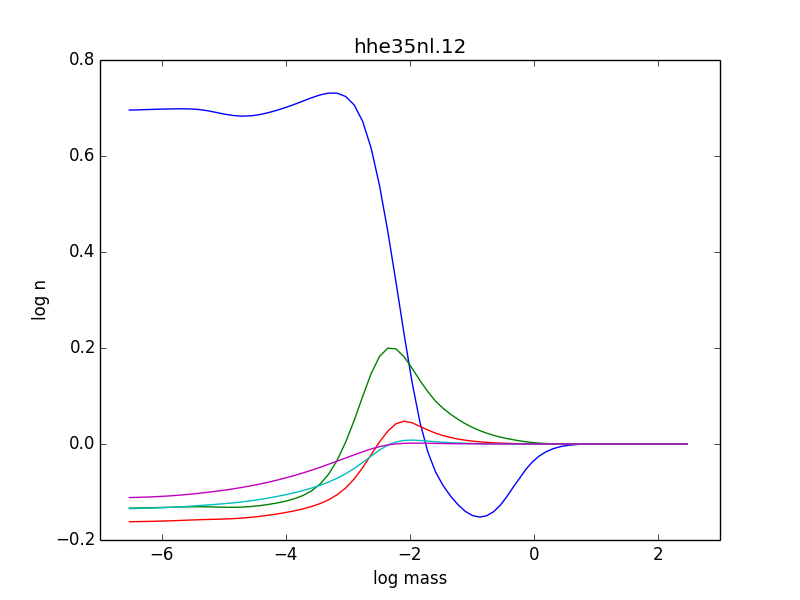}
\caption{A plot of (traditional) $b$-factors for the first five levels of hydrogen for the test H-He model {\tt hhe35nl}.}
\end{center}
\vspace{-1em}
\end{figure}
%-------------------------------------

\noindent $\bullet$
To plot the emergent spectrum for two models, say {\tt hhe35lt} and {\tt hhe35nl},
\begin{verbatim}
>>> tl.pflux(['hhe35lt','hhe35nl'])
\end{verbatim}
where, if no extension is specified, {\tt pflux} assumes {\tt .14}. One can also plot
relative differences (in percentage) of the second model from the first one, 
$(F_2/F_1-1)\times100$,
\begin{verbatim}
>>> tl.pflux(['hhe35lt','hhe35nl'],rel=True)
\end{verbatim}

\subsection{Plotting output from {\sc synspec}}
\label{guisyn}

Since the main output from {\sc synspec}, the files {\tt fort.7} (synthetic spectrum), and {\tt fort.17}
(theoretical continuum), are simple files with two columns -- wavelength versus flux, plotting the
results is trivial both in IDL as well as PYTHON. However, since the input to {\sc synspec} is
relatively simple and straightforward, we have created two packages that serve as a wrapper
to prepare the necessary input files, run the code, possibly perform rotational and instrumental 
convolutions, and plot the results.

\subsubsection{Synplot -- an IDL wrapper} {\sc synplot}  is an IDL wrapper to perform all the tasks outlined above, 
plus providing line identifications based on the output in {\tt fort.12}. The program is described in a separate document.
However, it has been developed to work with an older version of {\sc synspec}. 
This original version of {\sc synplot}, that works with {\sc synspec39}, together with the
the source code of {\sc synspec39} and the users's guide, can be downloaded from\\ [-6pt]

\noindent
{\tt https://www.as.arizona.edu/}$\widetilde{\,\,}\,${\tt hubeny/pub/synplot2.1.tar.gz}\\

There is a recent upgrade of {\sc synplot}, developed by one of us (YO), which is compatible 
with {\sc synspec54}, and which
works in IDL and GDL. The files are available at\\[-6pt]

\noindent
{\tt https://owncloud.iac.es/index.php/s/0CO8xQYTZXjJpxN}\\

The individual files are as follows:
\begin{description}
\item[{\tt synplot.pro}:]
 Is a wrapper for {\sc synspec}, quite analogous to the original {\sc synplot} mentioned above.
 It creates all the necessary input files and links and perform a calculation using {\sc synspec}. 
 The calculation can be a synthetic spectrum (absolute flux, normalized, convolved, LTE, NLTE, with 
 line identification, etc.), or an opacity table. The name of the parameters follows the coding using in this work. 
 Calling {\tt synplot} with no arguments prints the different options of the routine.  
Similarly to the original {\sc synplot}, it does not support all functionalities of {\sc synspec}, for instance
 it works with a specification of line lists using {\tt fort.55}, but not with the new input from {\tt fort.3}.

\item[{\tt loadopt.pro}:]
 Loads an opacity table produced by {\sc synspec} into IDL for testing and/or modifications. 
 The opacity table is loaded as a structure. The routine {\tt printopt} prints a preloaded opacity table, which 
 has to be in text format.

\item[{\tt plotspec}:]
The "save" option in {\sc synplot}, saves a series of files from a {\sc synspec} calculation. 
Those files are restored, and the spectrum is plotted with this routine. This routine can also print labels 
for the lines, both atomic and molecular, present  in the synthetic spectrum. It can also label lines of a particular 
element or a group of elements.

\item[{\tt lineid}$\!\!\!$]\verb|_|{\tt  annot.pro}, \verb|lineid_select.pro|:
These are slight modifications of the analogous routines present in the original {\sc synplot}, used for an
automatic identification and labelling of predicted lines (written originally by D. Lindler, NASA/GSFC).

\item[{\tt convol.pro}:]
Performs a convolution of the spectrum with the rotational and/or instrumental broadening.
\end{description}

\subsubsection{Synple -- a Python wrapper}
Recently, one of us has developed a new Python code, {\sc synple.py}. It is to some extent
analogous to {\sc synplot}, although there are differences. In the present version, it does not
allow to use all functionalities of {\sc synspec}, but again, we plan to upgrade it soon. 

Synple can take as input MARCS, Kurucz, Phoenix and Tlusty model atmospheres, and is designed to be useful to researchers that are not experts in stellar atmospheres and radiative transfer. It facilitates  performing massive spectral synthesis calculations for large grids, parallelizing the calculations over multiple cores, using the python libraries {\tt multiprocessing} or {\tt ray},  and multiple processors, setting up scripts to run {\sc Synspec} on distributed-memory 
computers.

The package can be downloaded from:\\ [-6pt]

\noindent
{\tt https://github.com/callendeprieto/synple}

%%%%%%%%%%%%%%%%%%%%%%%%%%%%%%%%%%%%%%%%%

\section{Downloading and installing the package}
\label{install}

The procedure is similar to that described in detail in Paper~I, Chap. 3, albeit
with some important differences. Here, we provide
a brief description as well, underlining possible differences from the 205 package.

Downloading the package proceeds in one or more steps, depending on the user's
needs. The first file is the main one, which needs to be downloaded in any case.
In addition, there are several files that are optional, The reasons for 
splitting the complete package into several files to download are the following: 
(i) The additional files are large, which may cause problems in downloading; 
(ii) long-time users may already have some or all of the
auxiliary files; and (iii) many applications do not require these files, so for some 
users they would represent an unnecessary waste of computer memory and downloading time.

The files can be downloaded form various sites. The original one, used before,
{\tt http://aegis.as.arizona.edu/}$\ldots$, is no longer operational. 
The present main site is the following:\\[-4pt]

   {\tt https://www.as.arizona.edu/}$\widetilde{\,\,}\,${\tt hubeny/tlusty208-package}\\

At this site, there are four files:
\begin{enumerate}
\item The main file, {\tt tl208-s54.tar.gz}; and three additional files:
\item Line lists -- {\tt linelist.tar.gz},
\item Iron data -- {\tt irondata.tar.gz}, and
\item Opacity tables -- {\tt optables.tar.gz} 
\end{enumerate}

We describe these files in turn.
\subsection{Main file}
The main file, which is sufficient for most purposes, is {\tt tl208-s54.tar.gz}.
It can be downloaded interactively using any web browser;
but we will use here the {\tt wget} command as an example, namely:\\[-8pt]

{\small
\noindent
{\tt wget https://www.as.arizona.edu/}$\widetilde{\,\,}\,${\tt hubeny/tlusty208-package/tl208-s54.tar.gz}\\[4pt]
}
Once it is obtained, one unpacks it, for instance as follows:
\begin{verbatim}
tar xvfz tl208-s54.tar.gz
\end{verbatim} 
That action creates the directory {\tt ./tl208-s54}, with several subdirectories.
Although the user can obviously choose any location for the files, we will use in the
following examples the directory tree that originates from the {\tt tl208-s54.tar} file
extracted in a specific directory, created by the user. To provide a useful guide for the
implementation of the package, we encourage the user to set up an environment variable,
say {\tt TL208}, that specifies such main package directory. If the user sets such directory as,
for instance, $\widetilde{\,\,}${\tt /tlusty208-package} (just one layer below the home directory), 
then it is useful to define {\tt TL208} as: \\[2pt]
{\tt setenv TL208 }$\widetilde{\,\,}${\tt /tlusty208-package} ~~ (in {\tt tcsh})  \\[2pt]
\noindent [or {\tt export TL208=\,}$\widetilde{\,\,}${\tt /tlusty208-package} ~~ (in {\tt bash})]. \\[2pt]
We will use this convention in the following text.
 
Directory {\tt ./tl208-s54} also contains several shell scripts that work
with the directory tree thus generated. In the following text, we will refer to this
directory tree as the {\em standard directory tree}.
If the user prefers, or has already created,
a different directory system, the provided script files have to be modified accordingly. 
In any case, the test cases can be run either individually, or all of them in one run. 

The standard directory tree has the following structure: The main directory 
{\tt ./tl208-s54} contains the shell scripts {\tt RTlusty} and {\tt RSynspec} for 
running {\sc tlusty}, and {\sc synspec}, respectively, and five subdirectories:
\begin{itemize}
\item {\tt tlusty} - which contains the source file for the current version of {\sc tlusty},
called {\tt tlusty208.f}, together with auxiliary files {\tt *.FOR};
\item {\tt synspec} - an analogous directory for {\sc synspec}. It contains the
source files, plus the utility programs {\sc rotin} and {\sc ngrid}.
\item {\tt data} - a collection of needed atomic data. This is  similar to the
subdirectory {\tt data} contained in the {\tt tlusty205} distribution described in Paper~I, Chap.\,3,
but is more extended.
It is described in more detail below.
\item{\tt tests} - The directory that provides the input and output files for various 
test cases for running  {\sc tlusty} and {\sc synspec}. It has two subdirectories:
\begin{itemize}
\item {\tt tests/tlusty} - contains the shell script {\tt Runtest} that runs all the {\sc tlusty} test cases, and five
subdirectories for the test cases, described in \S\,\ref{teststlus}.  The test cases are analogous to
the tests for {\sc tlusty205}, except for the tests of model construction using opacity tables, which are 
now different.
\item{\tt tests/synspec} - Analogously, the directory contains several subdirectories for
various test cases, described in \S\,\ref{testssyn}.
\end{itemize}
\item {\tt gui} - contains the graphical user interface programs described in \S\,\ref{gui}.
\end{itemize}

\subsubsection*{Contents of the ``{\tt ./tl208-s54/data}" directory}
\label{data}

The directory contains several kinds of files; some having just a historical significance and being
kept for comparison with older models.
\begin{enumerate}
\item Original atomic data files. They are essentially the files that were used for constructing grids  of NLTE
metal line-blanketed model atmospheres for O stars (Lanz \& Hubeny 2003) and B stars (Lanz \& Hubeny
2007). The file names consist of the name of the element and the ionization degree, possibly with
an information about the number of energy levels considered, and with the extension ``{\tt .dat}". 
For instance, \verb|ne2_138lev.dat|  is the data file for Ne~II with 138 energy levels included; this file was
used in a detailed study of neon by Cunha, Hubeny, \& Lanz (2006).
\item Extended atomic data files for selected neutral and once-ionized atoms, with essentially all
levels considered by the Opacity Project TopBASE dataset with updated observed energies from NIST. 
The filenames consists simply of name of the element
and the ionization stage (1 corresponds to the neutral atoms), with an extension ``{\tt .t}". For instance,
{\tt mg1.t} contains the data file for Mg~I, with a 71-level model atom (Allende Prieto et al. 2003).
\item Specific data files for photoionization cross-sections of iron-peak elements (remember, the bound-free
cross-section for other atoms/ions are contained in the basic ``{\tt *.dat}" or ``{\tt *.t}" files.). These files
have extension ``{\tt .rap}" (which stands for resonance-averaged profiles; for details refer to
Allende Prieto et al. 2003) -- for iron; or ``{\tt .72}" to ``{\tt .76}" -- for nickel.
\item Irwin-like partition function tables: \verb|irwin_orig.dat| -- the original Irwin (1980) data; and
\verb|irwin_bc.dat| -- a recent update based on results by Barklem \& Collet (2016).
\item Subdirectory {\tt EXOMOL} that contains 32 files for the partition functions of molecules listed in
\S\,\ref{eossyn}, namely AlO, C${}_2$, CH, CN, CO, CS, CaH, CaO, CrH, FeH, H${}_2$, HCl, HF, MgH, 
MgO, N${}_2$, NH, NO, NS, NaH,
OH, PH, SH, SiH, SiO, SiS, TiH, TiO, VO; and H${}_2$O,  H${}_2$S, and CO${}_2$.  
\item Tsuji-like molecular equilibrium data, that is, the fitting coefficients to evaluate the
equilibrium constants for all considered molecular species. The file \verb|tsuji.molec_orig| is the original
Tsuji (1973) table; while the file \verb|tsuji.molec_bc2| is the improved table based on Barklem \& Collet (2016) results.
\item Hydrogen line-broadening data: \verb|tremblay_old.dat| and {\tt tremblay.dat}; two variants of the
Tremblay \& Bergeron (2009) broadening data; the latter file contains data for more lines (first 20
lines of the Lyman and Balmer series, first 19 lines of the Paschen, and the first 10 lines of the 
Bracket series), in contrast to the old file, that contains data for only first 10 members of the Lyman and Balmer
series. There is also file {\tt lemke.dat} that contains Lemke (1997) line broadening data, and the file
{\tt hydprf.dat} with cross-sections by Schoening \& Butler (1989) that is kept 
for possible comparisons to older models.
\item Hydrogen quasi-molecular data, namely files {\tt laquasi.dat}, {\tt lbquasi.dat}, and {\tt lgquasi.dat} for the
cross-sections of the L$\alpha$, L$\beta$, and L$\gamma$ satellites, respectively. For details, refer to
Paper~I, \S\,12.1.2.
\item Helium line broadening data; files {\tt he1prf.dat}, and {\tt he2prf.dat}. For details and references, refer to 
Paper~I, Appendix~A. Briefly, for He~I, {\tt he1prf.dat} contains data for four triplet lines, 
$\lambda\lambda$ 4026, 4387, 4471, and 4921 \AA, after Barnard et al. (1974) and Shamey (1969).
For He~II, file {\tt he2prf.dat} contains data for the lines 2--3, 3--4 to 3--10, and 4--5 to 4--15,
after Schoening \& Butler (1989).
\item Collision-induced absorption (CIA) data for the four CIA processes:
H${}_2$--H${}_2$ collisions (\verb|CIA_H2H2.dat|),
H${}_2$--He collisions (\verb|CIA_H2HE.dat|), H${}_2$--H collisions (\verb|CIA_H2H.dat|), and H--He
collisions (\verb|CIA_HHE.dat|). The corresponding references are given in Table 1.
\item File {\tt linelist.test} that contains a sample atomic line list for performing simple tests
of {\sc synspec}.
\item Last, but not least, three important script files, called {\tt Linklist}, {\tt Linkiron}, and {\tt Linkopta},
which set the necessary symbolic links from the additional files (see below) to the {\tt data} directory, where
they can be easily accessed by {\sc tlusty} and/or {\sc synspec}.
\end{enumerate}

\subsection {Additional files}
\label{addf}

As explained above, extensive line lists, the data for iron lines used to generate
metal line-blanketed models, and the opacity tables for usage in {\sc tlusty}, 
are not a part the {\tt ./tl208-s54} directory.. For some purposes they may not be needed, or 
some users may perhaps already have these or analogous files. However, to run the {\sc tlusty}
and {\sc synspec} test cases described in \S\S\,\ref{teststlus} and \ref{testssyn} they are needed, 
and thus  should be downloaded separately to the main package directory, referenced through 
the environment variable {\tt TL208}, as described above.

In all three cases, upon extracting each file one gets a subdirectory that contains the appropriate files.
The users who already have these files, or some of them, have to provide the corresponding links
themselves.

\subsubsection{Line lists}
\label{linlists}
The tarball with the line lists, called {\tt linelists.tar.gz} can be obtained from the source 
listed above. It should then be  extracted 
\begin{verbatim}
tar xvfz linelist.tar.gz
\end{verbatim}
which generates a subdirectory {\tt ./linelist} which contains the three line lists described 
below, plus the programs {\sc list2bin}, and {\sc redulis}, and the script file {\tt Bin}. 
The two molecular line lists are already reduced, using the
utility program {\sc redulis} described above with a rejection parameter $s_{\rm reject} = -8.5$ --
see Appendix G. More extended molecular line lists are available upon request.
The line lists that are distributed in the present package are: 
\begin{itemize}
\item  {\tt gfATO.dat}, which is the atomic line list in the extended format (that is, containing
additional quantum numbers). It contains data for about $2.3\times 10^6$ lines between 18 \AA\  and 100 microns.
The list is based on Kurucz data\footnote{{\tt http://kurucz.harvard.edu/linelists.html}}, updated by NIST 
data\footnote{{\tt https://www.nist.gov/pml/atomic-spectra-database}} whenever available.

\item {\tt gfMOL.dat} - line list for 18 most important diatomic molecules, containing about
$6\times 10^6$ lines. As mentioned above, the data are based on Kurucz datasets for molecular lines.
\item{\tt gfTiO.dat}  - line list for TiO, containing data for about $8.3\times 10^6$ lines. This list is
based on EXOMOL data. A more extended list extracted from EXOMOL data, from which this
list was obtained, contains information for about $10^8$ lines. Again, more  detailed 
versions available upon request, but
we believe that for most practical purposes the present list is sufficient.
\end{itemize}

Since it is recommendable to work with binary files instead of text files, because
{\sc synspec} works significantly faster with them, the script {\tt Bin} generates the binary equivalents 
of the three lists.
Before using the script {\tt Bin}, one needs to compile  {\sc list2bin}, for instance as
\begin{verbatim}
gfortran -o list2bin.exe list2bin.f
\end{verbatim}

Once this is done, one may go to the standard ``data" directory, for example
{\tt ./tl208-source/data}, and use the script {\tt Linklist} to provide links from the directory where
the line lists are located to the ``data" directory. Before running {\tt Linklist}, one {\em has to set up 
the environment variable {\tt LINELIST} to the directory where the line lists are located}. For instance,
if the file {\tt linelist.tar.gz} was downloaded to and extracted in the main package directory, then one sets
the environment variable and runs the script as \\[4pt]
{\tt setenv LINELIST }\${\tt TL208/linelist  } (in {\tt tcsh}),\\[2pt]
%[or {\tt export LINELIST=}\verb|~|{\tt/linelist  } (in {\tt bash})]\\[2pt]
[or {\tt export LINELIST=}\${\tt/TL208/linelist  } (in {\tt bash})]\\[2pt]
{\tt Linklist}\\[4pt]
We stress that this is not necessary from the point of view of an operation of {\sc synspec}, but
it is advantageous to set up the test cases in an unambiguous way.

We also mention that the standard "{\tt data}" directory also
contains a  very short sample  line list  and therefore
some tests of {\sc synspec} can be done without
above-mentioned extensive line lists. Additional line lists in the {\sc tlusty}-compatible format,
as well as some atomic data files and various model atmospheres,
are also available.\footnote{https://owncloud.iac.es/index.php/s/6x5mRhXj7Ekyc9H}

\subsubsection{Iron data}
The data are the same as those mentioned in Paper~I, \S\,3.3. The users who already have
them do not need to download them again, only to link them to the current standard "{\tt data}"
directory. For the users who do not have them, and who intend to compute 
NLTE metal line-blanketed model atmospheres for early-type
stars, as shown on the test case presented in Paper~I, Chap. 6, in Paper~III, \S\,6.3,
and here in \S\,\ref{teststlus},
need to acquire these data, again from the source mentioned above. The corresponding 
file is {\tt irondata.tar.gz}, and its content is similar but more extended than that of the old file {\tt gf26.tar.gz}, 
It is extracted it as before,
\begin{verbatim}
tar xvfz irondata.tar.gz
\end{verbatim}

As a results, a subdirectory {\tt ./irondata} is created containing 
several files that represent Kurucz data files for iron levels and transitions, 
the files {\tt gf2600.gam} to {\tt gf2608.gam}, and {\tt gf26*.lin}  from which
{\sc tlusty} generates necessary parameters for iron superlevels and 
superlines - see Paper~II, \S\,3.6.
Analogously to the case of line lists, one should go to the standard ``data" directory, and set the links
to these files. Unlike the case of line lists, these links are now {\em mandatory}. The procedure is 
analogous: one needs to set the environment variable {\tt IRON} to the directory where the
downloaded iron data are located, and then run the script {\tt Linkiron}.
We stress that the users who do not intend to compute NLTE metal line-blanketed model atmospheres
do not need to download and store the iron data.

We stress that besides these data for iron ions that are used to construct superlevels and superlines
(Hubeny \& Lanz 1995; Hubeny \& Mihalas 2014), the standard ``data" directory also contains data for 
Fe~I and Fe~II where the usual treatment of levels and transitions is adopted.

\subsubsection{Opacity tables}
\label{addoptab}

Upon extracting the {\tt optables.tar.gz} tarball, one obtains 
a subdirectory called {\tt ./optables} that contains
two new opacity tables, \verb|optab11_7f.dat| -- the full table for the opacity of all species, and
\verb|optab11_7hmc.dat| -- the table that excludes the opacity of H, Mg, and Ca. 
These files can also be obtained by running one of the test
cases for {\sc synspec} -- see \S\,\ref{synoptab}, but are included for convenience, and also because
their evaluation is relatively time-consuming (taking about 27 and 31 minutes, respectively, on an iMac
with a 3.4 GHz Intel Core i5 CPU). Again, one then goes to the standard ``data'' directory, and uses the script 
{\tt Linkopta} to make symbolic links to these files in the standard
``data" directory. Again, in order to use the script {\tt Linkopta}, one first needs to set the environment
variable {\tt OPTABLES} to the directory where the opacity tables are located.

The files are used in the test cases for running {\sc tlusty} with the Opacity 
Tables, as described in \S\,\ref{tesop}.

%-----------------------------------------------------------------------

\subsection{Summary of implementation}
\label{sumim}

Here we summarize all the steps necessary for a successful implementation of {\sc tlusty} and {\sc synspec}.
For simplicity we assume that the reader needs all the available additional files. If not, the user should make the corresponding changes.
\begin{enumerate}

\item Create a convenient main package directory, and set up the environment variable {\tt TL208} that specifies
the name of the directory. For instance, if the user sets such directory as
$\widetilde{\,\,}${\tt /tlusty208-package} (just one layer below the home directory), 
then {\tt TL208} is set up as: \\[2pt]
{\tt setenv TL208 }$\widetilde{\,\,}${\tt /tlusty208-package} ~~ (in {\tt tcsh}), \\
or\\
{\tt export TL208=\,}$\widetilde{\,\,}${\tt /tlusty208-package} ~~ (in {\tt bash}).

\item Go to this directory, (i.e. {\tt cd }\${\tt TL208}), and download the main file, plus the additional files, there. 
This is done either interactively, going to

{\tt https://www.as.arizona.edu/}\verb|~|{\tt hubeny/tlusty208-synsp54} 

\noindent
and get the files {\tt tl208-s54.tar.gz}, {\tt linelist.tar.gz}, {\tt irondata.tar.gz}, and {\tt optables.tar.gz}\\
or, analogously, using command {\tt wget},\\
{\footnotesize
{\tt wget https://www.as.arizona.edu/}\verb|~|{\tt hubeny/tlusty208-synsp54/tl208-s54.tar.gz}}\\
\noindent
and similarly the other files.

\item Extract all the downloaded files\\
{\tt tar xvfz tl208-s54.tar.gz}\\
{\tt tar xvfz linelist.tar.gz}\\
{\tt tar xvfz irondata.tar.gz}\\
{\tt tar xvfz optables.tar.gz}
\item Set up the environment variables corresponding to all downloaded files  
\vskip-12pt
\begin{tabbing}
 xxxxxxxxxxxxxxxxxxxxxxxxxxxxxxxx \= xxxxxxxxxxxxxxxxxxxxxxxxxxxxxxxxxxxx \= \kill
 in {\tt tcshell}:      \>        in {\tt bash}:   \\
{\tt setenv TLUSTY }\${\tt TL208/tl208-s54}  \>    {\tt export TLUSTY=}\${\tt TL208/tl208-s54}\\
{\tt setenv LINELIST }\${\tt T208/linelist}  \>    {\tt export LINELIST=}\${\tt TL208/linelist}\\
{\tt setenv IRON }\${\tt TL208/irondata}  \>    {\tt export IRON=}\${\tt TL208/irondata}\\
{\tt setenv OPTABLES }\${\tt TL208/optables}    \>    {\tt export OPTABLES=}\${\tt TL208/optables}
\end{tabbing}
\item Go to the directory {\tt ./linelist} and compile program {\sc list2bin}, e.g. as\\
{\tt gfortran -o list2bin.exe list2bin.f}

\item Run the script {\tt ./Bin} in the same directory.\\
To be on the safe side, do {\tt ls} to verify the the ``{\tt *.bin}" files were indeed created.

\item Go now to the directory {\tt ../tl208-s54/data} (using the environment variable {\tt TLUSTY} the command
would simply be {\tt cd \$TLUSTY/data}) and run the three linking scripts:\\[-12pt]
\begin{verbatim}
./Linklist
./Linkiron
./Linkopta
\end{verbatim}
If the environment variables {\tt LINELIST}, {\tt IRON}, and {\tt OPTABLES} were set properly,
a bunch of new files (actually links) will appear in this directory, ({\tt gf*}, and {\tt optab*}).

\item  Go to the directory {\tt \${}TLUSTY/tlusty}, and compile {\sc tlusty}, e.g., as\\[2pt]
{\tt gfortran -fno-automatic -O3 -o tlusty.exe tlusty208.f}\\[2pt]
or, if needed, as\\[2pt]
{\tt gfortran -fno-automatic -mcmodel=large -o tlusty.exe tlusty208.f}\\[2pt]
For more details, see below.

\item Go to the directory {\tt ../synspec}, and compile {\sc synspec} analogously:\\[2pt]
{\tt gfortran -fno-automatic -O3 -o synspec.exe synspec54.f}\\[2pt]
or, similarly, as\\[2pt]
{\tt gfortran -fno-automatic -mcmodel=large -o synspec.exe synspec.f}\\[2pt]
Again, for details, see below.

\end{enumerate} 
\noindent
This essentially concludes the basic implementation. In order to perform the tests, one can simply do:
\begin{enumerate}
\item {\tt cd \${}TLUSTY/tests/tlusty}, and run all tests by simply issuing command\\
{\tt Runtest}\\
One then can go to the individual subdirectories and examine the results.
\item Analogously, one then can run the {\sc synspec} tests by\\
{\tt cd \${}TLUSTY/tests/synspec}, issue the command {\tt Runtest} there, and again examine the results 
in all subdirectories.
\end{enumerate}
A detailed explanation of the test cases is provided below.

%------------------------------------------------------------------------

\section{Compiling, testing,  and running TLUSTY}
\label{run}

\subsection{Source code files and compilation}

The program is
distributed as a single file that contains all subroutines,
plus several (8) small files referred to in the {\sc tlusty} source file by {\tt INCLUDE} 
statements, which declare most variables and arrays.
These can be changed to recompile the code and decrease or increase its
memory consumption, as explained in Paper III, \S\,\refcompcomp.  
All files should reside in the same directory.

The  compilation of the program is explained in more detail in Paper III,
\S\,\refcompcomp, where the compiler instructions on different
platforms are summarized. Here we only illustrate the compilation using
{\tt gfortran}, which is available on most Mac and LINUX platforms,
\begin{verbatim}
        gfortran -fno-automatic [-O3] -o tlusty.exe tlusty208.f   
\end{verbatim}
\noindent where the option {\tt "-fno-automatic"} indicates the static allocation 
of memory. The level-3 optimization ({\tt "-O3"}) should be switched on since it
improves the performance of the code considerably. For more details, and several
examples of compiling the program on different platforms, refer to Paper~III,
\S\,\refcompcomp.

Compared to the standard distribution of {\sc tlusty205}, the present one is set up
to allow for model atmospheres with a large number of explicit levels (MLEVEL=1134),
frequency points (MFREQ=135000), and a  relatively large number of depth points
(MDEPTH=100), which are the parameters in the current file {\tt BASICS.FOR}.
As a consequence, the executable file takes about 2.75 GB. Although
it is large, this is now acceptable for most modern computers. As explained in Papers~I and III,
if this presents a problem,
users who do not intend, for instance, to compute sophisticated NLTE metal
line-blanketed model atmospheres, similar to the grids OSTAR2002 and BSTAR2006
(Lanz \& Hubeny 2003, 2007) can decrease MLEVEL, or possibly MFREQ, to  
appropriate values.

Even for computers with  large enough memory, some compilers do limit the  individual
COMMON blocks to a certain size, the part of the memory known as the stack. 
In particular, the COMMON block CURDER, referred
to in the file {\tt MODELQ.FOR}, contains four 2-dimensional arrays {\tt MLEVEL} $\times$ {\tt MDEPTH},
which, for the present parameters, may be too large for certain compilers. In this case one 
needs to compile the program as\footnote{I.H. is indebted to Jano Budaj for alerting him about
this possibility.}
\begin{verbatim}
gfortran -fno-automatic -mcmodel=large -o tlusty.exe tlusty208.f
\end{verbatim}

\subsection{Test cases for {\sc tlusty}}
\label{teststlus}

As mentioned earlier, the standard {\tt tlusty} directory contains a subdirectory 
\${\tt TLUSTY/tests/tlusty},
which in turn contains five subdirectories for five different test cases:
\begin{itemize}
\item {\tt hhe} -- a simple H-He LTE and NLTE model atmosphere - see  
Paper~III, \S\,6.2. This test is analogous to that for {\sc tlusty205}, yet the results are
slightly different due to changes in setting the default values of some keyword
parameters. In particular, the present case takes into account Rayleigh scattering on
neutral hydrogen and helium, while the {\sc tlusty205} test did not. 
\item {\tt bstar} -- a NLTE metal line-blanketed model of a B stars, analogous to models
in the {\sc bstar2006} grid (Lanz \& Hubeny 2007); just a formal solution without
any global iterations -- see Paper~III, \S\,6.3. Again, the
results are slightly different for the same reasons as for the H-He models.
\item {\tt cwd} -- an LTE and NLTE model of a moderately cool DA (pure-hydrogen)
white dwarf - see Paper~III, \S\,6.5. The results are now appreciably different from the
analogous test performed for {\sc tlusty205} because of an improved evaluation of
the adiabatic gradient to describe the convective flux. Also, the keyword parameter 
file was somewhat modified with respect to that used with {\sc tlusty205}.
\item {\tt disk} - a simple H-He LTE and NLTE model of a vertical structure of an
accretion disk - see Paper~III, \S\,6.6. Similarly to the H-He test case, 
the results are here slightly different due to an automatic inclusion of the
Rayleigh scattering. 
\item {\tt optab} -- tests involving an opacity table, with both a full table and a partial table.
The full table mode is demonstrated on an LTE line-blanketed model of a G star.
This test case is similar to that discussed in  Paper~III, \S\,6.4; only 
here we use new, very simple, opacity tables, \verb|optab11_7f.dat| and \verb|optab11_7hmc.dat|, 
which can be obtained by running
one of the test cases for {\sc synspec} -- see \S\,\ref{synoptab}. These test opacity tables are 
included as well in the additional tarball, {\tt optables.tar.gz} -- see \S\,\ref{addoptab}.
The directory also contains input files for two new NLTE models, which will be described in more detail 
below.
\end{itemize}
All subdirectories contain (i) input data for a run, (ii) an  AAAREADME file with a short description of
the tests; (iii) a script file {\tt R1} to run all the tests, 
and (iv) a subdirectory named {\tt ./results} that
contains several important output files obtained
by running the test cases on an iMac (macOS Mojave, 10.14.2, with 3.4 GHz Intel Core i5 CPU)
done by one of us (I.H.), which are 
included to enable a comparison with user's runs for the purpose of checking the implementation
of the codes on the user's platform. 

\medskip

\noindent {\bf Important notes:}
\begin{itemize}
\item In order to use the provided script {\tt Runtest}, or the individual {\tt R1} scripts located in the
individual subdirectories, it is {\em mandatory to set up the environment variable} TLUSTY to the main
{\sc tlusty}/{\sc synspec} directory, as specified above.
\item The script {\tt Runtest}, located in the \${\tt TLUSTY/tests/tlusty} directory, is devised to run all the available tests
in a row. One can also do that manually, i.e. to go consecutively to the individual subdirectories and to
use scripts {\tt R1}, or to use the basics script {\tt RTlusty}.
\item The basic script for running {\sc tlusty}, called {\tt RTlusty}, performs most of the tasks needed to run the code.
It accepts up to two formal parameters:\\[2pt]
-- the core name of the model to be computed;\\
-- the core name of the starting model. If it is missing, then the model is assumed to be computed from scratch.\\[2pt]
The file with the core name and extension {\tt .5}, typically the standard input to {\sc tlusty}, must be present; otherwise
the script stops. Similarly, if the core name of the starting model is present then the file with this core
name plus extension {\tt .7} must be present as well.\\[2pt] 
The script then establishes a symbolic link to the {\tt ./data} directory, runs {\sc tlusty}, and copies 
some of the output files to the core name plus the corresponding extension. As explained in Papers~I and III,
these files are:\\
-- [core name]{\tt .6} -- standard output\\
-- [core name]{\tt .7} -- condensed model\\
-- [core name]{\tt .9} -- convergence log\\
-- [core name]{\tt .14} -- emergent flux $F_\lambda$\\
-- [core name]{\tt .69} -- timing\\

\item In some cases (disk models; using opacity tables), one needs to have additional input files,
for instance file {\tt fort.1} that sets the disk mode, or {\tt fort.15} that specifies the name of the opacity table.
One can still use the script {\tt RTlusty}, but before running it the user has to make sure that the additional input
files are copied or linked properly. The user is encouraged to look inside the script file {\tt Runtest} that provides
an actual demonstration how to use the script {\tt RTlusty}.

\item There is an analogous script {\tt RSynspec} for running {\sc synspec}, but this one is most useful for
computing synthetic spectra of hot and warm stars, where only an atomic line list is used. Its usage is
demonstrated  on some of the test cases in the directory \${\tt TLUSTY/tests/synspec}.
\end{itemize}

\subsubsection{Standard model atmospheres and disks}

As explained above, the subdirectories {\tt hhe}, {\tt cwd}, and {\tt bstar} contain three test cases for typical model
atmospheres, with increasing complexity, and the subdirectory {\tt disk} a very simple model of one annulus of
an accretion disk. These models are discussed in detail in Paper~III, Chap.\,6. There are also individual 
{\tt AAAREADME} files in each directory that summarize the basic properties of the test models.

\subsubsection{Tests involving an opacity table}
\label{tesop}

There are three test cases here. One involves computing an LTE model atmosphere from scratch using
a full opacity table ({\tt g55l}), and two NLTE models that use this LTE model as a starting one,
namely\\
-- {\tt g55nres}  -- a restricted NLTE model; that is a model in which the structure
(temperature, density, electron density) is held fixed, and one computes NLTE level populations 
of H, Mg, and Ca levels, using a reduced opacity table that excludes these elements. The model is
computed using the standard Complete Linearization/Accelerated Lambda Iteration (CL/ALI)
method (see Paper~II, Paper~III, and Hubeny \& Lanz 1995);\\
-- {\tt g55nryb}  -- a similar model that computes a full NLTE model in which departures form LTE are allowed for
H, Mg, and Ca, but where also the model structure is allowed to change as a response to NLTE effects
in these species. 

The background philosophy and adopted procedures to calculate these models were explained in \S\,\ref{ot/nlte}.
Here, we  describe the contents of the standard input file and the keyword parameter files in some detail.
We also stress that we use here relatively simple model atoms of Mg and Ca, in order to
provide a fast and inexpensive test run of {\sc tlusty}.  The resulting model is meant only as a test 
of performance of the code, not as a model that can be used for actual spectroscopic
diagnostics work, For such purposes, the user is encouraged to use more detailed model atoms, 
which are also available in the {\tt ./data} directory. Similar NLTE calculations, using much more
detailed model atoms, not only for Mg and Ca, but also for Na and K, were recently presented by Osorio et al. (2020).

\medskip
The LTE model, {\tt g55l}, is generated using the standard input file {\tt g55l.5}, with the following content:
\begin{verbatim}
--------------------------------------------------
5500. 4.5        ! TEFF, LOG G  
 T  T            ! LTE,  LTGRAY  (LTE model from scratch)
 'g55l.param'    ! name of file containing non-standard flags
*
* frequencies
*
  0                ! frequencies taken from the opacity table
*
* data for atoms
*
 30      ! NATOMS  (all first 30 species taken into account in EOS)
* mode abn modpf
    1   0.      0  ! H
    1   0.      0  ! He
    1   0.      0  ! Li
    1   0.      0  ! Be
    1   0.      0  ! B
    1   0.      0  ! C
    1   0.      0  ! N
    1   0.      0  ! O
    1   0.      0  ! F
    1   0.      0  ! Ne
    1   0.      0  ! Na
    1   0.      0  ! Mg
    1   0.      0  ! Al
    1   0.      0  ! Si
    1   0.      0  ! P
    1   0.      0  ! S
    1   0.      0  ! Cl
    1   0.      0  ! Ar
    1   0.      0  ! K
    1   0.      0  ! Ca
    1   0.      0  ! Sc
    1   0.      0  ! Ti
    1   0.      0  ! V
    1   0.      0  ! Cr
    1   0.      0  ! Mn
    1   0.      0  ! Fe
    1   0.      0  ! Co
    1   0.      0  ! Ni
    1   0.      0  ! Cu
    1   0.      0  ! Zn
*
* data for ions
*
* no explicit atoms, ions, levels
*iat   iz   nlevs  ilast ilvlin  nonstd typion  filei 
*
   0    0     0     -1      0      0    '    ' ' '
*
---------------------------------------------------
\end{verbatim}
The first line specifies $T_{\rm eff}$ and $\log g$; the second line stipulates that
an LTE model is going to be calculated, with no starting model, so that the present model
is calculated from scratch. The third line specifies the name of the keyword parameter file.
The reader is reminded that lines starting with an asterisk ({\tt *}) represent comments.
The next executive line, 0, says that the frequency points are taken from the opacity
table. The next block of lines sets all first 30 chemical species to MODE=1, i.e. they are 
taken into account for the equation of state, but not for opacities  (because 
all opacity is given through the opacity table). All abundances are the standard solar ones.
There are no explicit atoms, ions, or levels.

As specified in {\tt g55l.5}, the name of the file containing keyword parameters is  {\tt g55l.param},
that has the following content:
\begin{verbatim}
---------------------------------------------------
IOPTAB=-1,IFRYB=1,IFMOL=1,IDLST=0,IFRAYL=1
HMIX0=1,ITEK=50,IACC=50,IFRSET=30000
TAUFIR=1.e-7,TAULAS=1.0e2,TAUDIV=0.01
---------------------------------------------------
\end{verbatim}
which have the following meaning (see also Paper~III, \S\,7.10.3):\\ [2pt]
$\bullet\,$ IOPTAB=$-1$ -- sets the mode of evaluation of opacities to using 
the pre-calculated full opacity table, but with 
solving the equation of state and evaluating the thermodynamic parameters
on the fly.\\ [2pt]
$\bullet\,$ IFRYB=1 -- sets the global iteration method to the Rybicki scheme. 
As mentioned above, this scheme works very well for 
LTE models of cool stars, and in particular when convection is present. \\ [2pt]
$\bullet\,$ IFMOL=1 -- stipulates that molecules are included in the equation of state,
as is of course mandatory for such low-temperature models.\\ [2pt]
$\bullet\,$ IDLST=0  -- sets the proper treatment of the energy equation at the
lower boundary. This option has to be used in conjunction with IFRYB=1 or 2.\\ [2pt]
$\bullet\,$ IFRAYL=1 -- switches on an on-the-fly evaluation of the H, He, and H${}_2$
Rayleigh scattering opacity.\\ [2pt] 
$\bullet\ $ HMIX0=1 -- switches on the convection, with the mixing length equal
to 1 pressure scale height.\\ [2pt]
$\bullet\,$ ITEK=50, IACC=50 -- inhibits both Kantorovich and Ng acceleration
(that is, they would start at 50th iteration, but the total default number of global
iterations is 30). This option is safer when dealing with convective 
models.\\ [2pt]
$\bullet\,$ IFRSET=30000 -- sets the number of frequency points selected from
the opacity table, which  in this case has 100,000 frequencies.  For details, refer to
\S\,\ref{freq}.\\ [2pt]
$\bullet\,$ TAUFIR=1.e-7,TAULAS=1.0e2 -- sets the minimum and maximum
Rosseland optical depth when computing the starting LTE-gray model. In fact,
TAUFIR does not have to be set because 1.e-7 is the default anyway. The
model can easily be computed also with the default TAULAS (3.16e2), 
but one would
go to unnecessary large depths in the atmosphere, which are inconsequential
for the bulk of the atmospheric structure and the emergent radiation.\\ [2pt]
$\bullet\,$ TAUDIV=0.01 -- sets the division point for treating the radiative/convective 
equilibrium equation to 0.01. This setup is a usual choice if a model is computed using 
the Rybicki scheme.\\ [4pt]
Other keyword parameters, that were set for an analogous model considered in Paper~III, \S\,6.4,
are not necessary here because the default values of those parameters are sufficient.

Both NLTE models, {\tt g55nres}  and {\tt g55nryb},  treat the opacity of all species besides
H, Mg, and Ca,  through an opacity table, while the level populations of H, Mg, and Ca are 
treated in NLTE. As mentioned above, {\tt g55nres} is a so-called ``restricted NLTE" model 
where the temperature, density, and electron density are held fixed at the values for the input model
(in this case {\tt g55l}), and one solves for NLTE level populations of H, Mg, and Ca.

The standard input files for these two models are almost identical, differing only in the name of the 
keyword parameter file. For convenience, we show their content below:
\begin{verbatim}
--------------------------------------------------
5500. 4.5        ! TEFF, LOG G  
 F  F            ! LTE,  LTGRAY  (NLTE model)
 'nres.param'    ! or 'nryb.param' - keyword files
*
* frequencies
*
 50                ! additional frequencies 
*
* data for atoms
*
 30      ! NATOMS  (all first 30 species taken into account in EOS
*                   H, Mg, and Ca explicit)
* mode abn modpf
    2   0.      0  ! H - expl
    1   0.      0  ! He
    1   0.      0  ! Li
    1   0.      0  ! Be
    1   0.      0  ! B
    1   0.      0  ! C
    1   0.      0  ! N
    1   0.      0  ! O
    1   0.      0  ! F
    1   0.      0  ! Ne
    1   0.      0  ! Na
    2   0.      0  ! Mg - expl
    1   0.      0  ! Al
    1   0.      0  ! Si
    1   0.      0  ! P
    1   0.      0  ! S
    1   0.      0  ! Cl
    1   0.      0  ! Ar
    1   0.      0  ! K
    2   0.      0  ! Ca - expl
    1   0.      0  ! Sc
    1   0.      0  ! Ti
    1   0.      0  ! V
    1   0.      0  ! Cr
    1   0.      0  ! Mn
    1   0.      0  ! Fe
    1   0.      0  ! Co
    1   0.      0  ! Ni
    1   0.      0  ! Cu
    1   0.      0  ! Zn
*
* data for ions
*
*iat   iz   nlevs  ilast ilvlin  nonstd typion  filei 
*
   1   -1     1      0     0     0    ' H- ' 'data/hmin.dat'
   1    0     9      0     0     0    ' H 1' 'data/h1.dat'
   1    1     1      1     0     0    ' H 2' ' '
  12    0     41     0     0     0    'Mg 1' 'data/mg1.t'
  12    1     25     0     0     0    'Mg 2' 'data/mg2.dat'
  12    2     1      1     0     0    'Mg 3' ' '
  20    0     15     0    90     0    'Ca 1' 'data/ca1.t'
  20    1     18     0    90     0    'Ca 2' 'data/ca2.t'
  20    2     1      1     0     0    'Ca 3' ' '
   0    0     0     -1     0     0    '    ' ' '
*
---------------------------------------------------
\end{verbatim}

The restricted NLTE model, {\tt g55nres} has the keyword parameter file called
{\tt nres.param} which has the following content:
\begin{verbatim} 
---------------------------------------------------
IOPTAB=1,IFRYB=0,IFMOL=1,IFIXMO=1
IFRSET=30000
NLAMBD=4
---------------------------------------------------
\end{verbatim}
where\\ [2pt]
$\bullet\,$ IOPTAB=$1$ -- sets the mode of model construction to the hybrid NLTE/Opacity Table
-- see \S\,\ref{basic}.\\ [2pt]
$\bullet\,$ IFRYB=0 -- sets the standard CL/ALI linearization scheme (not the Rybickl scheme). This
is a default, so in fact it need not be specified. We include it  here for the sake of clarity.\\ [2pt]
$\bullet\,$ IFMOL=1 -- as before, sets the inclusion of molecules.\\ [2pt]
$\bullet\,$ IFIXMO=1 -- is a shorthand to specifying INHE=0, INRE=0, INPC=0 that stipulates that
the hydrostatic, radiative/convective, and charge equilibrium equations are not solved, and thus
the temperature, density and electron density are held fixed at the starting values.\\ [2pt] 
$\bullet\,$ IFRSET=30000 -- as before, sets the number of frequencies selected from the opacity table
to 30,000. \\ [2pt]
$\bullet\,$ NLAMBD=4 -- sets the number of ``Lambda" iterations of the formal solution to 4. This
somewhat improves the overall convergence behavior.\\ [2pt]

Finally, model {\tt g55nryb} computes s self-consistent model, where not only the level populations 
of H, Mg, and Ca are treated in NLTE, but also the self-consistent model structure (temperature,
density, electron density) is solved for. The corresponding keyword parameter file {\tt nryb.param}
has the following structure:
\begin{verbatim} 
---------------------------------------------------
IOPTAB=1,IFRYB=2,IFMOL=1,IDLST=0
ITEK=15,IACC=10,IFRSET=30000
HMIX0=1,TAUDIV=0.01
NLAMBD=8,CHMAX=1.e-5
---------------------------------------------------
\end{verbatim}
The meaning of the parameters is obvious, or was explained before. Notice that, as discussed
at length in \S\,\ref{ot/nlte}, it is important to use a large number of ``Lambda" iterations, NLAMBD (set here to 8)
and a relatively very low convergence parameter CHMAX (set here to 1.e-5). Also, since NLTE effects 
in H, Mg, and Ca do not not likely change the temperature structure significantly, it is now safe to switch on
the acceleration schemes - the Kantorovich acceleration in the 15th iteration -- ITEK=15, and the Ng
acceleration at the 10th iteration, ITEK=10. This could perhaps be done even earlier, but we did not explore this
issue any further since the computing time is already reasonably low in the present setup.

\section{Compiling, testing, and running SYNSPEC}
\label{testssyn}

\subsection{Source code files and compilation}

As for {\sc tlusty}, the program is
distributed as a single file that contains all subroutines,
plus several (6) small files referred to in the {\sc synspec} source file 
by {\tt INCLUDE} statements which declare most of variables and arrays.
These can be used to recompile the code and adjust
memory consumption, as explained in Paper III, \S\,\refcompcomp.  
All files should reside in the same directory.

The compilation of the program is explained in more detail in Paper III,
\S\,\refcompcomp, where the compiler instructions on different
platforms are summarized. Here we only show a compilation using
{\tt gfortran}, which is available on most Mac and LINUX platforms,
\begin{verbatim}
        gfortran -fno-automatic [-O3] -o synspec.exe synspec54.f   
\end{verbatim}
or, for some computers and/or compilers with limited size for the COMMON blocks,
\begin{verbatim}
gfortran -fno-automatic -mcmodel=large -O3 -o synspec.exe synspec54.f   
\end{verbatim}

\noindent Here, the option {\tt "-fno-automatic"} indicates the static allocation 
of memory. The level-3 optimization ({\tt "-O3"}) should again  be switched on since it
improves the performance of the code considerably. For more details, and several
examples of compiling the program on different platforms, refer to Paper~III,
\S\,\refcompcomp.

Analogously as for {\sc tlusty}, some arrays are quite big and may cause problems for
some computers and/or compilers. The critical arrays are those containing data for molecular
lines. There are several arrays {\tt MLINM0}$\times${\tt MMLIST}, located in the ``include"
file {\tt LINDAT.FOR}. Here, MLINM0 represents the maximum number of lines extracted from
a given molecular line list, at present set to MLINM0=9000000 and MMLIST, the maximum
number of molecular line lists, presently set to MMLIST=3. The users who do not intend
to work with molecular lines, can safely change the  parameter MLINM0=2 and MLINM=2,
which leads to a significant decrease of the memory taken by the executable file. 

On the  other hand, when working with larger molecular line lists, one should set MLINM0 and
MLINM to larger values. We have therefore provided the file \verb|LINDAT.FOR_big| which
can be used for the original, large, lists, but to run {\sc synspec} with these parameters one needs 
a computer with at least 8GB of RAM.

\subsection{Test cases for {\sc synspec}}
\label{testssyn}

There are six test cases for {\sc synspec}; three of them being analogous to those used
for version  51 and described in Paper~I, and three new cases that demonstrate the
construction of a simple opacity table, using {\sc synspec} for computing
spectra for models that were constructed using opacity tables, and for
computing pseudo-NLTE spectra.

The  relevant files are stored in the subdirectory \${\tt TLUSTY/tests/synspec}, which has
six subdirectories: 
\begin{itemize}
\item{\tt hhe} -- a sample synthetic spectrum for a simple H-He model, analogous to that considered
for {\sc synspec51} -- see Paper~I, \S\,5.7;
\item {\tt bstar} -- a sample synthetic spectrum for a NLTE metal line-blanketed model, again
analogous to that considered for {\sc synspec51} -- see Paper~I, \S\,5.7;
\item {\tt kurucz} -- sample synthetic spectra for two Kurucz models, similar, but extended with respect to those
considered for {\sc synspec51} -- see Paper~I, \S\,5.7;
\item {\tt hybrid} --  computing a synthetic spectrum for a hybrid Opacity Table/NLTE model atmosphere,
described in detail in \S\,\ref{hyb}.
\item {\tt pseudonlt} -- a test of computing a pseudo-NLTE spectrum,
where the input model is a Kurucz model with $T_{\rm eff}=5500$ K, $\log g = 4.5$,
but the NLTE level populations of H, Mg, and Ca, expressed through
the absolute $b^\ast$-factors, are taken from the previously computed model
{\tt g55nres} (one of the test cases for {\sc tlusty}). These $b^\ast$-factors are communicated to {\sc synspec}
in an input file named {\tt bfactors}, which is a copied from the output file {\tt fort.22} generated when computing the
model {\tt g55nres}.
The details are explained in \S\,\ref{pseudonlt}, 
\item {\tt optab} -- a test of constructing an opacity table, described in more detail in \S\,\ref{synoptab}.
\end{itemize}

Analogously to the test cases for {\sc tlusty},
all subdirectories contain (i) input data for a run, (ii) an AAAREADME file with a short description of
the tests; (iii) a script file {\tt R1} to run all the tests here, 
and (iv) a subdirectory named {\tt ./results} that
contains several important output files obtained
by running the test cases on an iMac (macOS Mojave, 10.14.2, with 3.4 GHz Intel Core i5 CPU)
done by one of us (I.H.), 
included to enable a comparison with user's runs for checking the installation on the user's platform. 

The directory \${\tt TLUSTY/tests/synspec} also contains a script file {\tt Runtest}, analogous to that for {\sc tlusty},
that runs all the tests in the first five subdirectories consecutively. 
The tests specified in the subdirectory {\tt optab} are not included
in {\tt Runtest}, They should be executed, if desired, separately. The reason is that the runs are somewhat
time-consuming, so they would make a quick testing of the implementation of the code
cumbersome.

\subsubsection{Tests of constructing an opacity table}
\label{synoptab}

An example of a construction of a simple, yet realistic, opacity table, is located in the directory
\${\tt TLUSTY/tests/synspec/optab}. This directory contains the input files needed for generating 
an opacity table
with 100,000 wavelength points, set logarithmically equidistant between 900 and 110,000 \AA, 
for 11 temperatures, spaced logarithmically between 3,000 and 15,000~K, and 7 densities, 
spaced between $10^{-12}$ and $10^{-6}$ g cm${}^{-3}$. This is specified in the file \verb|optab11_7f.2|,
located in the directory. This file should be copied to {\tt fort.2}, as explained in detail in \S\,\ref{infort2}. 
Additional input files located in this directory are {\tt tab.5}, a standard input file that in this case
specifies the continuum opacities, together with the corresponding keyword parameter file
{\tt tab.nst}, and \verb|optab11_7f.55|, which should be copied to {\tt fort.55} as explained in \S\,\ref{optabl}. 
As specified in the file {\tt fort.2},
the resulting opacity table is called \verb|optab11_7f.dat|. This file should be identical, or nearly
identical, to the file of the same name downloaded as part of the {\tt optables.tar} tarball. For convenience,
we show the corresponding {\tt fort.2}:
\begin{verbatim}
-----------------------------------
11   3000. 15000.
1 
7  1.e-12 1.e-6
100000  0  900  110000
'optab11_7f.dat'  0
------------------------------------
\end{verbatim}
Here we used the label ``f" in the name of the resulting file to signify the ``full" table (including opacities of
all species).

In this case it is not advantageous to use the script file {\tt RSynspec}, but rather to run a test
using the corresponding script file {\tt R1}, located in the directory as well.
Its first part looks like this:
\begin{verbatim}
ln -s -f $TLUSTY/data data
ln -s -f $TLUSTY/data/gfATO.bin gfATO.bin
ln -s -f $TLUSTY/data/gfMOL.bin gfMOL.bin
ln -s -f $TLUSTY/data/gfTiO.bin gfTiO.bin
cp optab11_7f.2 fort.2
cp optab11_7f.3 fort.3
cp optab11_7f.55 fort.55

$TLUSTY/synspec/synspec.exe <tab.5 > optab11_7f.6
cp fort.29 optab11_7f.29
cp fort.69 optab11_7f.69
cp fort.51 optab11_7f.51
cp fort.52 optab11_7f.52
cp fort.53 optab11_7f.53
cp fort.54 optab11_7f.54
\end{verbatim}
The input file \verb|optab11_7f.3| have to be copied to {\tt fort.3}, specifies the line lists to be used. We use
all line lists in binary format, which speeds up the calculations considerably. The file looks like this:
\begin{verbatim}
-----------------------------------
'gfATO.bin'     1
'gfMOL.bin'     1     8000.
'gfTiO.bin'     1     5000.
------------------------------------
\end{verbatim}
We stress that in order to be able to use this script that uses the binary line lists, one has to make sure
that they are properly created and linked to the ``data" directory, as explained in Sec.\,\ref{sumim}, items 4 and 5,
together with item 3.

One can also use the original text data,  using the input file {\tt fort.3} that looks like:
\begin{verbatim}
-----------------------------------
'gfATO.dat'     0
'gfMOL.dat'     0     8000.
'gfTiO.dat'     0     5000.    
------------------------------------
\end{verbatim}
in which case one needs to modify the  corresponding symbolic links as follows
\begin{verbatim}
ln -s -f $TLUSTY/data/gfATO.dat gfATO.dat
ln -s -f $TLUSTY/data/gfMOL.dat gfMOL.dat
ln -s -f $TLUSTY/data/gfTiO.dat gfTiO.dat
\end{verbatim}
In principle, one can mix text and binary files, for instance
\begin{verbatim}
-----------------------------------
'gfATO.dat'      0
'gfMOL.bin'      1     8000. 
'gfTiO.bin'      1     5000.
------------------------------------
\end{verbatim}

One can also use the old method that does not set up file {\tt fort.3}. In that case the commands are
\begin{verbatim}
ln -s -f $TLUSTY/data data
ln -s -f $TLUSTY/data/gfATO.dat fort.19
ln -s -f $TLUSTY/data/gfMOL.dat fort.20
ln -s -f $TLUSTY/data/gfTiO.dat fort.21
$TLUSTY/synspec/synspec.exe <tab.5 >tmp.log
\end{verbatim}
in which case the file {\tt fort.55} must contain the the association of the molecular line
lists with the input files unit 20 and 21. Assuming
that both line lists are of text format, the file {\tt fort.55} will look like this:
\begin{verbatim}
-------------------------------------
      -3       1       0
       1       0       0       0
       0       0       0       0       0
       1       0       0       0       0
       0       0       0
    900.   -110000.    50.00  2000.0   1.00000e-4  0.05
       2       20      21
-------------------------------------
\end{verbatim}
In the case that all line lists are in binary format, the file {\tt fort.55} has to be modified to
\begin{verbatim}
-------------------------------------
      -3       1       0
       1       0       0       0
       0       0       0       0       0
       1       0       0       1       0
       0       0       0
    900.   -110000.    50.00  2000.0   1.00000e-4  0.05
       2       20      21
-------------------------------------
\end{verbatim}
Here, the only change is the 4th number on the 4th line, INLIST, which is now set to 1.
But one must also link the binary versions of the line lists, {\tt gfATO.bin}, {\tt gfMOL.bin}, and
{\tt gfTiO.bin}, to {\tt fort.19}, {\tt fort.20}, and {\tt fort.21}, respectively.

We stress that any of these {\tt fort.55} examples can be when {\tt fort.3} is provided as well, since in that case
the quantity INLIST, and the information on the last line of input, are overridden by the input from {\tt fort.3}.
If one stores the line lists in a different directory than in the standard "{\tt data}" directory, 
or uses different line lists altogether,  one has to modify the links accordingly.

Finally, we emphasize that the above input files to generate an opacity table serve as a simple
example to test the performance of the code, but are NOT THE RECOMMENDED approach to generate actual 
production-run tables. In this case, one would need to consider, in addition to the species included in
the standard molecular line lists, data for H${}_2$O. To show how the input file {\tt fort.3}
will look like,  let us assume that that we have binary line list for water,  called
{\tt gfH2O.bin}, possibly with the Van der Walls broadening data from the 
EXOMOL tables\footnote{Such a table, which is not a part of the standard distribution, is also available upon request.}.
The file {\tt fort.3} is then 
\begin{verbatim}
-----------------------------------
'gfATO.bin'    1
'gfMOL.bin'    1     8000. 
'gfTiO.bin'    1     5000. 
'gfH2O.bin'    1     4000. 
------------------------------------
\end{verbatim}
Here, we assume that the TiO lines do not contribute above 5000 K, and the H${}_2$O lines are negligible above
4000 K, which seems to be a reasonable assumption. The line lists for these species are not even read above
these respective temperatures when evaluating the opacity table, which leads to considerable time
savings.

We could also generate an analogous table where the opacity of H, Mg, and Ca  is removed, so that it can
be used for constructing model atmospheres where H, Mg, and Ca are treated in NLTE, while the opacity
of all other species (atomic as well as molecular) is computed assuming LTE. The input files for
such table are the same, except {\tt fort.2} which now becomes:
\begin{verbatim}
-----------------------------------
11   3000. 15000.
1 
7  1.e-12 1.e-6
100000  0  900  110000
'optab11_7hmc.dat'  0
 1 0
12 0
20 0
------------------------------------
\end{verbatim}
This file is also located in the directory,
with the name \verb|optab11_7hmc.2|; the script file {\tt R1} copies it to {\tt fort.2}.

For convenience, we include here the standard input file {\tt tab.5}. It will also be useful
for a better explanation how to construct a standard input for computing synthetic spectra for the
hybrid Opacity Table/NLTE models.

\begin{verbatim}
----------------------------------------------------------------
6000. 4.0
 T  F              ! LTE,  LTGRAY
 'tab.nst'         ! name of file containing non-standard flags
*
* frequencies
*
 2000
*
* data for atoms
*
 30            ! NATOMS
* mode abn modpf
    2   0.      0  ! H
    2   0.      0  ! He
    1   0.      0  ! Li
    1   0.      0  ! Be
    1   0.      0  ! B
    2   0.      0  ! C
    2   0.      0  ! N
    2   0.      0  ! O
    1   0.      0  ! F
    2   0.      0  ! Ne
    1   0.      0  ! Na
    2   0.      0  ! Mg
    2   0.      0  ! Al
    2   0.      0  ! Si
    1   0.      0  ! P
    2   0.      0  ! S
    1   0.      0  ! Cl
    1   0.      0  ! Ar
    1   0.      0  ! K
    2   0.      0  ! Ca
    1   0.      0  ! Sc
    1   0.      0  ! Ti
    1   0.      0  ! V
    1   0.      0  ! Cr
    1   0.      0  ! Mn
    2   0.      0  ! Fe
    1   0.      0  ! Co
    1   0.      0  ! Ni
    1   0.      0  ! Cu
    1   0.      0  ! Zn
*
* data for ions
*
*iat   iz   nlevs  ilast ilvlin  nonstd typion  filei
*
   1   -1     1      0      0      9    ' H- ' 'data/hmin.dat'
   1    0     9      0      0      0    ' H 1' 'data/h1.dat'
   1    1     1      1      0      0    ' H 2' ' '
   2    0    24      0      0      0    'He 1' 'data/he1.dat'
   2    1     1      1      0      0    'He 2' ' '
   6    0    40      0      0      0    ' C 1' 'data/c1.dat'
   6    1    22      0      0      0    ' C 2' 'data/c2.dat'
   6    2     1      1      0      0    ' C 3' ' '
   7    0    34      0      0      0    ' N 1' 'data/n1.dat'
   7    1    42      0      0      0    ' N 2' 'data/n2_32+10lev.dat'
   7    2     1      1      0      0    ' N 3' ' '
   8    0    33      0      0      0    ' O 1' 'data/o1_23+10lev.dat'
   8    1    48      0      0      0    ' O 2' 'data/o2_36+12lev.dat'
   8    2     1      1      0      0    ' O 3' ' '
  10    0    35      0      0      0    'Ne 1' 'data/ne1_23+12lev.dat'
  10    1    32      0      0      0    'Ne 2' 'data/ne2_23+9lev.dat'
  10    2     1      1      0      0    'Ne 3' ' '
  12    0    41      0      0      0    'Mg 1' 'data/mg1.t'
  12    1    25      0      0      0    'Mg 2' 'data/mg2.dat'
  12    2     1      1      0      0    'Mg 3' ' '
  13    0    10      0      0      0    'Al 1' 'data/al1.t'
  13    1    29      0      0      0    'Al 2' 'data/al2_20+9lev.dat'
  13    2     1      1      0      0    'Al 3' ' '
  14    0    45      0      0      0    'Si 1' 'data/si1.t'
  14    1    40      0      0      0    'Si 2' 'data/si2_36+4lev.dat'
  14    2     1      1      0      0    'Si 3' ' '
  16    0    41      0      0      0    ' S 1' 'data/s1.t'
  16    1    33      0      0      0    ' S 2' 'data/s2_23+10lev.dat'
  16    2     1      1      0      0    ' S 3' ' '
  20    0    15      0      0      0    'Ca 1' 'data/ca1.t'
  20    1    18      0      0      0    'Ca 2' 'data/ca2.t'
  20    2     1      1      0      0    'Ca 3' ' '
  26    0    30      0      0      0    'Fe 1' 'data/fe1.dat'
  26    1    36      0      0     -1    'Fe 2' 'data/fe2va.dat'
   0    0                                      'data/gf2601.gam'
                                               'data/gf2601.lin'
                                               'data/fe2p_14+11lev.rap'
  26    2    50      0      0     -1    'Fe 3' 'data/fe3va.dat'
   0    0                                      'data/gf2602.gam'
                                               'data/gf2602.lin'
                                               'data/fe3p_22+7lev.rap'
  26    3     1      1      0      0    'Fe 4' ' '
   0    0     0     -1      0      0    '    ' ' '
*
----------------------------------------------------------------
\end{verbatim}
As this file stipulates, the bound-free opacities of the following atoms/ions are taken into account:
H${}^-$, H I, He I, C I, C II, N I. N II, O I, O II, Ne I, Ne II, Mg I, Mg II, Al I, Al II, Si I, Si II, S I, S II,
Ca I, Ca II, Fe I, Fe II, and Fe III.

The non-standard parameters file {\tt tab.nst} contains just one record, namely\\[2pt]
{\tt IFEOS=-1}\\[2pt]
that set the output of interesting equation-of-state parameters, as explained in \S\,\ref{eossyn}.

\subsubsection{Ordinary stellar synthetic spectra}

There are three test cases for computing ordinary stellar synthetic spectra, contained in subdirectories
{\tt hhe}, {\tt bstar}, and {\tt kurucz}, that we briefly describe below.\\

\noindent
$\bullet$ {\tt hhe} -- Contains a test of very simple synthetic spectrum for a NLTE H-He model,
called {\tt hhe35nl}, generated as a test case of {\sc tlusty} -- see \S\,\ref{teststlus}. The sample spectrum
is computed only between 1400 and 1410 \AA, using a test-case line list {\tt linelist.test} stored in
the standard ``data" directory. A similar spectrum was calculated as a test case of {\sc synspec51} -- see
Paper~I.\\

\noindent
$\bullet$ {\tt bstar} -- This is a similar test, but for computing a synthetic spectrum for a 
sophisticated NLTE metal line-blanketed modes atmosphere of a B-star, with 
$T_{\rm eff}=20,000$ K, and $\log g =4$. This is one of the models of the BSTAR2006 grid 
(Lanz \& Hubeny 2007). Again, only a very short
interval of spectrum, between 1400 and 1410 \AA, is computed as a test.\\

\noindent
$\bullet$ {\tt kurucz} -- This directory contains two tests with computations of synthetic spectra for  
a Kurucz model. The first one is a spectrum
for an early A-type star, $T_{\rm eff}=9,750$~K, and $\log g =4$, computed for the same short interval between
1400 and 1410 \AA. and using the test-case short atomic line list.

The second test is a synthetic spectrum for a solar-type star ($T_{\rm eff}=5,500$~K,  $\log g =4.5$),
between 2,500 and 7,500 \AA. Unlike the previous tests, this one uses the three full binary
line lists {\tt gfATO.bin}, {\tt gfMOL.bin}, and {\tt gfTiO.bin}. Similarly to the tests of generating an opacity table, one can
specify these line list either through the parameters in the file {\tt fort.55}, or through {\tt fort.3}; here we
chose the latter. In this test we also demonstrate setting thee parameter IFEOS to a negative value (IFEOS=-4),
which switches on the creation of the output files {fort.51} -- {\tt fort.54} that contain various interesting parameters 
and quantities, related to the equation of state (see \S\,\ref{eossyn}), at every fourth depth point.

\subsubsection{Synthetic spectrum for a hybrid Opacity Table/NLTE model atmosphere}
\label{hyb}

As explained in \S\,\ref{quasex}, the standard input file to be used for computing a synthetic
spectrum for a hybrid Opacity Table/NLTE model (such as the test model {\tt g55res} considered
in \S\,\ref{tesop}), has to be a modified file {\tt tab.5}, in which all the originally explicit species
except H. Mg, and Ca are considered as quasi-explicit (MODE=5). 
For convenience, we show here the
first part of the modified {\tt tab.5} file, called {\tt tabsp.5}:
\begin{verbatim}
----------------------------------------------------------------
6000. 4.0
 F  F              ! LTE,  LTGRAY
 ''            ! no keyword parameters
*
* frequencies
*
 2000
*
* data for atoms
*
 30                 ! NATOMS
* mode abn modpf
    2   0.      0  ! H
    5   0.      0  ! He
    1   0.      0  ! Li
    1   0.      0  ! Be
    1   0.      0  ! B
    5   0.      0  ! C
    5   0.      0  ! N
    5   0.      0  ! O
    1   0.      0  ! F
    5   0.      0  ! Ne
    1   0.      0  ! Na
    2   0.      0  ! Mg
    5   0.      0  ! Al
    5   0.      0  ! Si
    1   0.      0  ! p
    5   0.      0  ! S
    1   0.      0  ! Cl
    1   0.      0  ! Ar
    1   0.      0  ! K
    2   0.      0  ! Ca
    1   0.      0  ! Sc
    1   0.      0  ! Ti
    1   0.      0  ! V
    1   0.      0  ! Cr
    1   0.      0  ! Mn
    5   0.      0  ! Fe
    1   0.      0  ! Co
    1   0.      0  ! Ni
    1   0.      0  ! Cu
    1   0.      0  ! Zn
*
----------------------------------------------------------------
\end{verbatim}
The rest of the file remains unchanged. As is clearly seen here, H, Mg, and Ca are explicit (MODE=2),
because their level populations are provided by the input model atmosphere, and He, C, N, O, Ne, Al, Si,
S, and Fe are quasi-explicit (MODE=5), so that their bound-free opacity (obviously in LTE) will be included
(as it was in the original opacity table).

There are actually two test cases here, with the same input model atmosphere, and the same standard input.
The first computes a full synthetic spectrum between 6700 and 6800 \AA ~(with atomic and molecular lines),
and the second computes a theoretical continuum between 1500 and 7500 \AA, to demonstrate the
influence of the bound-free opacity of metals.

%%%%%%%%%%%%%%%%%%%%%%%%%%%%%%%%%%%%%%%%%%%%

\section{Summary and outlook}

We have presented and described new upgrades in the computer programs {\sc tlusty} and
{\sc synspec}, developed with the aim of providing powerful and flexible tools for computing
model stellar atmospheres and accretion disks, and analyzing stellar spectra. We would like to stress
that while we spent considerable effort in testing the codes for many physical conditions, they are
by no means error-free. In fact, we could paraphrase the disclaimer Bob Kurucz often uses, namely
``These codes are {\em guarantied} to contain errors". We strongly advice
any users against using these codes as black boxes, or taking results of these codes without a careful
inspection and checking for their physical soundness and consistency. The authors would also highly appreciate
receiving comments about possible bugs, and/or suggestion for improvements or changes that would
increase the accuracy and versatility of the codes.

Looking critically at the Section ``Outlook" of Paper~III, one can see that most of the projected future
developments of {\sc tlusty} have not yet materialized. So, we still have plenty of work to do for a foreseeable
future. We have however developed several improvements in {\sc synspec} that were not
envisaged previously. For both codes, the new developments described in this paper concern mostly
applications for cool stars.

Another significant improvement, on a purely coding level, is a transformation of {\sc tlusty}, version 205,
to FORTRAN~90, by Simon Preval (University of Leicester). That development, as well as the
present one, proceeded in parallel, and therefore the present versions of the codes still adhere to FORTRAN~77. 
We now plan to make analogous changes in {\sc tlusty208} to transfer it to FORTRAN~90,
and to pursue all future development and improvements on this language.

%%%%%%%%%%%%%%%%%%%%%%%%%%%%%%%%%%%%%%%%%%%%

\section*{Acknowledgements}
\addcontentsline{toc}{section}{Acknowledgements}
We thank Dr. Paul Barklem for advice on atomic and molecular data, 
Nikola Vitas for his help in testing the equation of state, and Jano Budaj and Krzysztof Gesicki
for pointing out some issues when compiling the programs on various platforms.
I.H. would like to acknowledge support that enabled his visit to IAC
in Tenerife from the Severo Ochoa program, awarded by the Government of Spain to the IAC.

%%%%%%%%%%%%%%%%%%%%%%%%%%%%%%%%%%%%%%%%%%%%

\bigskip
\bigskip

%===================================================================

\section*{Appendix A - Adopted sets of the standard solar abundances}
\addcontentsline{toc}{section}{Appendix A: Adopted sets of the standard solar abundances}

In Tables 3-5, we summarize the actual values of the standard solar abundances, used by
{\sc tlusty} and {\sc synspec}, described in \S\,\ref{outl}. The individual sets are labelled as
{\tt set0}, invoked by setting the parameter IABSET=0 (see \S\,\ref{outl});  {\tt set1}
by IABSET=1, and {\tt set2} by IABSET=2.

\begin{table}
\caption{Adopted sets of the standard solar abundances: H -- Zn} 
\begin{center}
\begin{tabular}[t]{||c||c|c||c|c||c|c||}
\hline
Element  & Set0 abs & Set0 log &  Set1 abs &  Set1 log &  Set2 abs  & Set2 log \\      
\hline
 H    &    1.00E+00  &    12.00  &    1.00E+00  &    12.00  &    1.00E+00  &    12.00  \\  
 He   &    8.51E-02  &    10.93  &    8.51E-02  &    10.93  &    1.00E-01  &    11.00  \\  
 Li   &    1.12E-11  &     1.05  &    1.82E-09  &     3.26  &    1.26E-11  &     1.10  \\  
 Be   &    2.40E-11  &     1.38  &    2.40E-11  &     1.38  &    2.51E-11  &     1.40  \\  
 B    &    5.01E-10  &     2.70  &    6.17E-10  &     2.79  &    5.00E-10  &     2.70  \\  
 C    &    2.45E-04  &     8.39  &    2.69E-04  &     8.43  &    3.31E-04  &     8.52  \\  
 N    &    6.03E-05  &     7.78  &    6.76E-05  &     7.83  &    8.32E-05  &     7.92  \\  
 O    &    4.57E-04  &     8.66  &    4.90E-04  &     8.69  &    6.76E-04  &     8.83  \\  
 F    &    3.63E-08  &     4.56  &    3.63E-08  &     4.56  &    3.16E-08  &     4.50  \\  
 Ne   &    6.92E-05  &     7.84  &    8.51E-05  &     7.93  &    1.20E-04  &     8.08  \\  
 Na   &    1.48E-06  &     6.17  &    1.74E-06  &     6.24  &    2.14E-06  &     6.33  \\  
 Mg   &    3.39E-05  &     7.53  &    3.98E-05  &     7.60  &    3.80E-05  &     7.58  \\  
 Al   &    2.34E-06  &     6.37  &    2.82E-06  &     6.45  &    2.95E-06  &     6.47  \\  
 Si   &    3.24E-05  &     7.51  &    3.24E-05  &     7.51  &    3.55E-05  &     7.55  \\  
  P    &    2.29E-07  &     5.36  &    2.57E-07  &     5.41  &    2.82E-07  &     5.45  \\  
 S    &    1.38E-05  &     7.14  &    1.32E-05  &     7.12  &    2.14E-05  &     7.33  \\  
 Cl   &    3.16E-07  &     5.50  &    3.16E-07  &     5.50  &    3.16E-07  &     5.50  \\  
 Ar   &    1.51E-06  &     6.18  &    2.51E-06  &     6.40  &    2.52E-06  &     6.40  \\  
 K    &    1.20E-07  &     5.08  &    1.20E-07  &     5.08  &    1.32E-07  &     5.12  \\  
 Ca   &    2.04E-06  &     6.31  &    2.19E-06  &     6.34  &    2.29E-06  &     6.36  \\  
 Sc   &    1.12E-09  &     3.05  &    1.41E-09  &     3.15  &    1.48E-09  &     3.17  \\  
 Ti   &    7.94E-08  &     4.90  &    8.91E-08  &     4.95  &    1.05E-07  &     5.02  \\  
 V    &    1.00E-08  &     4.00  &    8.51E-09  &     3.93  &    1.00E-08  &     4.00  \\  
 Cr   &    4.37E-07  &     5.64  &    4.37E-07  &     5.64  &    4.68E-07  &     5.67  \\  
 Mn   &    2.45E-07  &     5.39  &    2.69E-07  &     5.43  &    2.45E-07  &     5.39  \\  
 Fe   &    2.82E-05  &     7.45  &    3.16E-05  &     7.50  &    3.16E-05  &     7.50  \\  
 Co   &    8.32E-08  &     4.92  &    9.77E-08  &     4.99  &    8.32E-08  &     4.92  \\  
 Ni   &    1.70E-06  &     6.23  &    1.66E-06  &     6.22  &    1.78E-06  &     6.25  \\  
 Cu   &    1.62E-08  &     4.21  &    1.55E-08  &     4.19  &    1.62E-08  &     4.21  \\  
 Zn   &    3.98E-08  &     4.60  &    3.63E-08  &     4.56  &    3.98E-08  &     4.60  \\  
 \hline
\end{tabular}
\\[2.5pt]
{\small
Set0: Asplund et al. (2005), essentially the same as Grevesse et al. (2007);
Set1: Asplund et al. (2009);
Set2: Grevesse \& Sauval (1998) (former default).
``Abs" stands for $N_{\rm el}/N_{\rm H}$, while ``log" for $\log(N_{\rm el}/N_{\rm H})+12$.
}
\end{center}
\end{table}  

\begin{table}
\caption{Adopted sets of the standard solar abundances: Ga -- Nd} 
\begin{center}
\begin{tabular}[t]{||c||c|c||c|c||c|c||}
\hline
Element  & Set0 abs & Set0 log &  Set1 abs &  Set1 log &  Set2 abs  & Set2 log \\      
\hline
 Ga   &    7.59E-10  &     2.88  &    1.10E-09  &     3.04  &    1.35E-09  &     3.13  \\  
 Ge   &    3.80E-09  &     3.58  &    4.47E-09  &     3.65  &    4.27E-09  &     3.63  \\  
 As   &    1.95E-10  &     2.29  &    2.00E-10  &     2.30  &    2.34E-10  &     2.37  \\  
 Se   &    2.14E-09  &     3.33  &    2.19E-09  &     3.34  &    2.24E-09  &     3.35  \\  
 Br   &    3.63E-10  &     2.56  &    3.47E-10  &     2.54  &    4.27E-10  &     2.63  \\  
 Kr   &    1.91E-09  &     3.28  &    1.78E-09  &     3.25  &    1.70E-09  &     3.23  \\  
 Rb   &    3.98E-10  &     2.60  &    2.29E-10  &     2.36  &    2.51E-10  &     2.40  \\  
 Sr   &    8.32E-10  &     2.92  &    7.41E-10  &     2.87  &    8.51E-10  &     2.93  \\  
 Y    &    1.62E-10  &     2.21  &    1.62E-10  &     2.21  &    1.66E-10  &     2.22  \\  
 Zr   &    3.89E-10  &     2.59  &    3.80E-10  &     2.58  &    4.07E-10  &     2.61  \\  
 Nb   &    2.63E-11  &     1.42  &    2.88E-11  &     1.46  &    2.51E-11  &     1.40  \\  
 Mo   &    8.32E-11  &     1.92  &    7.59E-11  &     1.88  &    9.12E-11  &     1.96  \\  
 Tc   &    1.02E-22  &    -9.99  &    1.02E-22  &    -9.99  &    1.00E-24  &   -12.00  \\  
 Ru   &    6.92E-11  &     1.84  &    5.62E-11  &     1.75  &    6.61E-11  &     1.82  \\  
 Rh   &    1.32E-11  &     1.12  &    1.15E-11  &     1.06  &    1.23E-11  &     1.09  \\  
 Pd   &    4.90E-11  &     1.69  &    4.47E-11  &     1.65  &    5.01E-11  &     1.70  \\  
 Ag   &    8.71E-12  &     0.94  &    1.58E-11  &     1.20  &    1.74E-11  &     1.24  \\  
 Cd   &    5.89E-11  &     1.77  &    5.13E-11  &     1.71  &    5.75E-11  &     1.76  \\  
 In   &    3.98E-11  &     1.60  &    5.75E-12  &     0.76  &    6.61E-12  &     0.82  \\  
 Sn   &    1.00E-10  &     2.00  &    1.10E-10  &     2.04  &    1.38E-10  &     2.14  \\  
 Sb   &    1.00E-11  &     1.00  &    1.02E-11  &     1.01  &    1.10E-11  &     1.04  \\  
 Te   &    1.55E-10  &     2.19  &    1.51E-10  &     2.18  &    1.74E-10  &     2.24  \\  
 I    &    3.24E-11  &     1.51  &    3.55E-11  &     1.55  &    3.24E-11  &     1.51  \\  
 Xe   &    1.86E-10  &     2.27  &    1.74E-10  &     2.24  &    1.70E-10  &     2.23  \\  
 Cs   &    1.17E-11  &     1.07  &    1.20E-11  &     1.08  &    1.32E-11  &     1.12  \\  
 Ba   &    1.48E-10  &     2.17  &    1.51E-10  &     2.18  &    1.62E-10  &     2.21  \\  
 La   &    1.35E-11  &     1.13  &    1.26E-11  &     1.10  &    1.58E-11  &     1.20  \\  
 Ce   &    3.80E-11  &     1.58  &    3.80E-11  &     1.58  &    4.07E-11  &     1.61  \\  
 Pr   &    5.13E-12  &     0.71  &    5.25E-12  &     0.72  &    6.03E-12  &     0.78  \\  
 Nd   &    2.82E-11  &     1.45  &    2.63E-11  &     1.42  &    2.95E-11  &     1.47  \\  
 \hline
\end{tabular}
\\[2.5pt]
{\small
Set0: Asplund et al. (2005), essentially the same as Grevesse et al. (2007);
Set1: Asplund et al. (2009);
Set2: Grevesse \& Sauval (1998) (former default).
``Abs" stands for $N_{\rm el}/N_{\rm H}$, while ``log" for $\log(N_{\rm el}/N_{\rm H})+12$.
}
\end{center}
\end{table}

\begin{table}
\caption{Adopted sets of the standard solar abundances: Pm -- U} 
\begin{center}
\begin{tabular}[t]{||c||c|c||c|c||c|c||}
\hline
Element  & Set0 abs & Set0 log &  Set1 abs &  Set1 log &  Set2 abs  & Set2 log \\      
\hline
 Pm   &    1.02E-22  &    -9.99  &    1.02E-22  &    -9.99  &    1.00E-24  &   -12.00  \\  
 Sm   &    1.02E-11  &     1.01  &    9.12E-12  &     0.96  &    9.33E-12  &     0.97  \\  
 Eu   &    3.31E-12  &     0.52  &    3.31E-12  &     0.52  &    3.47E-12  &     0.54  \\  
 Gd   &    1.32E-11  &     1.12  &    1.17E-11  &     1.07  &    1.17E-11  &     1.07  \\  
 Tb   &    1.91E-12  &     0.28  &    2.00E-12  &     0.30  &    2.14E-12  &     0.33  \\  
 Dy   &    1.38E-11  &     1.14  &    1.26E-11  &     1.10  &    1.41E-11  &     1.15  \\  
 Ho   &    3.24E-12  &     0.51  &    3.02E-12  &     0.48  &    3.16E-12  &     0.50  \\  
 Er   &    8.51E-12  &     0.93  &    8.32E-12  &     0.92  &    8.91E-12  &     0.95  \\  
 Tm   &    1.00E-12  &     0.00  &    1.26E-12  &     0.10  &    1.35E-12  &     0.13  \\  
 Yb   &    1.20E-11  &     1.08  &    8.32E-12  &     0.92  &    8.91E-12  &     0.95  \\  
 Lu   &    1.15E-12  &     0.06  &    1.26E-12  &     0.10  &    1.32E-12  &     0.12  \\  
 Hf   &    7.59E-12  &     0.88  &    7.08E-12  &     0.85  &    5.37E-12  &     0.73  \\  
 Ta   &    6.76E-13  &    -0.17  &    7.59E-13  &    -0.12  &    1.35E-12  &     0.13  \\  
 W    &    1.29E-11  &     1.11  &    4.47E-12  &     0.65  &    4.79E-12  &     0.68  \\  
 Re   &    1.70E-12  &     0.23  &    1.82E-12  &     0.26  &    1.86E-12  &     0.27  \\  
 Os   &    2.82E-11  &     1.45  &    2.51E-11  &     1.40  &    2.40E-11  &     1.38  \\  
 Ir   &    2.40E-11  &     1.38  &    2.40E-11  &     1.38  &    2.34E-11  &     1.37  \\  
 Pt   &    4.37E-11  &     1.64  &    4.17E-11  &     1.62  &    4.79E-11  &     1.68  \\  
 Au   &    1.02E-11  &     1.01  &    6.31E-12  &     0.80  &    6.76E-12  &     0.83  \\  
 Hg   &    1.35E-11  &     1.13  &    1.48E-11  &     1.17  &    1.23E-11  &     1.09  \\  
 Tl   &    7.94E-12  &     0.90  &    5.89E-12  &     0.77  &    6.61E-12  &     0.82  \\  
 Pb   &    1.00E-10  &     2.00  &    1.10E-10  &     2.04  &    1.12E-10  &     2.05  \\  
 Bi   &    4.47E-12  &     0.65  &    4.47E-12  &     0.65  &    5.13E-12  &     0.71  \\  
 Po   &    1.02E-22  &    -9.99  &    1.02E-22  &    -9.99  &    1.00E-24  &   -12.00  \\  
 At   &    1.02E-22  &    -9.99  &    1.02E-22  &    -9.99  &    1.00E-24  &   -12.00  \\  
 Rn   &    1.02E-22  &    -9.99  &    1.02E-22  &    -9.99  &    1.00E-24  &   -12.00  \\  
 Fr   &    1.02E-22  &    -9.99  &    1.02E-22  &    -9.99  &    1.00E-24  &   -12.00  \\  
 Ra   &    1.02E-22  &    -9.99  &    1.02E-22  &    -9.99  &    1.00E-24  &   -12.00  \\  
 Ac   &    9.77E-03  &     9.99  &    1.02E-22  &    -9.99  &    1.00E-24  &   -12.00  \\  
 Th   &    1.15E-12  &     0.06  &    1.15E-12  &     0.06  &    1.20E-12  &     0.08  \\  
 Pa   &    1.02E-22  &    -9.99  &    1.02E-22  &    -9.99  &    1.00E-24  &   -12.00  \\  
 U    &    3.02E-13  &    -0.52  &    2.88E-13  &    -0.54  &    3.24E-13  &    -0.49  \\  
 \hline
\end{tabular}
\\[2.5pt]
{\small
Set0: Asplund et al. (2005), essentially the same as Grevesse et al. (2007);
Set1: Asplund et al. (2009);
Set2: Grevesse \& Sauval (1998) (former default).
``Abs" stands for $N_{\rm el}/N_{\rm H}$, while ``log" for $\log(N_{\rm el}/N_{\rm H})+12$.
}
\end{center}
\end{table}  

%================================================================

\section*{Appendix B- Mean molecular weight and total particle density}
\addcontentsline{toc}{section}{Appendix B: Mean molecular weight and total particle density}

In the context of {\sc tlusty}, the equation of state is used as a relation between gas pressure, $P_g$,
and mass density, $\rho$. Since one assumes an ideal gas behavior, 
$P_g = kNT$, with $k$ being the Boltzmann constant, $N$ the total particle number density, and $T$ the temperature, the equation of state is essentially a relation between $N$ and $\rho$.
It is written as
\begin{equation}
\label{rho}
\rho = (N-n_{\rm e}) \mu m_H,
\end{equation}
where $\mu$ is the mean molecular weight, and $m_H$ the mass of hydrogen atom. 
In version 205 of {\sc tlusty}, the mean molecular weight was assumed to be given by 
(see Paper~II, Eq.  83),
\begin{equation}
\mu = \frac{\sum_I A_I (m_I/m_H)}{\sum_I A_I}.
\end{equation}
where $A_I$ is the abundance of species $I$, by number relative to hydrogen,
and $m_I$ the mass of an atom of species $I$. In the absence of molecules,
$\mu$ depends only on the assumed chemical abundances, and is therefore
independent of atmospheric structure, and thus remains constant during the run of
{\sc tlusty}.

When molecules are present, Eq. (\ref{rho}) remains valid, but the mass density is
given by 
\begin{equation}
\rho = \sum_i n_i m_i,
\end{equation}
where $i$ runs over all species, atomic and molecular, with $n_i$
being the actual number density of particles $i$ and $m_i$ their mass.
Consequently, the mean molecular weight is no longer fixed, but igiven by
Eq. (\ref{rho}) that is now viewed as a definition relation for $\mu$,
\begin{equation}
\label{mu}
\mu = (\rho/m_H)/(N-n_{\rm e}).
\end{equation}
Consequently, either $\mu$ or $N$ has to be part of the basic structural
quantities to be stored as model output. A natural choice, in view of $N$ being among the
basic state parameters anyway, is to chose $N$ as the additional output parameter.

%================================================================

\section*{Appendix C - Thermodynamic parameters and convection}
\addcontentsline{toc}{section}{Appendix C: Thermodynamic parameters and convection}

{\sc tlusty208} allows the evaluation of the adiabatic gradient either through the
internal energy and its derivatives, or through the specific entropy and its derivatives.
This is controlled by the keyword parameter IFENTR -- see Paper~III, \S\,12.3.2, and in \S\,\ref{convec}
of this document. In version 205, 
the default value of this parameter is IFENTR=0, which sets an evaluation of the adiabatic gradient
through the internal energy -- Eqs. (346) -- (348) of Paper~II. Equation (349) of Paper~II
for the adiabatic gradient expressed through the specific entropy is incorrect, and Eq. (350)
is formally correct, but not advantageous for numerical evaluation.

In the present version 208, the default is switched to IFENTR=1, that is to using a specific entropy
as a basic quantity to evaluate the adiabatic gradient (although one can still set
IFENTR=0 to recover the previous treatment). The reason is that in the case of entropy, one needs
to evaluate fewer partial derivatives numerically than in the case of internal energy, which is
presumably less prone to numerical inaccuracies.

The adiabatic gradient and the specific heat are given by 
\begin{equation}
\nabla_{\rm ad} = - \frac{(\partial S/\partial P)_{T}}{(\partial S/\partial T)_{P}}\,
\frac{P}{T},
\end{equation}
and
\begin{equation}
c_P = T (\partial S/\partial T)_{P},
\end{equation}
where the specific entropy is given by
\begin{equation}
\label{entr}
S = \frac{k}{\rho} 
\sum_i N_i \left[\frac{3}{2}\ln(kT m_i) + \ln(U_i/N_i) + kT^2 \frac{\partial\ln U_i}{\partial\ln T} 
+ s_0 \right],
\end{equation}
where the summation extends over all components $i$ of the gas, that is, over neutral and ionized atoms,
molecules, and electrons. Here $k$ is the Boltzmann constant, $\rho$ the mass density, $T$  
the temperature, $m_i$ the mass 
of the species $i$ expressed in units of the hydrogen atom mass, and $N_i$ and $U_i$ are the
number density and the partition function of species $i$, respectively. The constant $s_0$ is
given by
\begin{equation}
s_0 = (3/2) \ln(2 \pi m_H/h^2) = 103.973
\end{equation}
where $m_H = 1.67333\times 10^{-24}$ is the mass of hydrogen atom in grams, and $h$ is the
Planck constant.

%================================================================

\section*{Appendix D - Improved upper boundary condition for the radiative transfer equation}
\addcontentsline{toc}{section}{Appendix D: Improved upper boundary condition for the radiative transfer equation}

We have developed an improved upper boundary condition for the radiative transfer equation. Actually,
it was already mentioned in the original paper on {\sc tlusty} [Hubeny (1988), Eq. (3.8)], but in 
subsequent versions it was deemed unimportant and disabled. It turned out that in the context of
cool stars it plays a role in determining the correct temperature at the surface.

The traditional boundary condition is based on the assumption that there is no incoming radiation
at the surface, or that there is a prescribed irradiation from an external source, e.g., in the case of a close
binary, or a planetary atmosphere. Here we consider the case of no prescribed external
irradiation. so that the incoming radiation is
\begin{equation}
\label{ubc1}
I_\nu^-(\mu) = 0,
\end{equation}
where $I_\nu^-$ denotes $I_\nu(\mu)$ at frequency $\nu$ for $\mu<0$; 
$\mu$ is the cosine of the angle between the direction of 
photon propagation and the normal to the surface. However, Eq. (\ref{ubc1}) is strictly valid
only at optical depth $\tau=0$, while in reality the region above the first depth point, with column mass $m_1$,
is not empty and can thus generate some radiation coming from directions with $\mu<0$ at the uppermost
depth point. It can be approximated, assuming that the source function is constant above $m_1$, as (dropping the
frequency subscript)
\begin{equation}
I^-(\tau_1,\mu) = \int_0^{\tau_1} S(t)\, e^{-t/\mu} dt/\mu = S(\tau_1) (1 - e^{-\tau_1/\mu}).
\end{equation}
Assuming that the opacity is a linear function of  $m$ for $m<m_1$, then
\begin{equation}
\tau_1 = (1/2) \chi(m_1) m_1/\rho_1,
\end{equation}
where $\chi$ is the absorption coefficient (per cm), and $\rho_1\equiv \rho(m_1)$ is the mass density at the
uppermost point.
The discretized upper boundary condition for the Feautrier variable $j$ [where 
$j(\mu)\equiv (1/2)[I(\mu) + I(-\mu)]$ for $\mu >0$ -- for details, refer e.g., 
to Hubeny \& Mihalas (2014), \S\,11.6] can be written as 
\begin{equation}
j_1 \left[ \frac{2\mu^2}{\Delta\tau_{3/2}^2} + \frac{2\mu}{\Delta\tau_{3/2}} +1 \right] - j_2
\frac{2\mu^2}{\Delta\tau_{3/2}^2}  = S_1 \left[ 1 + \frac{2\mu}{\Delta\tau_{3/2}} \left(1-e^{-\tau_1/\mu}\right) \right],
\end{equation}
where $S$ is the source function.
Here, the last term in the square bracket on the right-hand side is the additional term due to the
contribution of the upper layers to the radiation intensity at the uppermost point of the atmosphere.

By integrating over $\mu$ one gets a zero-order moment equation, which is one of the basic structural equations
to describe an atmosphere,
\begin{equation}
J_1 \left[\frac{2 f_1}{\Delta\tau_{3/2}^2} + \frac{2 g_1}{\Delta\tau_{3/2}} +1\right] - J_2 \frac{2 f_2}{\Delta\tau_{3/2}^2} =
S_1 [1 + 2 Q(\tau_1)/\Delta\tau_{3/2}]
\end{equation}
where $f$ and $g$ are the variable Eddington factors, and
\begin{equation}
Q(t) \equiv \int_0^1 (1-e^{-t/\mu}) \mu\, d\mu = \frac{1}{2} - E_3(t),
\end{equation}
where $E_3(t)$ is the third exponential integral. 

The reason why this correction is particularly significant in the context of cool star model atmospheres
is that the monochromatic opacity can be vastly different for different frequencies. For instance, the opacity
in the Lyman continuum and in the Lyman lines is many orders of magnitude larger than the opacity elsewhere
(since hydrogen is predominantly in the ground state), so that even if one selects a relatively low smallest
Rosseland optical depth used to construct the initial LTE-gray model, the optical depth in the Lyman lines and the
Lyman continuum may be quite large. For instance, in our test models for a solar type star (see \S\,\ref{teststlus})
we chose the Rosseland optical depth at the first depth point, the keyword parameter TAUFIR, equal to $10^{-7}$,
yet the monochromatic 
optical depth close to the Lyman limit is about 700 (!), and the optical depth for all frequencies up to the L$\alpha$
line is still larger than or close to unity. 
It is thus clear that the contribution of layers above $m_1$ cannot be neglected.

%================================================================

\section*{Appendix E - Treatment of hydrostatic equilibrium within the Rybicki scheme}
\addcontentsline{toc}{section}{Appendix E: Treatment of hydrostatic equilibrium within the Rybicki scheme}
 
We stress that when using the Rybicki scheme as a global linearization procedure, only the radiative
transfer and the radiative/convective equilibrium equations are solved simultaneously, and thus only the 
mean intensities of radiation and the temperature follow directly from the iteration process. Other
structural equations are to be solved, and hence the updated values of other structural parameters 
are to be determined, separately.

We have already described our treatment of the kinetic equilibrium equations in the case of computing NLTE
models - see \S\,\ref{ot/nlte}. Here we briefly outline our adopted treatment of the hydrostatic equilibrium
equation, and the way of updating the gas pressure and consequently the total particle number density.
The next step is to solve the equation of state to determine number densities of all atoms, ions, molecules, 
and electrons.

While the equation of state is local, and thus its treatment is the same for stellar atmospheres and
disks, the hydrostatic equilibrium equation differs for these two cases, so we will discuss stellar atmospheres and 
disks separately.

\subsection*{Stellar atmospheres}
In this case the adopted procedure depends on whether the radiation pressure is neglected
(as may be the case for cool stars), or is taken into account.  In the case
of negligible radiation pressure, there is in fact no procedure necessary. The total pressure is equal
to the gas pressure, and is given by
\begin{equation}
P_{\rm gas} = P_{\rm tot} = m g,
\end{equation}
where $m$ is the column mass and $g$ the gravity acceleration, assumed constant throughout  the atmosphere for
our 1-D plane-parallel models. Since $m$ is taken as the basic geometrical coordinate, the hydrostatic equilibrium
is fulfilled identically. The total particle number density is then given, assuming the ideal gas equation of state, by
\begin{equation}
\label{nkt}
N = P/kT,
\end{equation}
where $k$ is the Boltzmann constant and $T$ the temperature.

If the radiation pressure is taken into account, then the hydrostatic equilibrium equation reads
\begin{equation}
\label{heqr}
dP_{\rm gas}/dm = gm - dP_{\rm rad}/dm.
\end{equation} 
For the purpose of determining the gas pressure, we need to estimate the response of
the gradient of the radiation pressure to the new temperature. 
This is estimated as follows: Assuming that $P_{\rm rad} \propto T^4$
(i.e., it follows its equilibrium value, although it does not have to be equal to it exactly), then
\begin{equation}
\label{pradnew}
P_{\rm rad}^{\rm new} =  P_{\rm rad}^{\rm old} (1+ 4\,\Delta T/T^{\rm new}),
\end{equation}
where $P_{\rm rad}^{\rm old}$ is the radiation pressure from the previous iteration, and 
$\Delta T = T^{\rm new} - T^{\rm old}$ is the change of temperature determined by the Rybicki scheme. 
We also note that the
gradient of the radiation pressure is generally given by
\begin{equation}
dP_{\rm rad}/dm \equiv g_{\rm rad} = (4\pi/c)\int_0^\infty\!\! K_\nu d\nu = (4\pi/c)\int_0^\infty\!\! \chi_\nu H_\nu d\nu,
\end{equation}
where $H_\nu$ and $K_\nu$ are the first and the second moment of the specific intensity of radiation,
respectively, and $\chi_\nu$ is the total absorption (extinction) coefficient. An update of $g_{\rm rad}$ follows
the same relation as $P_{\rm rad}$ in Eq. (\ref{pradnew}). The discretized form of Eq. (\ref{heqr}) is then,
denoting $P_{\rm gas}(m_d)$ as $P_d$,
\begin{eqnarray}
\label{patm}
P_1 &=& m_1 [g - g_{{\rm rad},1}  (1+ 4\Delta T_1/T_1)] \\
P_d &=& P_{d-1} + g(m_d - m_{d-1}) - g_{{\rm rad}, d} (1 + 4\Delta T_d/T_d],\quad d=2,\ldots,ND.\nonumber 
\end{eqnarray}
Equation (\ref{patm}) is a simple recurrence relation, so the determination of the new gas pressure is straightforward
and simple.

\subsection*{Accretion disks}
The situation is more complicated in the case of disks, because the gravity acceleration is proportional 
to the vertical distance from the central plane, $z$,
\begin{equation}
\label{qdef}
g(z) = Q z, \quad Q\equiv GM_\ast/R^3,
\end{equation}
where G is the gravitational constant, $M_\ast$ is the mass of the central star, and $R$ the radial distance from
the central star. We use here for simplicity a non-relativistic expression for $Q$. 
For a general relativistic case, which can also be treated by {\sc tlusty}, refer to Paper~II, \S\,2.2.1. In any case,
$Q$ can be viewed as one of the basic input parameters for a given annulus. 

In {\sc tlusty}, the column mass is a basic geometrical coordinate, while $z$ is taken as one of the
structural parameters to be determined by the model. It is given by the so-called $z$-$m$ relation.
\begin{equation}
\label{zm}
dm = -\rho dz,
\end{equation}
where $\rho$ is the mass density. The minus sign follows from the convention that $z$ is taken as 0 at 
the central plane and increases upward, while $m$ increases from the top downward. 

To update the gas pressure, density, and the vertical distance for a new $T$,
we use an
approach suggested by Hubeny (1990). Here the dependence of $g$ on $z$ is accounted for by
differentiating Eq. (\ref{heqr}) once more over $m$, and using Eq. (\ref{zm}) to obtain a second-order equation
for $P_{\rm gas}$, namely
\begin{equation}
\label{p2}
\frac{d^2\! P_{\rm gas}}{dm^2} = - \frac{Q}{\rho} - \frac{d^2\! P_{\rm rad}}{dm^2}.
\end{equation}
We then express $\rho$ using a relation 
\begin{equation}
P_{\rm gas} = c_s(T)^2 \rho = \gamma T  \rho.
\end{equation}
Here, the first equality expresses the mass density through the isothermal sound speed, while the
second equality takes the square of the sound speed proportional to the temperature. Equation (\ref{p2})
is then written as (writing $P_{\rm gas}$ simply as $P$)
\begin{equation}
\label{prad22}
P \frac{d^2 P}{dm^2} = -\gamma T Q - \frac{d^2 P_{\rm rad}}{dm^2}.
\end{equation}
where the quantities $\gamma$ and $d^2 P_{\rm rad}/dm^2$ are taken from the previous iteration step.
We have experimented with using Eq. (\ref{pradnew}) to update the radiation pressure in Eq. (\ref{prad22}), 
but this procedure proved to be unstable in some cases. In fact, the reason why using 
$d^2P_{\rm rad}/dm^2$ from the previous iteration step is preferrable
follows from the observation that this quantity does not change very much from iteration to iteration. It is seen
by expressing
\begin{equation}
\frac{d^2 P_{\rm rad}}{dm^2} = \frac{d g_{\rm rad}}{dm} = 
\frac{4\pi}{c} \frac{d}{dm}\left( \int_0^\infty\!\! \chi_\nu H_\nu d\nu \right)
\approx \frac{4\pi}{c} \frac{d(\chi_{\rm Ross} H)}{dm} 
\end{equation}
where $\chi_{\rm Ross}$ is the Rosseland mean opacity, and $H$ the total flux. $H$ does not change from iteration
to iteration, and $\chi_{\rm Ross}$ changes only a little, so our approach of keeping the second derivative of
the radiation pressure from the previous iteration step is quite reasonable.

The lower boundary condition follows from expressing the pressure at depth $D-1$ ($D$ being the 
index of the last depth point corresponding to the midplane) through a Taylor expansion (writing $m_D$
as $M$)
\begin{equation}
\label{hubc}
P(m_{D-1}) = P(M) + (m_{D-1} - M) P^\prime (M) + (1/2) (m_{D-1} - M)^2 P^{\prime\prime}(M),
\end{equation}
where $P^\prime \equiv dP/dm$. Here, $P^\prime(M)=0$ because of the symmetry of the disk around the midplane.
The lower boundary condition follows from substituting  the second derivative of $P$ 
from Eq. (\ref{p2}) to Eq. (\ref{hubc}). 

The upper boundary condition is more complicated. It was  derived by Hubeny (1990); here we present only the
final result:
\begin{equation}
\rho_1 = \frac{m_1}{H_g f_1}.
\end{equation}
where
\begin{equation}
f_1 = f\left(\frac{z_1-H_r}{H_g}\right)\!, \quad {\rm with}\quad\quad f(x) \equiv \frac{\sqrt{\pi}}{2} \exp(x^2)\, {\rm erfc}(x),
\end{equation}
and
\begin{equation}
H_g = \left(\frac{2\gamma_1 T_1}{Q}\right)^{1/2}\!\!\!, \quad {\rm and} \quad H_r = \frac{g_{{\rm rad}, 1}}{Q}.
\end{equation}
Upon discretizing, one obtains a set of non-linear algebraic equations for $P_d$, viz,
\begin{eqnarray} 
\label{trip}
&P_1 - \frac{m_1}{f_1} \left(\frac{\gamma T_1 Q}{2}\right)^{1/2} = 0, \nonumber \\
&a_d P_{d-1}P_d  + c_d P_{d+1}P_d  - b_d P_d^2  + q_d P_d  + \gamma_d T_d Q = 0 , \quad d=2,\ldots,D-1, \nonumber \\
&P_{D-1} P_D \frac{2}{\Delta m_{D-1/2}^2} - P_D^2 \frac{2}{\Delta m_{D-1/2}^2} + \gamma_D T_D Q = 0,
\end{eqnarray}
where $q_d \equiv (d^2 P_{\rm rad}/dm^2)_d$, and
\begin{eqnarray}
a_d &=& \frac{1}{\Delta m_{d-1/2}\Delta m_d}, \\
c_d &=& \frac{1}{\Delta m_{d+1/2}\Delta m_d},\\
b_d &=& a_d + c_d,
\end{eqnarray}
for $d=2,\ldots, D-1$, with
\begin{equation}
\Delta m_{d\pm 1/2} = |m_d - m_{d\pm 1}|, \quad \Delta m_d = (m_{d+1/2} - m_{d-1/2})/2.
\end{equation}
Set of non-linear equations (\ref{trip}) is solved by the standard Newton-Raphson method. 

\subsection*{Equation of state}

Having determined the gas pressure corresponding to the new temperature that was determined by the current
iteration of the global linearization (Rybicki) scheme, the subsequent procedure is the same for both atmospheres and disks.
One first determines the total particle number density from the known temperature and the gas pressure using
Eq. (\ref{nkt}). Then, one solves iteratively for the number densities of all components of the gas, as described in
Paper~II, \S\,2.7, with updates in the molecular equation of state presented here in \S\,\ref{eostlu}.

%===================================================================

\section*{Appendix F - Working with EXOMOL line broadening parameters}
\addcontentsline{toc}{section}{Appendix F: Working with EXOMOL line broadening parameters}
Originally, we have generated H${}_2$O line data based on data from the Kurucz website{\footnote
{http://kurucz.harvard.edu/molecules/h2o/}}, which contains data from Partridge \& Schwenke (1997). 
Since now more modern data are available, in particular from the
EXOMOL project{\footnote{http://www.exomol.com/}},
we plan to employ all their data to generate {\sc synspec}-compatible line lists.
So far, we have prepared line  lists for H${}_2$O 
based on EXOMOL data{\footnote{http://www.exomol.com/data-rtypes/linelist/H2O/1H2-16O.POKAZATEL/}}  
(Barton et al. 2017a,b), including
improved line broadening parameters, which is usually the most uncertain part of any molecular line list.

In their formalism, the pressure broadening is described by a Lorentzian,
\begin{equation}
f_L(\bar\nu,\bar\nu_0, \bar\gamma) = \frac{\bar\gamma}{\pi} \frac{1}{(\bar\nu - \bar\nu_0)^2 + \bar\gamma^2},
\end{equation}
where $\bar\nu$ is the actual wavenumber, $\bar\nu_0$ is the wavenumber of the line center, and the $\bar\gamma$ is 
the Lorentzian half-width, expressed in wavenumbers. 
Since {\sc synspec} works in terms of frequencies, denoted here as $\nu$ without bar, then the profile coefficients in both 
units should satisfy
\begin{equation}
f(\nu)\, d\nu = f(\bar\nu)\, d\bar\nu, \quad {\rm where}\quad \bar\nu=\nu/c,
\end{equation}
and thus 
\begin{equation}
f_L(\nu) = f_L(\bar\nu) (d\bar\nu/d\nu) = f_L(\bar\nu)/c.
\end{equation}
Consequently,
\begin{equation}
f_L(\nu) = 
\frac{\bar\gamma}{\pi}\, \frac{1}{[(\nu - \nu_0)/c]^2 + \bar\gamma^2}\,\frac{1}{c} =
\frac{c \bar\gamma}{\pi}\,\frac{1}{(\nu-\nu_0)^2 + (c\bar\gamma)^2},
\end{equation}
and so the broadening parameter used by {\sc synspec}, denoted by $\Gamma_{\rm vdw}$ is
\begin{equation} 
\Gamma_{\rm vdW}= c \bar\gamma.
\end{equation}

The pressure broadening coefficients of Barton et al. (2017a,b) are given by\footnote{In fact, 
the correct expression is Eq. (2) of Barton et al. (2017b); while there is a typo in Eq. (4) of Barton et al. (2017a).}
\begin{equation}
\label{gam1}
\bar\gamma = \sum_b \bar\gamma_{{\rm ref},b}\,  \left( \frac{T_{\rm ref}}{T}\right)^{\!\!n_b} \left(\frac{P_b}{P_{\rm ref}}\right) 
\end{equation}
where $T$ is the temperature, $T_{\rm ref}= 296$ K is the reference temperature, 
$P_{\rm ref} = 1\, \rm{bar} = 10^{6}$ in cgs units is the reference pressure, 
$\bar\gamma_{{\rm ref},b}$ and $P_b$ are  the reference broadening parameter
and the reference partial pressure, respectively, of the broadener $b$.  The broadeners are typically H${}_2$ and He.
EXOMOL tables provide $\bar\gamma_{{\rm ref},b}$ for these two considered broadeners $b$. 

{\sc synspec} works in terms of number densities, related to the partial pressures as
\begin{equation}
\label{partprs}
P_b = N_b kT,
\end{equation} 
and, therefore, the pressure broadening parameter to be used 
by {\sc synspec} based on the EXOMOL calculations, is then given by
\begin{equation}
\Gamma_{\rm vdw} = 10^{-6}\, ckT\, \sum_b (296/T)^{n_b}\, \bar\gamma_{{\rm ref},b}\, N_b.
\end{equation}
Finally, the complete profile is given through the usual Voigt function  $H(a,x)$ that represents a convolution 
of the Lorentz profile for the radiation and Van der Waals (pressure) broadening, and the Doppler (thermal)
broadening.
Here
\begin{equation}
a= (\Gamma_{\rm rad}/4\pi + \Gamma_{\rm vdw})/\Delta\nu_D,  \quad\quad x=(\nu-\nu_0)/\Delta\nu_D,
\end{equation}
where $\Delta\nu_D$ is the Doppler width, 
\begin{equation}
\Delta\nu_D = \frac{\nu_0}{c} \sqrt{\frac{2kT}{M} + v_{\rm tb}^2}.
\end{equation}
where $M$ is the mass of the radiating particle, and $v_{\rm tb}$ the microturbulent velocity.

%===================================================================

\section*{Appendix G - Reducing the number of lines in a molecular line list}
\addcontentsline{toc}{section}{Appendix G: Reducing the number of lines in a molecular line list}
Within {\sc synspec}, weak lines are eliminated by computing the ratio of the opacity in the lines center
and the opacity in the continuum,. If the ratio is smaller than a chosen parameter (RELOP; typically taken as 
$10^{-4}$), the line is rejected. But its parameters were still read from the line list. As explained earlier,
it is desirable to reasonably eliminate weak lines from the line list from the outset, even before they are
read by {\sc synspec}. We outline a procedure below that works reasonably well for this purpose.

The opacity in the line center of a line between levels $l$ and $u$ of the molecular species ``mol"
is proportional to
\begin{equation}
\kappa(\nu_0) \propto (N_{\rm mol}/U_{\rm mol}) f_{lu} g_{lu} \exp(-E_l/kT)/\Delta\nu_D,
\end{equation}
where $N$ is the number density  and $U$ the partition function, $f_{lu}$ the oscillator strength,
$g_l$ and $E_l$ are the statistical weight and the energy of the lower level, respectively, and
$\Delta\nu_D$ the Doppler width. The strength parameter of a line $l \leftrightarrow u$,
defined as a logarithm of the line strength (without additive constants). can then be written as
\begin{equation}
s_{lu} = \log(N/U) + \log(gf) - 0.625 E_l/T + \log\lambda_0,
\end{equation}
(we skipped subscript ``mol" at $N$ and $U$),
where $E_l$ is assumed to be expressed in cm${}^{-1}$. Here, we have neglected the dependence of the
Doppler width on the mass of the molecule, which is a reasonable approximation since the true thermal
velocity is often smaller than the microturbulent velocity, so that $\Delta\nu_D \propto \nu_0$, and thus
$1/\Delta\nu_D \propto \lambda_0$.

The strength parameter $s$ still depends on the temperature (through the term $E_l/T$), as well as on
density, through $\log(N_{\rm mol}/U_{\rm mol})$. In order to avoid the need of having tailored
limited line lists for different temperatures and densities, we adopt the following procedure:
We select a ``characteristic" molecular species, for instance H${}_2$O, at a characteristic wavelength,
say 1$\mu$. We then find, by extensive testing, what is the minimum strength parameter of a line
of a characteristic species around this characteristic wavelength. In other words. it is necessary
to find what is $s_{lu}$ of a line to contribute to the total opacity
in an appreciable way, or, analogously, that the synthetic spectrum computed with and without this
line being included differs by less than some chosen degree of accuracy, say 0.1\%. We formulate
a rejection criterion for a characteristic species in such a way that the line is rejected if
\begin{equation}
\label{rej}
s \equiv \log gf - 0.625 E_l/T_{\rm char} + \log(\lambda/\lambda_{\rm char}) < s_{\rm reject} .
\end{equation}
For instance, selecting $T_{\rm char}=3000$ K and $\lambda_{\rm char}=1\mu$, the value
of $s_{\rm reject}$ that gave a reasonable accuracy of the predicted spectrum was about $-9$. Selecting
lower values of $s_{\rm reject}$ yields a limited line list with more lines, but without yielding a significant
difference in the predicted spectra.

To construct an analogous rejection criterion for a list that contains lines of several molecular species, we modify
Eq. (\ref{rej}) to read
\begin{equation}
s \equiv \log(N/U)_{\rm max} - \log(N/U)_{\rm char} + \log gf - 0.625 E_l/T_{\rm char} + 
\log(\lambda/\lambda_{\rm char}) < s_{\rm reject},
\end{equation}
where $\log(N/U)_{\rm char}$ is the $N/U$ value for the characteristic species at the characteristic temperature, 
and for the density where its concentration at this temperature is highest (typically $\rho\approx 10^{-6}$ g cm${}^{-3}$).
For water, and for the above specified $T$ and $\rho$ this value is $\log(N/U)^{\rm char}\approx 10$.
In order to find  $\log(N/U)_{\rm max}$, we first compute the values of $N/U$ for a given molecule and for a number
of temperatures and densities, and set the value of $T_{\rm char}$ to be the temperatures where 
$N/U$ reaches the maximum,
and take this $N/U$ as $(N/U)^{\rm max}$. On this way, we can account for the fact that various molecules
have various concentrations and various partition functions under different conditions, and that they are not
spuriously disregarded by choosing the temperature and density where they are not formed in significant
amounts.

It should be kept in mind that this procedure is an approximate one, so the rejection parameter
$s_{\rm reject}$ should be chosen conservatively, i.e. to be reasonably small. Our tests showed
that the value $s_{\rm reject} \approx -8.5$ is a good and conservative choice for diatomic molecules.  
When computing synthetic spectra using a full list for diatomic molecules (with about 79.6 million lines),
and a list reduced with $s_{\rm reject} = -8.5$ (with just 6 millions lines), the relative difference between
the spectra for both a cool M-star model and a solar-type model, is around or less than 0.2\%.
The same conclusion can be drawn for TiO and water.

%%%%%%%%%%%%%%%%%%%%%%%%%%%%%%%%%%%%%%%%

\section*{References}
\addcontentsline{toc}{section}{References}

\def\reference{\par \leftskip20pt \parindent-20pt\parskip4pt}
\noindent

\reference Allende Prieto, C., Lambert, D. l.. Hubeny, I., \& Lanz, T., 2003, ApJS, 147, 363.

\reference Asplund, M., Grevesse, N., \& Sauval, A., 2005, ASP Conf. Ser. 336, 25.

\reference Asplund, M., Grevesse, N.,  Sauval, A., \& Scott, P., 2009, Ann. Rev. Astron. Astrophys., 47, 481.

\reference Barklem, P. S., 2016, Astron. Astrophys. Rev., 24, 9.

\reference Barklem, P.S., \& Collet, R., 2016, Astron. Astrophys., 588, 96.  

\reference Barnard, A. J., Cooper, J., \& Smith, E. W., 1974, JQSRT, 14, 1025.

\reference Barton, E. J., Hill, C., Yurchenko, S. N., Tennyson, J., Dudaryonok, A. S., \& Lavrentieva, N. N., 
2017a, JQSRT, 187, 453 (BHYTDL).

\reference Barton, E. J., Hill, C.,  Czurylo, M., Li, H. Y., Hyslop, A, Yurchenko, S. N.,\&  Tennyson, J., 2017b, 
JQSRT, 203, 490.

\reference Bell, K. L., 1980, J. Phys. B, 13,1859.

\reference Borysow, A., Jorgensen, U.G., \& Fu, Y., 2001, JQSRT 68, 235

\reference Cunha, K., Hubeny, I., \& Lanz, T. 2006, ApJ, 647, L143.

\reference Grevesse, N., Asplund, M., \& Sauval, A., 2007, Space Sci. Rev., 130,105. 

\reference Grevesse, N., \& Sauval, A., 1998, Space Sci. Rev., 85, 161. 

\reference Griem, H., 1974, Spectral Line Broadening by Plasmas, New York: Academic Press. 

\reference Gustafsson, B., Edvardsson, B., Eriksson, K., Jorgensen, U. G., Nordlund, \AA, \& Plez, B., 2008,
Astron. Astrophys., 486, 951.

\reference Gustafsson, M., \& Frommhold, L., 2001, ApJ 546, 1168.

\reference Gustafsson, M., \& Frommhold, L., 2003, Astron. Astrophys., 400, 1161.

\reference Hubeny, I., 1988,  Computer Phys. Comm., 52, 103.

\reference Hubeny, I. 1990, ApJ, 351, 632.

\reference Hubeny, I., Hummer, D.G., \& Lanz, T. 1994, Astron. Astrophys., 282, 151.

\reference Hubeny, I., \& Lanz, T. 2017a, A Brief Introductory Guide to
TLUSTY and SYNSPEC, arXiv:1706.01859  (Paper~I).

\reference Hubeny, I., \& Lanz, T. 2017b, TLUSTY User's Guide II:
Reference Manual, arXiv:1706.01935   (Paper~II).

\reference Hubeny, I., \& Lanz, T. 2017c, TLUSTY User's Guide III:
Operational Manual, arXiv:1706.01937   (Paper~III).

\reference Hubeny, I., \& Mihalas, D., 2014, Theory of Stellar Atmospheres,
Princeton: Princeton Univ. Press.

\reference Husser, T. O., Wende von Berg, S., Dreizler, S., Homeier, D., Reiners, A., Barman, T., 
\& Hauschildt, P. H., 2013, Astron, Astrophys, Suppl., 553, 6.

\reference Irwin, A.W. 1981, ApJS, 45, 621.

\reference Jorgensen, U.G., Hammer, D., Borysow, A., \& Falkesgaard, J., 2000, Astron. Astrophys.,  361, 283

\reference Kurucz, R.L. 1970, SAO Spec. Rep. 309.

\reference Lanz, T., \& Hubeny, I., 2003, ApJS, 146, 417.

\reference Lanz, T., \& Hubeny, I., 2007, ApJS, 169, 83.

\reference Lemke, M., 1997, A\&AS, 122, 285.

\reference M\'esz\'aros, Sz., Allende Prieto, C., Edvardsson, B., Castelli, F., Garcia P\'erez, A. E., 
Gustafsson, B., Majewski, S. R., Plez, B., Schiavon, R., Shetrone, M.,  \& de Vicente, A., 2012, AJ, 144, 120.

\reference Osorio, Y., Allende Prieto, C., Hubeny, I., M\'esz\'aros, Sz., \& Shetrone, M., 2020,
Astron. Astrophys., 637, 17.

\reference Partridge, H., \& Schwenke, D. W., 1997, J. Chem. Phys., 106, 4618.

\reference Rybicki, G. B., \& Hummer, D. G., 1991, Astron. Astrophys., 245, 171.

\reference Rybicki, G. B., \& Hummer, D. G., 1992, Astron. Astrophys., 262, 209.

\reference Schoening, T., \& Butler, K. 1989, A\&AS, 78, 51.

\reference Shamey, L.1969, PhD thesis, University of Colorado.

\reference Tremblay, P.-E., \& Bergeron, P., 2009, ApJ, 696, 1755.

\end{document}